\documentclass[a4paper,fleqn,usenatbib]{mnras}
\usepackage{newtxtext,newtxmath}

\usepackage[T1]{fontenc}
\usepackage{ae,aecompl}

\usepackage{graphicx}	
\usepackage{amsmath}	
\usepackage{amssymb}	
\usepackage[compatibility=false]{caption}
\usepackage{subcaption}
\usepackage{array}

\title[47 Tuc X9 -- optical spectrum and timing]{HST spectrum and timing of the ultra-compact X-ray binary candidate 47 Tuc X9}

\author[V. Tudor et al.]{V. Tudor,$^1$\thanks{E-mail: vlad.tudor@postgrad.curtin.edu.au}
J. C. A. Miller-Jones,$^{1}$
C. Knigge,$^{2}$
T. J. Maccarone$^{3}$,
T. M. Tauris,$^{4,5}$
\newauthor
A. Bahramian$^{6,7}$,
L. Chomiuk$^{6}$,
C. O. Heinke$^{7}$,
G. R. Sivakoff$^{7}$,
J. Strader$^{6}$,
\newauthor
R. M. Plotkin,$^{1}$
R. Soria$^{1,8}$,
M. D. Albrow$^{9}$,
G. E. Anderson$^{1}$,
M. van den Berg$^{10,11}$,
\newauthor
F. Bernardini$^{12,13}$,
S. Bogdanov$^{14}$,
C. T. Britt$^{6}$,
D. M. Russell$^{13}$,
D. R. Zurek$^{15}$
\\
\\
$^{1}$International Centre for Radio Astronomy Research -- Curtin University, GPO Box U1987, Perth, WA 6845, Australia
\\
$^{2}$Department of Physics and Astronomy, University of Southampton, Southampton, SO17 1BJ, UK
\\
$^{3}$Department of Physics, Texas Tech University, Box 41051, Lubbock, TX 79409--1051, USA
\\
$^{4}$Max-Planck-Institut f\"{u}r Radioastronomie, Auf dem H\"{u}gel 69, 53121 Bonn, Germany
\\
$^{5}$Argelander-Institut f\"{u}r Astronomie, Universit\"{a}t Bonn, Auf dem H\"{u}gel 71, 53121 Bonn, Germany
\\
$^{6}$Department of Physics and Astronomy, Michigan State University, East Lansing, MI 48824, USA
\\
$^{7}$Department of Physics, University of Alberta, CCIS 4--183, Edmonton, AB T6G 2E1, Canada
\\
$^{8}$National Astronomical Observatories, Chinese Academy of Sciences, Beijing 100012, China
\\
$^{9}$Department of Physics and Astronomy, University of Canterbury, Private Bag 4800, Christchurch, New Zealand
\\
$^{10}$Harvard-Smithsonian Center for Astrophysics, 60 Garden Street, Cambridge, MA 02138, USA
\\
$^{11}$Anton Pannekoek Institute for Astronomy, University of Amsterdam, Science Park 904, 1098 XH Amsterdam, The Netherlands
\\
$^{12}$Istituto Nazionale di Astrofisica -- Osservatorio Astronomico di Roma, via Frascati 33, 00040 Monteporzio Catone (Roma), Italy
\\
$^{13}$New York University Abu Dhabi, PO Box 129188, Abu Dhabi, UAE
\\
$^{14}$Columbia Astrophysics Laboratory, Columbia University, 550 West 120th Street, New York, NY 10027, USA
\\
$^{15}$Department of Astrophysics, American Museum of Natural History, Central Park West and 79th Street, New York, NY 10024--5192, USA
}

\date{Accepted XXX. Received YYY; in original form ZZZ}

\pubyear{2017}

\begin{document}
\label{firstpage}
\pagerange{\pageref{firstpage}--\pageref{lastpage}}
\maketitle


\begin{abstract}
To confirm the nature of the donor star in the ultra-compact X-ray binary candidate 47\,Tuc~X9, we obtained optical spectra (3,000--10,000{\AA}) with the {\it Hubble Space Telescope} / {\it Space Telescope Imaging Spectrograph}. We find no strong emission or absorption features in the spectrum of X9. In particular, we place $3\sigma$ upper limits on the H\,$\alpha$ and He\,{\sc ii} $\lambda 4686$ emission line equivalent widths $\rm -EW_{\rm H \alpha} \lesssim 14$\,{\AA} and $\rm -EW_{He \textsc{ii}} \lesssim 9$\,{\AA}, respectively. This is much lower than seen for typical X-ray binaries at a similar X-ray luminosity (which, for $L_{\rm 2-10\,keV} \approx 10^{33}-10^{34}$\,erg\,s$^{-1}$ is typically $\rm -EW_{\rm H \alpha} \sim 50$\,{\AA}). This supports our previous suggestion (by Bahramian et al.) of an H-poor donor in X9. We perform timing analysis on archival far-ultraviolet, $V$ and $I$-band data to search for periodicities. In the optical bands we recover the seven-day superorbital period initially discovered in X-rays, but we do not recover the orbital period. In the far-ultraviolet we find evidence for a 27.2 min period (shorter than the 28.2 min period seen in X-rays). We find that either a neutron star or black hole could explain the observed properties of X9. We also perform binary evolution calculations, showing that the formation of an initial black hole / He-star binary early in the life of a globular cluster could evolve into a present-day system such as X9 (should the compact object in this system indeed be a black hole) via mass-transfer driven by gravitational wave radiation.
\end{abstract}




\section{Introduction}

The high interaction rates in globular clusters, especially at their centres, lead to the efficient formation of exotic binaries like accreting compact objects -- cataclysmic variables (white dwarfs) and X-ray binaries (neutron stars and black holes; \citealp{1975ApJ...199L..93C}). Soon after the first X-ray missions, it was recognised that X-ray transients in globular clusters were disproportionately associated with neutron stars, with no confirmed black holes \citep{1995A&A...300..732V}, a fact that remains true to this day \citep{2014ApJ...780..127B}. In contrast, about a third of the X-ray transients in the rest of the Galaxy contain black holes \citep{2015MNRAS.453.3918M, 2016ApJS..222...15T}. Originally it was proposed that due to mutual gravitational interactions that all black holes would have been ejected from globular clusters \citep{1993Natur.364..421K, 1993Natur.364..423S}. However, there are several recent indications that a number of black holes may still exist within clusters, based on X-ray observations of extragalactic clusters \citep{2007Natur.445..183M, 2010ApJ...721..323S, 2010ApJ...725.1805B, 2010ApJ...712L...1I, 2011MNRAS.410.1655M}, radio/X-ray observations of Galactic clusters \citep{2007A&G....48e..12M, 2012Natur.490...71S, 2013ApJ...777...69C, 2015MNRAS.453.3918M}, as well as theoretical simulations \citep{2013MNRAS.430L..30S, 2015ApJ...800....9M, 2016MNRAS.462.2333P}. The reason why these black holes are so elusive could perhaps be due to the nature of the X-ray binaries they form, rather than their number. If the majority of accreting black holes in globular clusters have very low-mass, possibly degenerate, donors, they will have short, faint outbursts, making them undetectable by current and past all-sky X-ray surveys \citep{2014MNRAS.437.3087K}. Indeed, the dynamics in globular clusters are thought to effectively produce ultra-compact X-ray binaries \citep{1987ApJ...312L..23V, 1996ApJ...472L..97D, 2000ApJ...530L..21D, 2005ApJ...621L.109I, 2010ApJ...717..948I}, in which a black hole or neutron star accretes matter from a H-poor donor in a system with a very short orbital period ($\lesssim 1$ hour; \citealp{1986ApJ...304..231N}).

In globular clusters, binaries comprised of a white dwarf secondary and a neutron star or black hole primary most frequently form via exchange encounters or physical collisions \citep{2008MNRAS.386..553I, 2010ApJ...717..948I}. Once the white dwarf fills its Roche lobe, rapid mass transfer is initiated. If stable, the binary becomes an ultra-compact X-ray binary \citep{2012A&A...537A.104V, 2017MNRAS.470L...6S}. Due to the orbit shrinking via gravitational wave emission, a neutron star accreting from a He white dwarf spends $10^7$--$10^8$ years as a persistent source, after which it becomes transient, undergoing occasional outbursts \citep{2012A&A...537A.104V}. During this mass-transfer stage, mass loss from the donor leads to orbital expansion and thus a decrease in the mass-accretion rate.

A number ($\sim$20) of confirmed and candidate ultra-compact X-ray binaries have been discovered in the Galaxy
(\citealp{2007A&A...465..953I}, \citealp{2010NewAR..54...87N}). Most of these Galactic sources produce thermonuclear X-ray bursts, implying accretion on to a neutron star, and persistently accrete at greater than 1\% of the Eddington luminosity. Their optical spectra lack Balmer emission or absorption features, with the most prominent emission lines being from He, C, N and O \citep{2004MNRAS.348L...7N, 2006MNRAS.370..255N, 2006A&A...450..725W}, on top of a hot, faint continuum from a small accretion disc. In general, the C and O emission lines are faint, with an equivalent width (line to continuum ratio) $\rm -EW \lesssim 4$\,{\AA}. In 4U\,0614+091 ($L_{\rm 2-10\,keV} \approx 10^{36}$\,erg\,s$^{-1}$; \citealp{2010ApJ...710..117M}), however, the Bowen florescence C\,{\sc iii} feature at $\lambda \approx 4650$\,{\AA} has $\rm -EW \approx 10$\,{\AA} \mbox{\citep{2006MNRAS.370..255N}}. The N{\sc iii} feature at $\lambda \approx 4515$\,{\AA} is similarly strong ($\rm -EW = 4-15$\,{\AA}) in four out of five ultra-compacts studied by \mbox{\citet{2006MNRAS.370..255N}} (including three persistent sources, and one near the peak of the outburst at $L_{\rm X} \approx 0.01\,L_{\rm Edd}$). The identification of these metal lines, together with an absence of Balmer lines, is an avenue for selecting new candidate ultra-compact binaries. In H-rich X-ray binaries, \citet{2009MNRAS.393.1608F} found that the equivalent width of the H\,$\alpha$ line is stronger at lower X-ray luminosities, offering a diagnostic tool for the presence of H in the disc. 

So far, one strong extragalactic black hole ultra-compact candidate has been found in the elliptical galaxy NGC\,4472. This system exceeds the Eddington luminosity for a $10\,M_\odot$ object \citep{2007Natur.445..183M}, and shows a strong [O{\sc iii}] $\lambda 5007$ emission complex ($\rm -EW > 34$\,{\AA}; \citealp{2014ApJ...785..147S}), probably originating from an accretion-powered outflow \citep{2008ApJ...683L.139Z}. In comparison, the upper limit on H\,$\beta$ emission is much fainter ($\rm -EW < 1$\,{\AA}; \citealp{2014ApJ...785..147S}), pointing to a H-deficient donor, most likely a C/O white dwarf \citep{2014ApJ...785..147S}.

The first Galactic candidate for an ultra-compact system with a black hole accretor is the source X9 \citep{2015MNRAS.453.3918M}, in the massive globular cluster 47 Tucanae ($M = (6.5 \pm 0.5)\times 10^{5}\,M_\odot$; \citealp{2015AJ....149...53K}) located at a distance of $4.53 \pm 0.08$\,kpc \citep{2016ApJ...831..184B}, with foreground reddening $E(B-V) = 0.04 \pm 0.02$ \citep{2007A&A...476..243S}. Originally proposed to be a cataclysmic variable \citep{1992Natur.360...46P, 2001Sci...292.2290G}, X9 has recently been found to be a source of strong radio emission (relative to the X-ray luminosity), making it a black hole candidate \citep{2015MNRAS.453.3918M}. Its relatively high X-ray luminosity ($L_{\rm 0.5-10\,keV} \approx 10^{33} - 10^{34}$\,erg\,s$^{-1}$), broadband FUV--optical--infrared spectrum, and C\,{\sc iv} emission lines \citep{2008ApJ...683.1006K} suggested a degenerate donor \citep{2015MNRAS.453.3918M}. There is some evidence from narrowband {\it Hubble Space Telescope} ({\it HST}) colours for weak H\,$\alpha$ emission with $\rm -EW \approx 10$\,{\AA}, but both the presence of H\,$\alpha$ emission and its exact EW depend strongly on the model assumed for the underlying continuum \citep{2015MNRAS.453.3918M}. Based on X-ray timing and spectroscopy, \citet{2017MNRAS.467.2199B} suggested a short orbital period ($P_{\rm orb}=28.2$\,min), and identified oxygen lines (<1\,keV, similar to the neutron star ultra-compact 4U\,1626--67; \citealp{2007ApJ...660..605K}) that support the presence of a C/O white dwarf donor. The previous suggestions of periodicities (of low significance) in photometric studies with {\it HST} ($P_{\rm orb} \approx 6$ hours; \citealp{1992Natur.360...46P}, or $\approx 3.5$ hours; \citealp{2003ApJ...596.1197E}) are likely to be spurious \citep{2017MNRAS.467.2199B}. \citet{2017MNRAS.467.2199B} concluded that 47 Tuc X9 might be more likely to host a black hole accretor, on account of its radio and X-ray behaviour, although we stress that the nature of the accreting object is still unknown.

In this work, we report on spectroscopic observations of X9 between 3,000 and 10,000{\AA} with the {\it Space Telescope Imaging Spectrograph} ({\it STIS}; \citealp{1998PASP..110.1183W}) mounted on {\it HST}. Firstly, we constrain the abundance of H and He of the donor (Section~\ref{sec:compare}), and its physical properties (Section~\ref{sec:donor}). Secondly, we search for periodicities in the far-ultraviolet (FUV) and optical bands in the archival {\it HST} data previously used by \citet{2002ApJ...579..752K, 2003ApJ...599.1320K, 2008ApJ...683.1006K} and \citet{2003ApJ...596.1177E, 2003ApJ...596.1197E}, to search for the orbital and superorbital modulations seen in the X-ray band, as suggested by \citet{2017MNRAS.467.2199B}. Based on the broadband spectral and timing results, along with the known C\,{\sc iv} lines in the system, we attempt to constrain the nature of the accretor (Sections~\ref{sec:mdot}, \ref{sec:source_emission}, \ref{sec:civ}, \ref{sec:x-spec}, \ref{sec:orb_mod}, \ref{sec:sorb}). Lastly, we calculate an evolutionary sequence for a black hole + He-star binary to examine whether it could eventually lead to an ultra-compact X-ray binary, and thus to X9 (should the accretor indeed prove to be a black hole; Section~\ref{sec:simul}).

\section{Observations \& Data analysis}

\subsection{Spectroscopy}

We observed X9 with {\it STIS} on 16 June 2016, using the CCD detector and the 52$^{\prime\prime}\times$0.1$^{\prime\prime}$ E1 slit position. The program (GO--14203) was allocated a total of four orbits, acquiring a total of five and six (for cosmic-ray rejection) low-resolution spectra ($R \sim 500$) with the G430L (dispersion $\Delta \lambda = 2.73$ {\AA}\,pixel$^{-1}$) and G750L ($\Delta \lambda = 4.92$ {\AA}\,pixel$^{-1}$) gratings, respectively. The observations were not dithered. After each of the G750L observations, we observed fringe flats to improve the removal of fringes (caused by interference between incident and reflected beams in the CCD) on the redward side of the spectra \citep{2003PASP..115..218M}. A summary of the science observations is provided in Table~\ref{tab:obs}.

We retrieved the calibrated 2-D spectra from the Mikulski Archive for Space Telescopes (MAST). These data frames were reduced with the {\it STIS} pipeline (CALSTIS v3.4), using the calibration files closest in time to the science observations. We processed these spectra using the recommended PyRAF (v2.1.14; \citealp{1993ASPC...52..173T}) and Space Telescope Science Data Analysis System (STSDAS v3.17) tools. We recalibrated the wavelength solution using the GO-wavecal exposures, removed the fringes on each of the G750L spectra above 6000\,{\AA} using their associated flat files, and merged the individual spectra of each grating into a G430L stack and a G750L stack. We extracted the two stacked spectra using the task {\tt x1d}, adopting a three pixel aperture (corresponding to 0.15\arcsec). We subtracted the background using two emission-free regions adjacent to X9. We smoothed the background in these two regions using a median boxcar filter with a 9 pixel window. We inferred the sky values at the position of X9 by taking the mean of the two background regions, and we fitted the resulting sky spectrum with a fourth order polynomial (as a function of wavelength).

The extracted spectrum still contains a relatively high number of deviant data points, which can be distinguished from line emission or absorption because their point spread function is undersampled. The presence of such data points can be explained as hot and cold pixels that could not be removed when stacking individual spectra because the observations have not been dithered. Since these deviant points have not been flagged by the {\tt x1d} task as being of poor data quality, we need to remove them using a different strategy. First, we smoothed the extracted 1D spectrum by convolving it with a Gaussian with FWHM = 7 pixels. We removed from the original spectrum those data points that deviated by more than 4 standard deviations from the smoothed spectrum. Less than 1\% of the data were removed in this process, and we manually checked that no emission or absorption lines found in typical X-ray binaries were removed.

\begin{table}
	\centering
	\caption{Summary of {\it HST} / {\it STIS} observations on 16 June 2016, using the 52$^{\prime\prime}\times$0.1$^{\prime\prime}$E1 slit.}
	\begin{tabular}{llll}
	\hline
	\multicolumn{1}{l}{Gratings} & \multicolumn{1}{l}{Dataset} & \multicolumn{1}{l}{\begin{tabular}[l]{@{}l@{}}Start\\(UT)\end{tabular}} & \multicolumn{1}{l}{\begin{tabular}[l]{@{}l@{}}Exp\\(s)\end{tabular}} \\
	\hline
	\hline \\[-8pt]	
	G430L & OCXP02010 & 06:56 & $2\times661$ \\
	G430L & OCXP02020 & 07:42 & $3\times1052$ \\
	G750L & OCXP02030 & 09:17 & $3\times1078$\\
	G750L & OCXP02050 & 10:52 & $3\times1051$ \\
	\hline
	\end{tabular}
	\label{tab:obs}
\end{table}

\subsection{Calibration checks}

To verify the quality of the calibration, we used the spectrum of a star within 0.5 pixels of the centre of the slit near X9 (J002404.30--720457.7, 0.4$^{\prime\prime}$ NE of X9), which we extracted with a similar procedure as for X9 using the same sky regions.

To find the spectral class of the check star, we used BT-Settl model grids of theoretical spectra \citep{2012RSPTA.370.2765A} with surface gravity log\,$g$ = 4.5, overall metal abundance [M/H] = --0.5 and [M/H] = --1.0, and reddening $E(B-V) = 0.04$. We convolved these with the spectral resolution of {\it STIS}. We searched for the spectral template that best fit the observed spectrum by using a $\chi^2$ minimization approach that left the flux normalisation and radial velocity as free parameters. For each model spectrum, we formed a grid of possible radial velocities (from --300 to 300\,km\,s$^{-1}$ in steps of 50 km\,s$^{-1}$; 47 Tuc has a heliocentric radial velocity of --18\,km\,s$^{-1}$ and a central velocity dispersion of 11\,km\,s$^{-1}$, \citealp{1996AJ....112.1487H}).

We found the best-matching template for the check star was a $T$ = 6,600\,K, $M \approx 0.6\,M_\odot$ dwarf. The 3,000--8,500\,{\AA} flux of such a star is consistent with the measured flux in our spectrum, thus confirming that flux calibration has been correctly performed. The spectrum at wavelengths longer than $\approx$ 8,500\,{\AA} was systematically brighter than indicated by the spectral template (where the background dominates), so we ignored this part of our spectra in the remaining analysis of the check star and X9. A comparison of the observed check star spectrum to the template in the 3,000--8,500\,{\AA} range (which covers the G430L and G750L gratings) results in $\chi_\nu^2 = 4.8$, indicating that the error bars produced by CALSTIS are underestimated by a factor $k \approx \sqrt{\chi_\nu^2} \approx 2.2$. The required rescaling is similar for G430L and G750L. The spectral template might be a poor match for the star, so we take the value of $k = 2.2$ as an upper limit to the rescaling factor. As another method of estimating the rescaling factor, we also fit power-law functions to line-free spectral regions (from visually inspecting the observed and template spectra): 4,600--4,800\,{\AA} on G430L, and 6,700--8,500\,{\AA} on G750L. For G430L, we find $k = 1.5$, and for G750L, $k = 1.9$. We chose to rescale the error bars by these factors. After rescaling, we measure the equivalent width of the H\,$\alpha$ line in absorption, and its radial velocity, for this check star. We find $\rm EW_{H\,\alpha} = 3 \pm 1$\,{\AA}, and a radial velocity $V_{\rm r} = 210 \pm 150$\,km\,s$^{-1}$. We estimate a radial velocity shift of up to 100\,km\,s$^{-1}$ in this check star owing to its misaligned position in the slit, which might explain the large radial velocity estimate.

\subsection{Timing analysis}

To investigate the orbital ($P \approx 28$ min) and superorbital ($P \approx 7$ days) modulations found in the X-ray band \citep{2017MNRAS.467.2199B}, we performed timing analyses on archival FUV and optical {\it HST} observations of 47~Tuc, previously reported by \citet{2008ApJ...683.1006K}, \citet{2000ApJ...545L..47G} and \citet{2001ApJ...559.1060A}. 

The FUV data, taken as part of a campaign to study exotic objects in 47~Tuc, were taken with the FUV-MAMA detector on the {\it STIS} spectrograph (program GO--8279). The observations were composed of six distinct epochs, with the first occurring almost a year before the other five. We only included the spectroscopic data (there are only two photometric data points per epoch), which was performed in slitless mode, and we summed the entire flux in the spectrum of X9 in each exposure. We analysed the entire data set jointly, which favours the detection of highly coherent signals. Importantly, the exposure times were typically 600s, approximately one-third of the $\approx 28$ min orbital period seen in the X-rays. This will smooth out the amplitude and significance of any signal at such a short period.

The optical data, taken with the F555W and F814W filters of the Wide Field Planetary Camera 2, were obtained over a period of 8.3 days, using 160\,s exposures (program GO--8267). We reanalysed the differential time series photometry reported by \citet{2001ApJ...559.1060A}, who refer to X9 as PC1-V47, which they classify as a variable with a 6.39 day period.

Time conversion to Barycentric Julian Dates in the Barycentric Dynamical Time standard was done using the online calculator from \citet{2010PASP..122..935E}. To search for significant periodicities in the FUV and optical datasets, we used the Generalised Lomb-Scargle algorithm \citep{2009A&A...496..577Z}, as implemented in AstroML (which accounts for heteroscedastic errors; \citealp{astroML}). We evaluated the statistical significance via randomization (i.e. keeping the magnitudes of each measurement, and shuffling their observation times); errors on the period were estimated via bootstrapping (i.e. selecting random samples from the original dataset). In both cases, 10,000 fake data sets were used. In analysing the optical data, we allowed for a $\sigma_m = 0.085$ magnitude intrinsic dispersion (accounting for source variability), which brings $\chi^2_\nu \approx 1$ for a sine wave fit to the folded light curve (with the period fixed to the most significant peak in the power spectrum).

\section{Results}
 
\subsection{Continuum} 
 
\begin{figure*}
	\includegraphics[width=\textwidth]{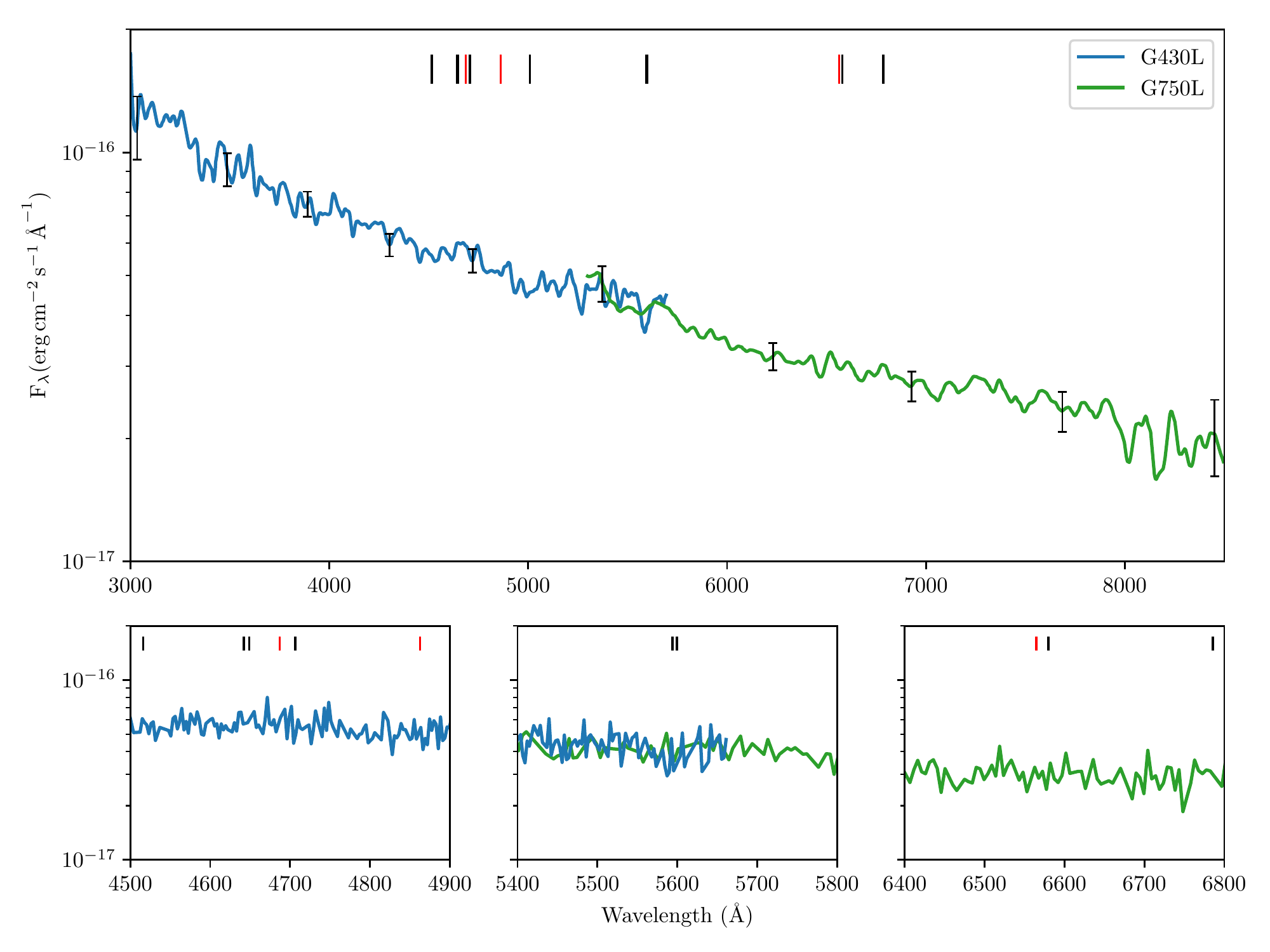}
	\caption{{\it Top}: The smoothed optical spectrum of X9 (convolved with a Gaussian with FWHM = 7 pixels). We show only some of the error bars for clarity. {\it Bottom:} Zoomed-in regions of the non-smoothed spectrum. The vertical ticks above the spectrum indicate the wavelengths of the H\,$\alpha$ $\lambda$6564, H\,$\beta$ $\lambda$4862 and He\,{\sc ii} $\lambda$4687 lines (red ticks) typical of ordinary X-ray binaries, and the brightest C/N/O emission features (black ticks) in ultra-compact systems (with emission $\rm -EW >3$\,{\AA} in at least one of the three systems studied by \citealp{2004MNRAS.348L...7N}), which may be expected in X9. No obvious emission lines are visible. The upper limits on the EW (in emission and absorption) for these lines are listed in Table~\ref{tab:ew}.}
	\label{fig:x9_spec}
\end{figure*}

The optical spectrum of X9 is featureless and blue (Figure~\ref{fig:x9_spec}), with the colour $(B - R) = 0.3$ (Vega magnitudes; obtained by convolving the spectrum through {\it HST} filter throughput curves\footnote{http://www.stsci.edu/hst/acs/analysis/throughputs}) and $F_\nu \propto \nu^{-0.2}$ ($F_\lambda \propto \lambda^{-1.8}$; obtained by fitting a reddened power-law over the 3,000 to 8,500\,{\AA} range). The blue colour of the source, determined from broadband data, has previously been reported in the literature \citep{1992Natur.360...46P, 2008ApJ...683.1006K}, and we measure the same $(B - R)$ value as \citet{2015MNRAS.453.3918M} from quasi-simultaneous photometric measurements taken in 2002.

\subsection{Line significance}
\label{sec:lines}
No strong lines can be readily identified in the 3,000--8,500\,{\AA} spectrum of the optical counterpart of X9. We measure the colour $(H\,\alpha - B) = -0.3$ (Vega magnitudes; obtained by convolving the spectrum through {\it HST} filter throughput curves), which is similar to that reported by \citet{2015MNRAS.453.3918M}. We have estimated the upper limits on the strongest H and He lines seen in typical X-ray binaries \citep{1996MNRAS.282.1437S}, and also on the C/N/O lines/complexes of other ultra-compacts that have been reported by \citet{2004MNRAS.348L...7N} to have emission $\rm -EW > 3$\,{\AA}. To estimate these upper limits, we first use a power-law model, \texttt{redden\,$\times$\,pow} in {\sc Xspec} (v12.9.1; \citealp{1996ASPC..101...17A}) to fit the continuum in a 4\,$\times$\,FWHM$_{\rm C{\textsc{iv}}}$ (FWHM$_{\rm C{\textsc{iv}}} = 3600$\,km\,s$^{-1}$; see Section~\ref{sec:civ}) window centred on a given wavelength (the central wavelengths are listed in Table~\ref{tab:ew}), ignoring the central 1.09\,$\times$\,FWHM$_{\rm C{\textsc{iv}}}$ (corresponding to 80\% of a Gaussian line emission) to avoid line emission. Secondly, we re-add to the data set the previously ignored central window, and add to the model a Gaussian component (\texttt{agauss}) centred on the given wavelength, with FWHM $= 3600$\,km\,s$^{-1}$. We fit the line model by fixing the continuum level, line FWHM, and central wavelength, and letting only the normalisation of the Gaussian (the total line flux) be free to vary. Using the {\tt error} task, we take the upper limit for the line flux to correspond to $\Delta \chi^2 = 9.0$ ($3\sigma$). The equivalent width of the Gaussian component is measured with the \texttt{eqwidth} task. No significant line emission was found. Line upper limits are reported in Table~\ref{tab:ew}. The central wavelength of a line could be offset from the assumed position by less than 1\,px. The absolute wavelength calibration for STIS is accurate to within 0.2\,px \citep{1998PASP..110.1183W}, and the central velocity dispersion of 47 Tuc is 11\,km\,s$^{-1}$, or $< 0.1$\,px \citep{1996AJ....112.1487H}. Any line offset from the assumed centres should therefore have minimal impact on the reported EW upper limits. The FWHM of an optical line, however, could be lower than that of the C{\sc iv} line. Repeating the fitting procedure by letting the FWHM free to vary between the spectral resolution of the instrument and the C{\sc iv} FWHM (3600\,km\,s$^{-1}$), we derive similar upper limits (within 10\%). This demonstrates the robustness of the EW upper limits presented in Table~\ref{tab:ew}.

\begin{table}
	\centering
	\caption{Limits for the EWs of selected lines, assuming a FWHM = 3600\,km\,s$^{-1}$}
	\begin{tabular}{llcc}
	\hline
	Species & \shortstack{Line$^{\rm a}$\\(\AA)} & \shortstack{EW$^{\rm b}$ \\(absorption, \AA)} & \shortstack{$\rm -EW^{\rm b}$\\(emission, \AA)}\\
	\hline
	\hline
He\,{\sc ii} & 4687 & $2.5$ & $8.8$ \\
H\,$\beta$ & 4862 & $5.9$ & $6.4$ \\
H\,$\alpha$ & 6564 & $8.5$ & $13.7$ \\
\hline
N\,{\sc iii} & 4516 & $7.8$ & $3.7$ \\
N\,{\sc iii}$^{\rm c}$ & 4641 & $4.6$ & $6.9$ \\
C\,{\sc iii}$^{\rm c}$ & 4648 & $3.7$ & $8.7$ \\
O\,{\sc ii} & 4706 & $4.0$ & $7.8$ \\
$[$O\,{\sc iii}$]$ & 5008 & $12.0$ & $1.2$ \\
O\,{\sc iii} & 5593 & $12.3$ & $8.0$ \\
O\,{\sc v} & 5599 & $11.3$ & $9.2$ \\
C\,{\sc ii} & 6579 & $8.7$ & $13.4$ \\
C\,{\sc ii} & 6785 & $8.3$ & $17.9$ \\
	\hline
	\multicolumn{3}{l}{\bf Notes.} \\
	\multicolumn{3}{l}{$^{\rm a}$ Vacuum wavelengths}\\
	\multicolumn{3}{l}{$^{\rm b}$ 3$\sigma$ limits}\\
	\multicolumn{3}{l}{$^{\rm c}$ Strongest lines in Bowen blend}
	\end{tabular}
	\label{tab:ew}
\end{table}

\subsection{Timing}

\subsubsection{Optical}
\label{sec:timing}
\begin{figure}	
	\includegraphics[width=\columnwidth]{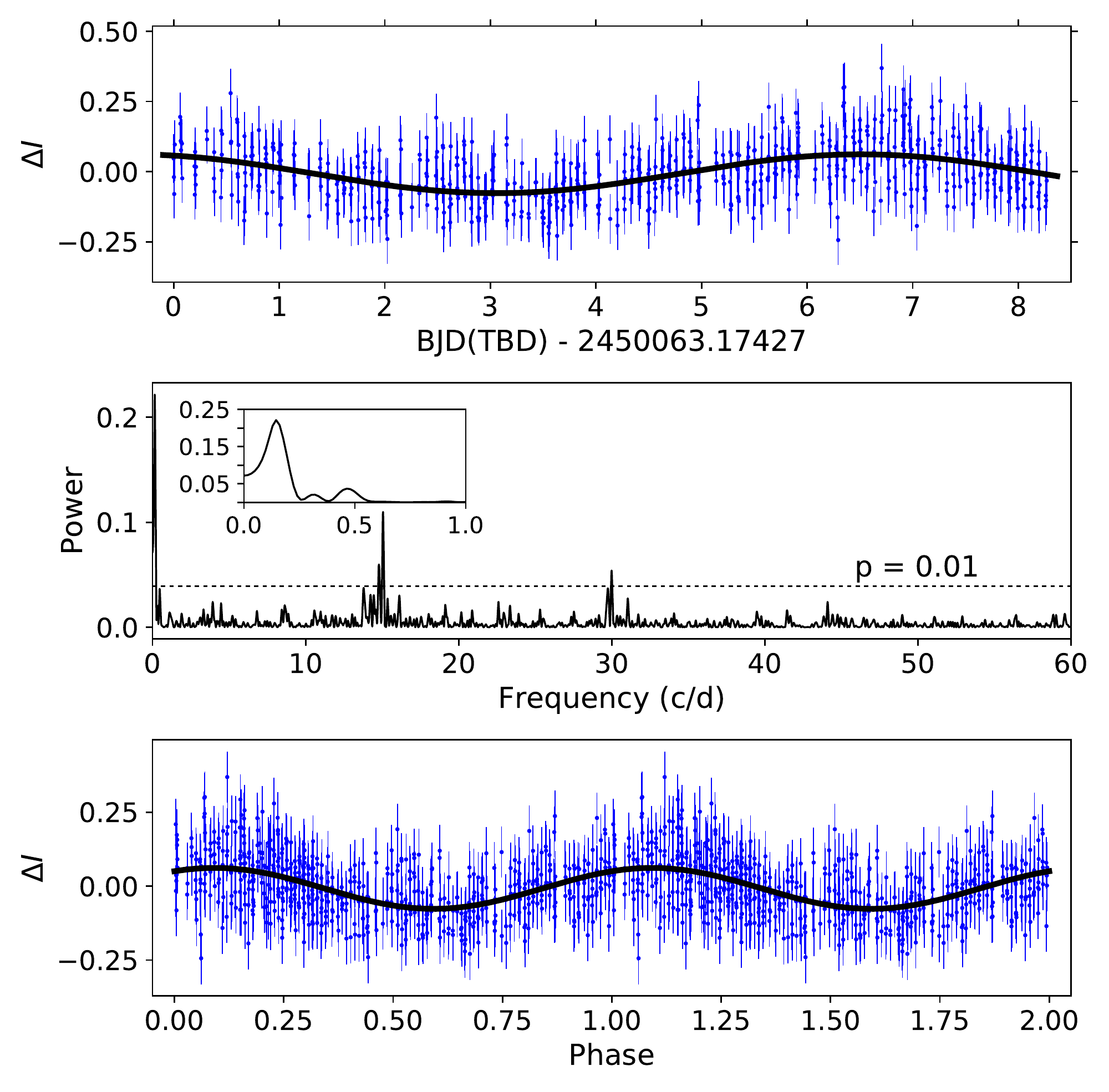}
	\caption{Timing analysis on the $I$-band data of \citet{2000ApJ...545L..47G}. {\it Top:} normalised light curve showing the sine-wave signal (black curve) retrieved from the power spectrum and folded light curve. {\it Middle:} power spectrum (horizontal dotted line indicating the false alarm probability $p = 0.01$). The most significant peak ($P = 0.14$ cycles day$^{-1}$ = 7 days) coincides with the superorbital period reported by \citet{2017MNRAS.467.2199B}, while the peaks at $\approx$ 15 and $\approx$ 30 cycles day$^{-1}$ are likely to be associated with the orbital period of the HST and its first harmonic. No orbital modulation ($P = 51$ cycles day$^{-1}$) is detected in this dataset. {\it Bottom:} the light curve folded on the most significant period, revealing a superorbital modulation with a semi-amplitude of 5\%. The timing analysis on the $V$-band data gives similar results.}
	\label{fig:i-timing}
\end{figure}

We detect the seven-day period (0.14 cycles day$^{-1}$) from \citet{2017MNRAS.467.2199B} with very high significance in both $V$ and $I$ bands (Figure~\ref{fig:i-timing}). The bootstrap simulations show the period estimate is fully consistent with the X-ray-based one. The optical data spans just over one cycle of this period, so, taken alone, its persistence would be questionable. However, coupled with its detection in the X-ray band at different epochs, we consider this superorbital period to be real and stable. The semi-amplitude of the signal is $\approx 0.05$ mag (i.e. about 5\% of flux). Thus, in the optical band, this seven-day modulation has an amplitude two orders of magnitude smaller than the corresponding X-ray signal (which has a semi-amplitude of a factor of about three).

Unfortunately, due to the long stretch of time between the optical observations (July 1999) and the X-ray observations when the superorbital phase was well known (Jan 2006; \citealp{2017MNRAS.467.2199B}), we cannot determine if the optical and X-ray modulations vary in phase with each other.

We also detect significant peaks at $\approx$ 15 and $\approx$ 30 cycles day$^{-1}$. The former are most likely associated with {\it HST}'s orbit ($\approx$ 96 min), and the latter are probably associated with its first harmonic. The signal that \citet{2003ApJ...596.1197E} tentatively mentioned in their paper corresponds to the peak near 6 cycles day$^{-1}$ (i.e. $\approx$ 4 hours). However, this has low significance (false alarm probability $p > 0.1$) in the $V$-band power spectrum and is almost absent in the $I$-band data. In addition, we do not find evidence for power around 28.2 min in either band. The 3$\sigma$ upper limits (in the $V$ and $I$ bands) for this signal have semi-amplitudes of $< 0.025$ mag (2\% of the flux). No signal at half that period (as expected for ellipsoidal variations) is detected either.

\subsubsection{FUV}

\begin{figure*}
	\includegraphics[width=\textwidth]{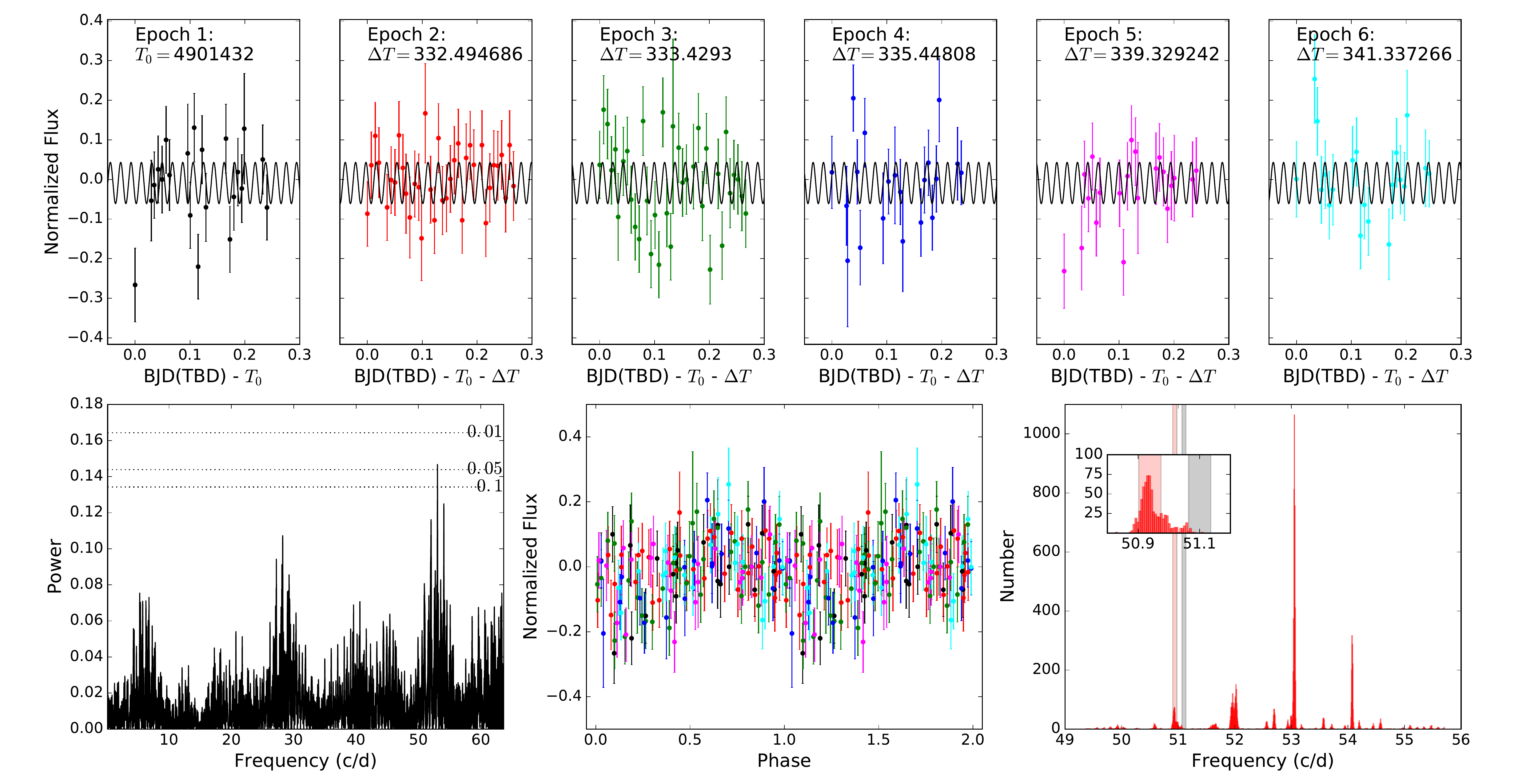}
	\caption{Timing analysis on the FUV data of \citet{2008ApJ...683.1006K}. {\it Top:} normalized light curves for the individual epochs; the associated error bars have been increased by $\sigma_{\rm m} = 0.085$ to account for source flickering. The parameters of the shown sine-wave curve (black) are those found through the power spectrum and folded light curve. {\it Bottom left:} power spectrum (horizontal dotted lines indicating false-alarm probabilities, the strongest signal just exceeds the $2\sigma$ level); {\it bottom middle:} folded light curve (on the most significant period -- 27.2 min, with a semi-amplitude of $\approx 5$\%), with different epochs shown in the same colors as in the top panel; {\it bottom right:} the result of the bootstrapping simulations, showing what periods are plausible, highlighting (gray region) the suggested orbital period based on X-ray data ($28.18 \pm 0.02$ min; \citealp{2017MNRAS.467.2199B}). The period measured from the FUV data (53 cycles day$^{-1}$; $27.2$ min) is close to, but not statistically consistent with the X-ray period.}
	\label{fig:fuv_timing}
\end{figure*}

The strongest peak in the FUV power spectrum occurs at a frequency of 53 cycles day$^{-1}$, i.e. $27.2$ min (false alarm probability $p < 0.05$; Figure~\ref{fig:fuv_timing}). The second strongest peak in the power spectrum (28 cycles day$^{-1}$) may not be caused by any aliases of the orbital period of X9 or of {\it HST}. It is possible that this weaker signal is partly responsible for the flickering observed in individual FUV epochs, and may be drowning out the orbital signal of X9. The light curve phased on the 27.2 min period shows a smooth varying signal (similar to the orbital modulation in X-rays), with a semi-amplitude of about 5\%. We consider this a lower limit, since the long exposure times imply that a significant part of the true signal in each exposure has been averaged. All individual epochs show the same signal in the folded light curve, showing the detected periodicity cannot be associated with variability between the epochs. If the FUV signal is real, the discrepancy between the FUV and X-ray periods is significant. Of the 10,000 bootstrap simulations, only $\approx$ 0.1\% return a peak frequency in the range $28.18 \pm 0.02$ min, determined from X-ray observations. Even though this 27.2 min signal is only significant at the 2$\sigma$ level,  we further investigate it in subsequent discussions (Sections~\ref{sec:orb_mod}, \ref{sec:sorb}) due to its proximity to the X-ray signal.

\section{Discussion}

\subsection{Comparison with other systems}
\label{sec:compare}

\subsubsection{H\,$\alpha$ EW vs X-ray luminosity}

\begin{figure}
	\includegraphics[width=\columnwidth]{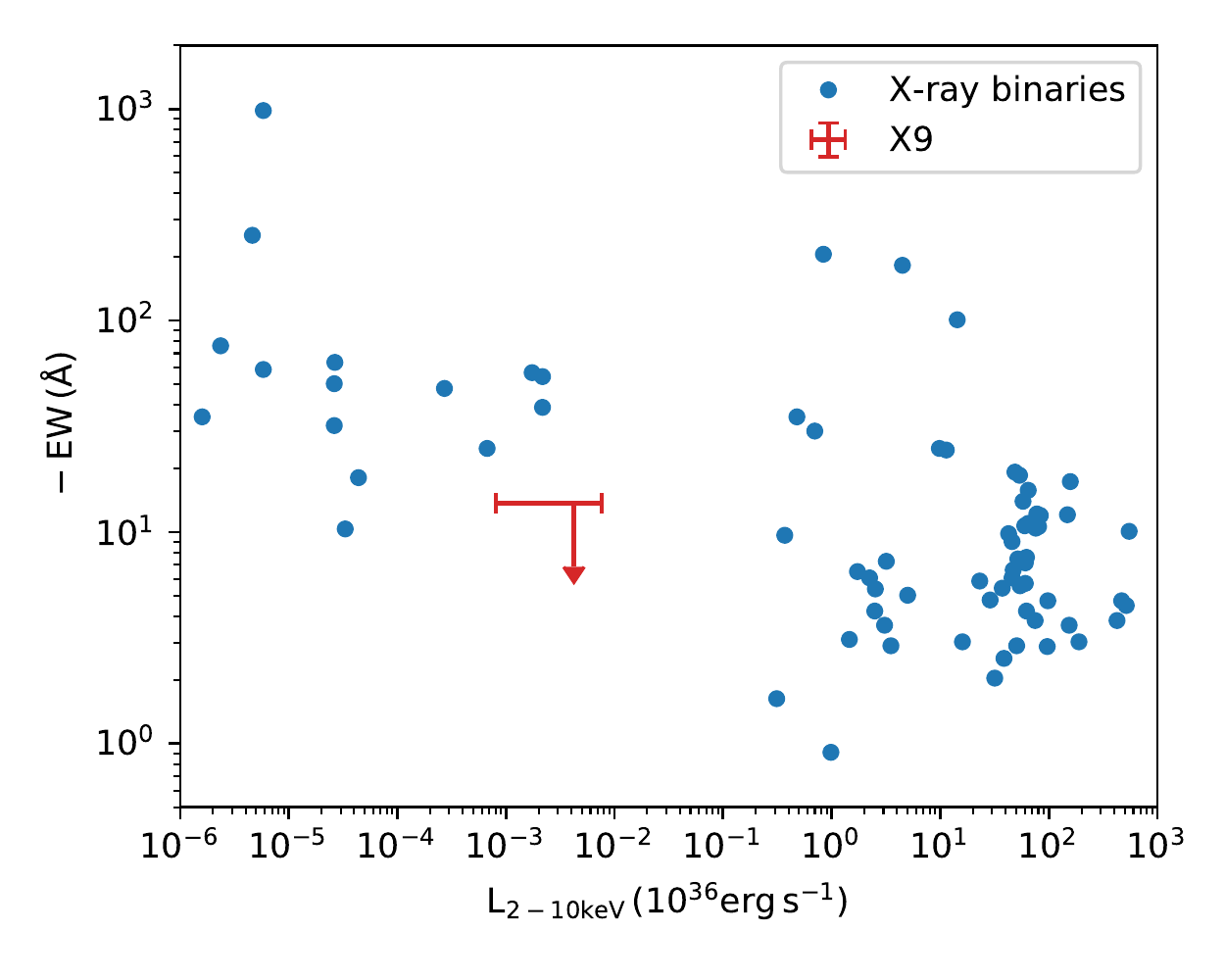}
	\caption{The empirical relationship between the H\,$\alpha$ EW and X-ray luminosity for X-ray binaries (including black holes and neutron stars) from \citet{2009MNRAS.393.1608F}. The upper limit on the H\,$\alpha$ emission in X9 is lower than the emission seen from typical X-ray binaries at a similar X-ray luminosity, as expected for a H-poor donor.}
	\label{fig:ew_lx}
\end{figure}

The trend between the EW of the H\,$\alpha$ line in emission and the X-ray luminosity of X-ray binaries reported by \citet{2009MNRAS.393.1608F} can be used to estimate the expected EW of the H\,$\alpha$ line for X9. Taking into account the X-ray variability of X9 ($L_{\rm X} = 10^{33} - 10^{34}$\,erg\,s$^{-1}$; \citealp{2017MNRAS.467.2199B}), we would expect $\rm -EW_{H\,\alpha} \sim 50$\,{\AA} (Figure~\ref{fig:ew_lx}). However, we found $\rm -EW_{H\,\alpha} < 14$\,{\AA} (3$\sigma$ upper limit). The large scatter ($\gtrsim 1$ dex) of $\rm EW_{H\,\alpha}$ measurements for a given $L_{\rm X}$ in the well-sampled $L_{\rm X} = 10^{30}-10^{32}$\,erg\,s$^{-1}$ and $L_{\rm X} = 10^{35}-10^{39}$\,erg\,s$^{-1}$ ranges suggests a similar scatter may be expected in the range $L_{\rm X} = 10^{32}-10^{34}$\,erg\,s$^{-1}$ (which only contains five $\rm EW_{H\,\alpha}$ measurements). Still, our upper limit on the H\,$\alpha$ line strengthens the case for a H-deficient system, and thus for a C/O white dwarf donor.

\subsubsection{Optical spectral index}

We compare the optical spectrum of X9 with the spectra of other ultra-compact candidates: 4U\,0614+091, XTE\,J0929--314, 4U\,1626--67, XB\,1916--05 \citep{2006MNRAS.370..255N}, 2S\,0918--549 and 4U\,1543--624 \citep{2004MNRAS.348L...7N}. The spectral indices ($F_\nu \propto \nu^{\alpha}$) in these sources are $\alpha = -1$ to 0. For X9, we measure a similarly blue continuum, with $\alpha = -0.2$. Most emission lines in these systems are faint ($\rm -EW < 10$\,{\AA}). In particular, the strongest emission line in the optical spectrum of 2S\,0918--549 is C\,{\sc ii} $\lambda6785$ with $\rm -EW \approx 4$\,{\AA}), indicating that a lack of bright optical emission lines ($\rm -EW \gtrsim 10$\,{\AA}) in ultra-compact X-ray binaries may be a common occurrence. In 4U\,0614+091, however, the C\,{\sc iii} feature at $\lambda \approx 4648$\,{\AA} has $\rm -EW \approx 10$\,{\AA}, which, if present, would have been marginally detected in X9.

\subsection{Nature of donor}
\label{sec:donor}

The lack of H and He in the optical spectrum reported in this work, coupled with the C\,{\sc iv}\,$\lambda\lambda1548,1550$ emission and lack of He\,{\sc ii}\,$\lambda 1640$ in the FUV spectrum \citep{2008ApJ...683.1006K}, and the detection of oxygen features in the soft X-ray spectra \citep{2017MNRAS.467.2199B}, indicate the donor is likely a C/O white dwarf.

\begin{figure}
	\includegraphics[width=\columnwidth]{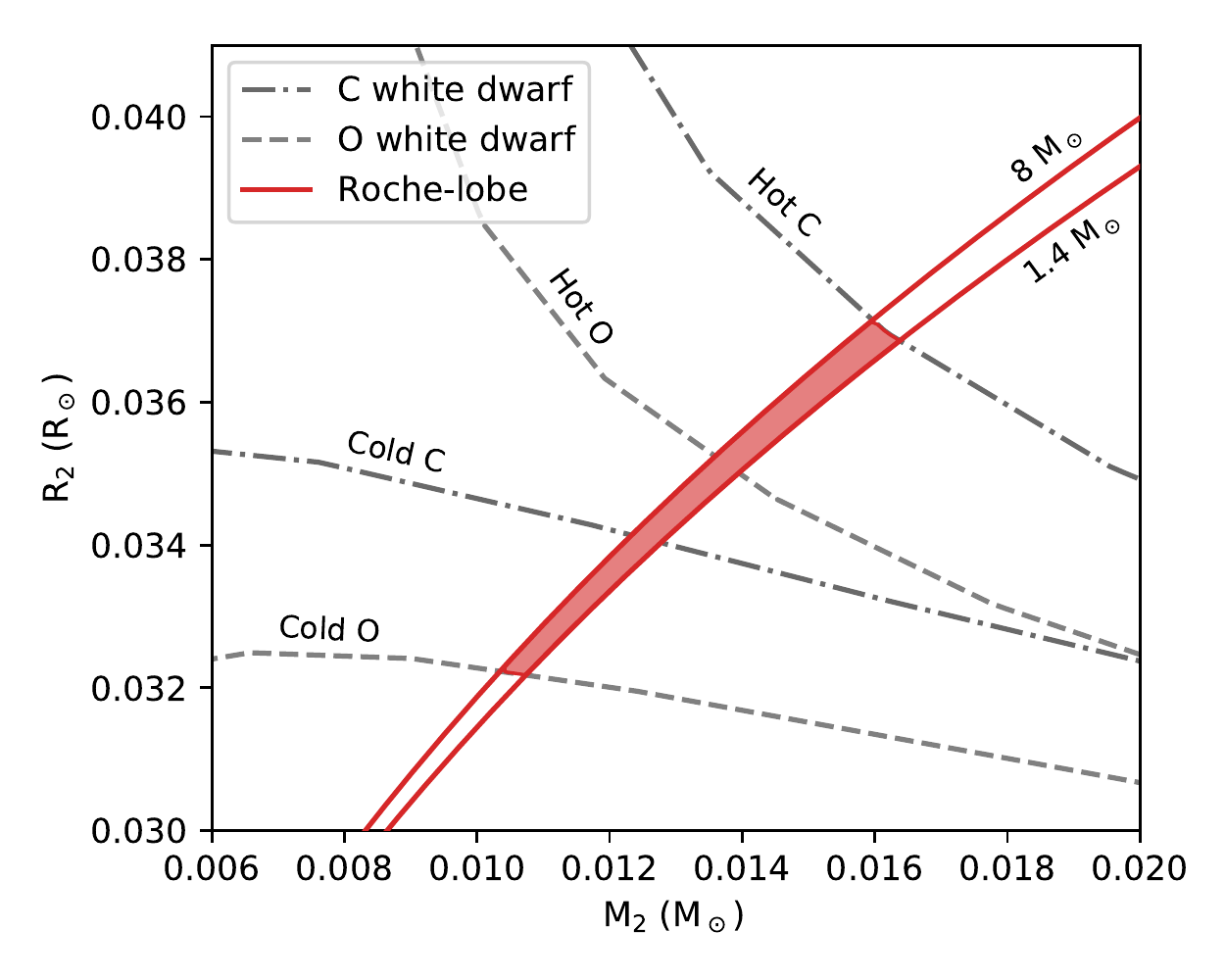}
	\caption{The mass--radius relationships (gray curves) for cold ($T=10^4\,K$, lower curves) and hot ($T=3 \times 10^6\,K$, upper curves) O (dashed curve) and C (dot-dashed) low-mass white dwarfs \citep{2003ApJ...598.1217D}, and the mass-radius relationship (red curves) for a Roche-lobe filling donor in a 28.2 minute orbital period with a $1.4\,M_\odot$ (lower curve) or $8\,M_\odot$ (upper curve) primary. The donor is likely to have a mass and radius between $M_2 = 0.010\,M_\odot$, $R_2 = 0.032\,R_\odot$ (for a cool O white dwarf), and $M_2 = 0.016 \,M_\odot$, $R_2 = 0.037\,R_\odot$ (for a hot C white dwarf).}
	\label{fig:mr_wd}
\end{figure}

In Figure~\ref{fig:mr_wd} we show the mass-radius relationship for degenerate stars of C and O composition based on the equation of state derived by \citet{2003ApJ...598.1217D}, together with the mass-radius relationship for a Roche-lobe-filling companion to both a neutron star and black hole, using the \citet{1983ApJ...268..368E} relation. A $M_2 \approx 0.01\,M_\odot$ C/O white dwarf donor in an X-ray binary is likely to have a core temperature in the range $T_{\rm core} = 10^5 - 10^6$\,K \citep{2002ApJ...577L..27B, 2003ApJ...598..431N}. For an orbital period of 28.2 minutes, the mass and radius of a C/O white dwarf are likely to lie between $M_2 = 0.010\,M_\odot$, $R_2 = 0.032\,R_\odot$ (for a cool, $T_{\rm core} = 10^4\,K$ O white dwarf), and $M_2 = 0.016 \,M_\odot$, $R_2 = 0.037\,R_\odot$ (for a hot, $T_{\rm core} = 3 \times 10^6\,K$ white dwarf), with a weak dependence on the mass of the primary (for $M_2 \ll M_1$ and a given orbital period, the density of the secondary is approximately constant; \citealp{1983ApJ...268..368E}). The Roche lobe of the secondary can accommodate a smaller and lighter donor if it contains a significant presence of Ne or Mg.

\begin{figure}
	\includegraphics[width=\columnwidth]{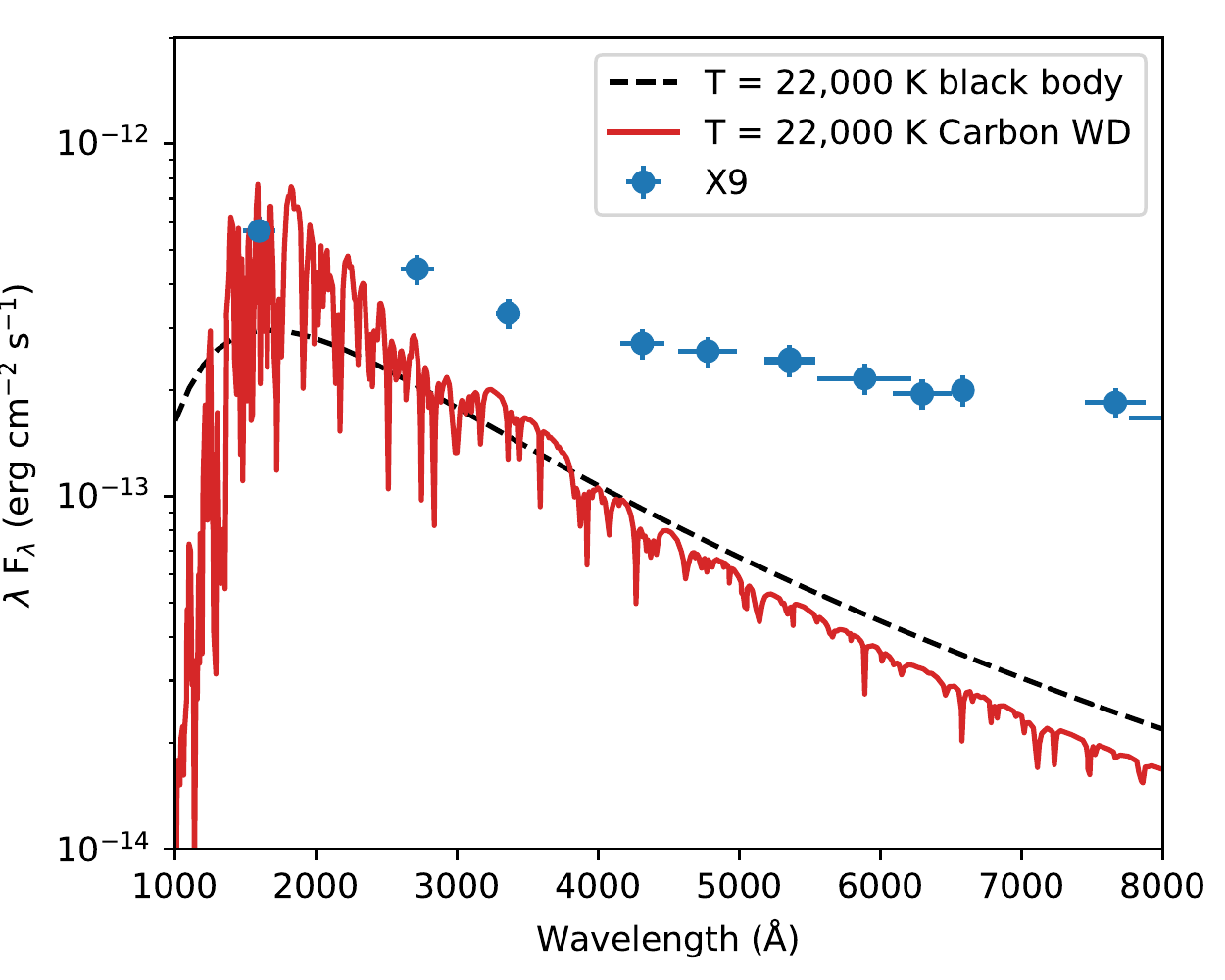}
	\caption{Model spectrum for a carbon-atmosphere white dwarf of $T = 22,000$\,K \citep{2008ApJ...683..978D}, scaled to $R_{2} = 0.035 R_\odot$ and $D = 4.53$\,kpc, compared with the spectrum of X9. The donor in X9 is likely to be cooler than $T \approx 22,000$\,K, otherwise it would start dominating the broadband spectrum of X9 in the UV.}
	\label{fig:temp_wd}
\end{figure}

Since we do not see evidence of emission from the surface of the donor (no absorption lines, and X9 flickers across all wavelengths), we can also estimate the upper limit on the effective temperature of the white dwarf donor when we know its size. For this purpose, we overlay the template spectrum of a carbon white dwarf \citep[$\log g = 8$, $T = 22,000$\,K;][]{2008ApJ...683..978D}, scaled to a radius $R_2 = 0.035\,R_\odot$ and distance $D = 4.53$\,kpc, over the UV-optical spectrum of X9 (Figure~\ref{fig:temp_wd}). We found that the donor in X9 likely has an extremely low mass. Using $M_2 \approx 0.013\,M_\odot$ (the middle value of the allowed range), results in a surface gravity $\log g = 5.5$. This surface gravity is almost three orders of magnitude lower than that of the template white dwarf spectrum, and X9 contains additional atomic species (in particular oxygen). Thus, the template spectrum we have used should not be taken as a quantitative match to the donor in X9. We also overlay a black body spectrum with the same parameters for comparison. We find that the donor cannot be much hotter than $T = 22,000$\,K, since otherwise it would start diluting the flickering activity seen in the FUV and optical bands.

\subsection{Mass accretion rate}
\label{sec:mdot}

Next, we investigate whether the mass accretion rate inferred from the X-ray luminosity is compatible with the theoretical prediction of mass transfer driven by gravitational wave radiation in the cases of a neutron star or black hole.

The conversion of mass to radiation is determined by the radiative efficiency ($\eta$), such that $L_{\rm bol} = \eta \dot{M} c^2$. For black holes, we assume $\eta = 0.1$ in the soft state (independent of mass accretion rate), and $\eta \propto \dot{M}$ in the hard state and quiescence, with a continuous transition between the two regimes \citep{2008NewAR..51..733N}. For neutron stars, we assume $\eta = 0.1$ regardless of mass transfer rate or spectral state (the innermost stable circular orbit for a neutron star is similar in size to its radius). The hard-to-soft transitions of X-ray binaries (neutron stars and black holes) occur at Eddington ratios $L_{\rm bol,tr}/L_{\rm Edd} = 0.004 - 0.25$, with an average of $L_{\rm bol,tr}/L_{\rm Edd} = 0.02$ \citep{2003A&A...409..697M, 2013ApJ...779...95K}. We therefore adopt $L_{\rm bol,tr}/L_{\rm Edd} = 0.02$ for X9, which could be uncertain by an an order of magnitude. The bolometric luminosity is then given by:
\begin{equation}
\label{eq:lbol}
L_{\rm bol} = \eta_0 \frac{\dot{M}^2}{\dot{M}_{\rm tr}} c^2.
\end{equation}
\citet{2017MNRAS.467.2199B} found the 0.5--10\,keV luminosity of X9 to vary in the range $L_{\rm X} = (2-7) \times 10^{33}$\,erg\,s$^{-1}$ on the seven-day period. Assuming a cut-off power-law for the X-ray spectrum, with $\Gamma = 1.1$ and $E_{\rm fold} = 50$\,keV \citep{2017MNRAS.467.2199B}, the bolometric correction for X9 is $L_{\rm 0.5-10\,keV}/L_{\rm 0.5-200\,keV} = 0.13$. Therefore, its bolometric (0.5--200\,keV) luminosity varies between $L_{\rm bol} = (15-54)\times 10^{33}$\,erg\,s$^{-1}$, owing to the seven-day superorbital modulation \citep{2017MNRAS.467.2199B}. Using Equation~\ref{eq:lbol}, for a black hole (assuming a typical mass $M_1 = 8\,M_\odot$; \citealp{2010ApJ...725.1918O}) this range corresponds to a radiative efficiency $\eta_{\rm BH} = (3-5)\times 10^{-3}$ and a mass accretion rate through the disc $\dot{M}_{\rm BH} = (9-18) \times 10^{-11}\,M_\odot$\,yr$^{-1}$. If X9 instead hosts a neutron star ($M_1 = 1.4\,M_\odot$, $R_1 = 10$\,km), the accretion rate ranges between $\dot{M}_{\rm NS} = (0.3-1.0) \times 10^{-11}\,M_\odot$\,yr$^{-1}$. Note that these values assume the luminosity variations are due to variations in mass accretion rate. If instead the variability is caused by geometric effects, the mass accretion rates derived above are lower limits only.

The mass transfer rate for conservative mass transfer driven by gravitational wave radiation is given by \citet{1975ctf..book.....L}, \citet{1993ARA&A..31...93V}:
\begin{equation}
\label{eq:mdot_gr}
\frac{\dot{M}_2}{M_2} = 2 \frac{\dot{J}}{J} \frac{1}{n_{\rm R} + 5/3 - 2M_2/M_1},
\end{equation}
where $n_{\rm R}$ describes the change of the donor's radius as its mass changes, and $J$ and $\dot{J}$ are the angular momentum of the system and the angular momentum loss rate respectively, given by \citet{1964PhRv..136.1224P}:
\begin{equation}
\label{eq:jdot_gr}
\frac{\dot{J}}{J} = -\frac{32}{5} \frac{G^3}{c^5} \frac{M_{\rm 1}M_{\rm 2}(M_{\rm 1} + M_{\rm 2})}{a^4},
\end{equation}
where $a$ is the orbital separation; $a_{\rm BH} = 4.3 \times 10^{10}$\,cm or $a_{\rm NS} = 2.4 \times 10^{10}$\,cm using Kepler's third law for a 28.2 min orbital period. For a C/O white dwarf donor of mass $M_2 = 0.010 - 0.016\,M_\odot$, $|n_{\rm R}| < 0.1$ \citep{2003ApJ...598.1217D}, making it negligible in Equation~\ref{eq:mdot_gr}. For a black hole primary ($M_1 = 8\,M_\odot$), $\dot{M}_{2, {\rm BH}} = (5-11) \times 10^{-11}\,M_\odot$\,yr$^{-1}$ (with higher rate for a more massive donor). For a neutron star, we obtain a mass transfer rate in the range $\dot{M}_{2, {\rm NS}} = (1-4) \times 10^{-11}\,M_\odot$\,yr$^{-1}$. We stress these estimates are based on the assumption of conservative mass transfer.

If X9 is a transient system in quiescence, the mass lost by the donor should be larger than the rate of mass transfer reaching the primary ($|\dot{M}_2| > \dot{M}$), such that matter accumulates in the disc. The mean mass accretion rates estimated above from the X-ray luminosity and conservative mass transfer are compatible both with a neutron star or black hole.

\subsection{Source of optical emission}
\label{sec:source_emission}

The sources of optical/UV continuum and line (C\,{\sc iv}) emission could originate from the accretion disc (viscous dissipation or X-ray reprocessing), the impact point of the accretion stream on the disc, the donor star (intrinsic or reprocessed spectrum), and/or the optically-thin part of the jet. The corona, base of the jet, or an inner advection-dominated accretion flow (ADAF) might also contribute UV-blue photons. It is thought, however, that the optical emission (disregarding the companion star) in X-ray binaries is most often dominated by X-ray reprocessing in the disc \citep{1990A&A...235..162V, 2006MNRAS.371.1334R, 2016ApJ...826..149B}.

Before we investigate the likely scenario of disc irradiation, we discuss the other potential sources of optical emission. In the absence of outflows, an ADAF is expected to have a sharp peak at $\nu_{\rm peak} \sim 10^{15} (M/M_\odot)$\,Hz \citep{1994ApJ...428L..13N,1999ApJ...520..298Q}. Since we do not observe such a peak in the optical/UV spectrum (Figure~\ref{fig:x9_wholespec}), an ADAF is unlikely to be dominating in these bands. The radio flux density of X9 is at the level of $S_\nu \approx 30$\,$\mu$Jy \citep{2015MNRAS.453.3918M, 2017MNRAS.467.2199B}. Assuming a flat radio spectrum, this corresponds to $\sim 50$\% of the measured optical flux, so it is possible that the jet could contribute to the optical emission. The jet in the hard state, however, is thought to turn over at frequencies lower than $\nu_{\rm break} \lesssim 10^{14}$\,Hz $\approx 10^{4}$\,{\AA} \citep{2013MNRAS.429..815R}, making it unlikely to dominate the optical emission, especially towards bluer wavelengths. The jet could, however, contribute on the red side of the spectrum.

The maximum luminosity that may be reached by the stream-impact point is estimated as the release of gravitational energy as matter falls from the first Lagrangian point ($R_{L1}$) to the minimum impact radius on the disc (half the circularization radius, $R_{\rm circ}$) \citep{2001ApJ...557..304M}:
\begin{equation}
L_{\rm impact} < G M_1 \dot{M}_2 \left( \frac{1}{0.5 R_{\rm circ}} - \frac{1}{R_{L1}} \right).
\end{equation}
Using the formulations for $R_{\rm circ}$ from \citet[their Equation 13]{1988ApJ...332..193V} and for $R_{\rm L1}$ from \citet[their Equation C.3]{2012A&A...537A.104V}, and taking $M_1 = 8\,M_\odot$, $\dot{M} = 10^{-10}\,M_\odot$ (the maximum allowed from gravitational wave radiation for $M_2 = 0.016\,M_\odot$), the maximum stream-impact luminosity is $L_{\rm impact} < 3 \times 10^{32}$\,erg\,s$^{-1}$. This value is a factor of $\sim$3 lower than the mean monochromatic optical/FUV luminosity ($L = \lambda F_{\lambda} \approx 1 \times 10^{33}$\,erg\,s$^{-1}$), meaning that it is unlikely to be produced at the impact point. The bulk of the disc and possibly the jet are most likely the sources of optical emission in X9. Next, we explore how bright the reprocessed emission in X9 should be, by comparing X9 to sources where reprocessing is known to occur.

\subsubsection{Disc irradiation}

In their optical/X-ray correlation study, \citet{2016ApJ...826..149B} focused on the black hole and neutron star X-ray binaries V404 Cyg and Cen\,X--4. They found that the correlations between the total optical (R or V band) and X-ray (3--9\,keV) luminosities in the two systems have different correlation indices and normalisations, V404 Cyg being more optically bright than Cen\,X--4 at the same X-ray luminosity. However, when only the reprocessed optical and bolometric X-ray flux (0.5--200\,keV) are considered, the two sources have virtually indistinguishable optical/X-ray correlations in outburst and quiescence. This suggests the optical/X-ray correlation for reprocessed X-rays could be universal across neutron stars and black holes with different mass ratios. 

To compare X9 with V404\,Cyg and Cen\,X--4, we follow the steps laid out by \citet{2016ApJ...826..149B}, and calculate the 0.5--200\,keV luminosity and normalise the optical flux to that of Cen\,X--4. Since there is no soft X-ray component reminiscent of emission from the surface of a neutron star, we do not subtract any surface contribution from the X-ray luminosity. Because the inclination of X9 is poorly constrained ($i \gtrsim 30^\circ$; see Section \ref{sec:orb_mod}), but comparable to V404\,Cyg ($i = {67^\circ}^{+3}_{-1}$; \citealp{2010ApJ...716.1105K}) and Cen\,X--4 (${32^\circ}^{+8}_{-2}$; \citealp{2014MNRAS.440..504S}), we do not correct for inclination effects.

\begin{figure}
	\includegraphics[width=\columnwidth]{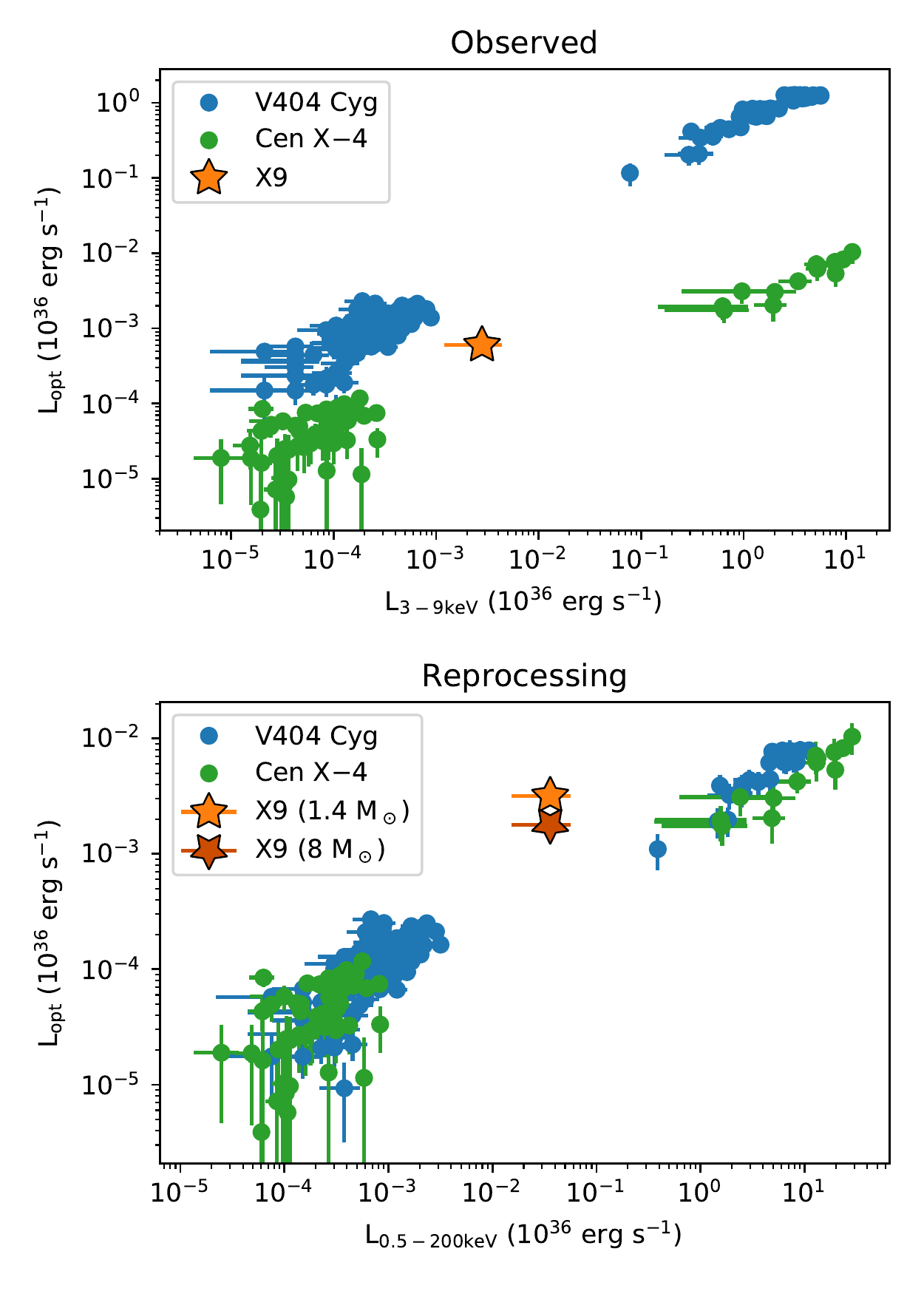}
	\caption{{\it Top}: the relationship between the observed optical and X-ray luminosities of the accreting neutron star Cen\,X--4 and black hole V404\,Cyg, in comparison to X9. {\it Bottom}: for the same sources, the optical and X-ray luminosities associated with reprocessing only (removed contributions from the donor and jet for Cen\,X--4 and V404\,Cyg), normalised to the primary mass ($M = 1.4\,M_\odot$) and orbital period ($P = 0.629$ days) of Cen\,X--4. This Figure was adapted from \citet{2016ApJ...826..149B}. The optical emission of X9 is brighter than expected when compared to V404\,Cyg and Cen\,X--4, indicating more efficient reprocessing of X-rays, or the significant contribution of a jet at optical wavelengths.}
	\label{fig:lxlo_corr}
\end{figure}

\begin{figure*}
	\includegraphics[width=\textwidth]{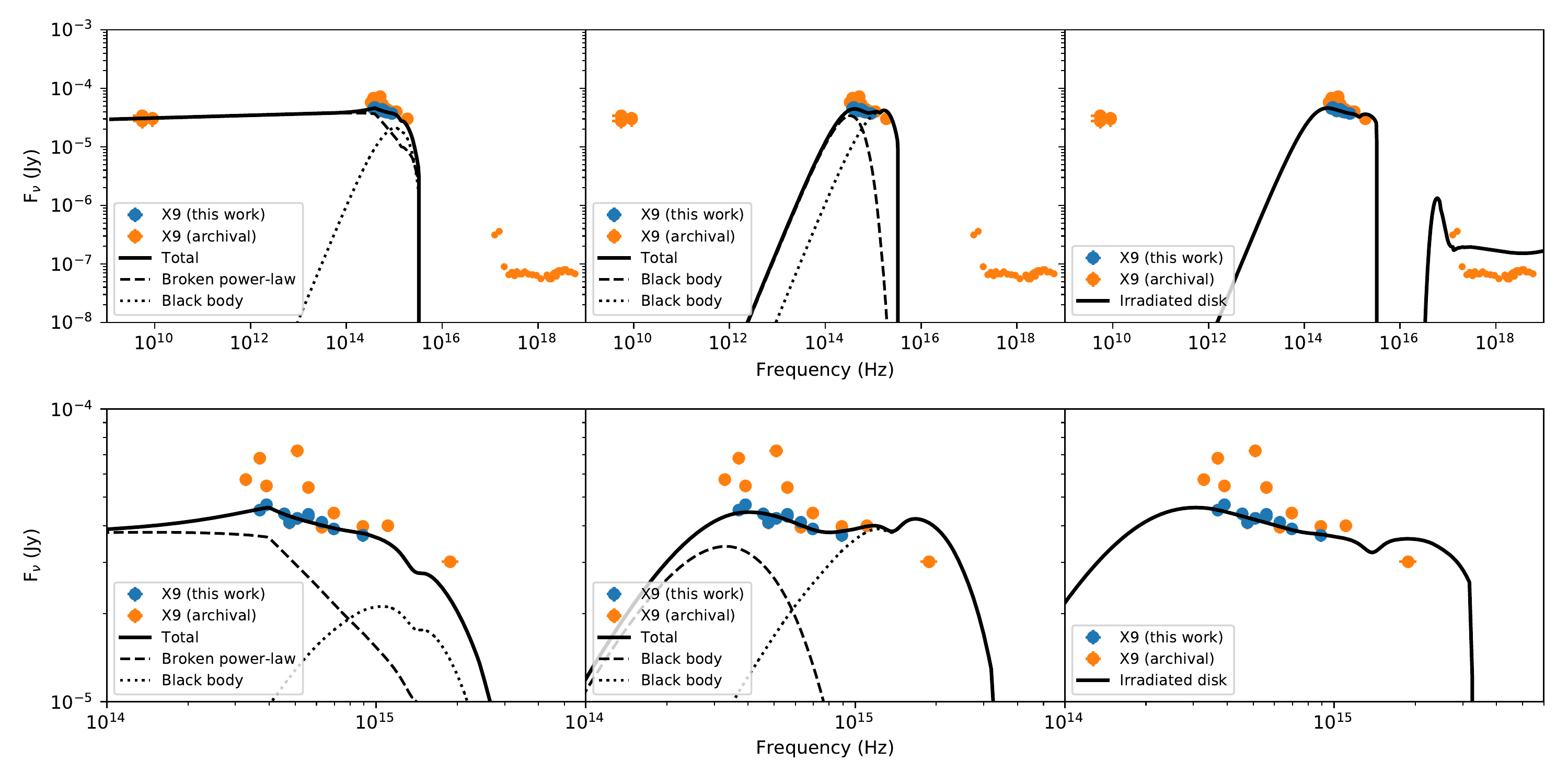}
	\caption{Observed broadband spectrum for X9 (radio data from \citealp{2015MNRAS.453.3918M} and \citealp{2017MNRAS.467.2199B}, archival FUV--infrared data from \citealp{2008ApJ...683.1006K}, and optical data from this work, displayed as broadband fluxes calculated from our spectrum, X-ray data from \citealp{2017MNRAS.467.2199B}), with superimposed models. For fitting the models, we only used the optical spectroscopic data from this work. In the {\tt power-law + black body} model, we also included the radio data in the fit, and for the {\tt diskir} model we assumed a 1--10\,keV power-law spectrum ($\Gamma = 1.1$), with $L_{\rm 1-10\,keV} = 10^{34}$\,erg\,s$^{-1}$, representing the likely intrinsic X-ray luminosity of X9 (see Section~\ref{sec:sorb}), which is brighter than the apparent luminosity. {\it Left:} a broken power-law + black body, where the jet dominates up to the optical band; {\it middle:} two black bodies, with emission from a large, cool disc, and a small, hot inner disc or stream/disc impact point; {\it right}: an irradiated disc. {\it Bottom}: The same spectrum and models, zoomed-in on the optical data.}
	\label{fig:x9_wholespec}
\end{figure*}

The reprocessed emission from the accretion disc depends on its size (and hence on the orbital period and the mass of the primary, via Kepler's third law). Therefore, to make an effective comparison, the optical luminosity of X9 ($L_{\rm opt} \approx 6 \times 10^{32}$\,erg\,s$^{-1}$, calculated from the $V$-band flux) needs to be normalised to that of Cen\,X--4 accounting for its mass and orbital period using $L_{\rm opt} \propto M_1^{1/3} P_{\rm orb}^{2/3}$ (for $M_2 \ll M_1$; \citealp{1994A&A...290..133V, 2016ApJ...826..149B}). Moreover, one must also take into account the bolometric luminosity (which irradiates the disc). Taking the mass to be either $1.4 M_{\odot}$ or $8 M_{\odot}$, we find the normalised optical monochromatic luminosities to be $L_{\rm opt} \approx 3.2 \times 10^{33}$\,erg\,s$^{-1}$ and $L_{\rm opt} \approx 1.8 \times10^{33}$\,erg\,s$^{-1}$ for a neutron star and black hole respectively. The normalised optical luminosity is a factor of three (for a black hole) to six (for a neutron star) larger than those of V404\,Cyg and Cen\,X--4 at similar bolometric luminosities ($L_{\rm opt} \approx 5 \times 10^{32}$\,erg\,s$^{-1}$; Figure~\ref{fig:lxlo_corr}), suggesting the disc reprocesses more of the irradiating X-ray flux (if the disc has a lower albedo, or a larger aspect ratio $H/R$ than typical X-ray binaries), or the jet contributes significantly in the optical band.

In typical X-ray binaries, it is usually assumed the albedo of the disc to grazing X-rays is $A \gtrsim 0.9$ \citep{1996A&A...314..484D}, where the albedo determines what fraction of the incident X-rays is reprocessed to optical emission ($L_{\rm repro} \propto (1 - A)$). As O-rich gas effectively absorbs X-rays, the effective albedo of the disc could be lower than that of Cen\,X--4. Still, the reflection signatures in the X-ray spectrum of X9 \citep{2017MNRAS.467.2199B} show the disc albedo is significant, and we expect a value in the range of $A_{X9} = (0.1-1.0) \times A_{\rm Cen X-4}$, implying the reprocessed optical emission in X9 would have been a factor of $\lesssim 10$ fainter if it had solar metallicity. As long as the inclination of the system $i \lesssim 70^\circ$ (when emission from the heated face of the disc dominates over emission from the edge of the disc), the minimum fractional disc contribution to the optical light is $\gtrsim 1/6$ for a neutron star, and $\gtrsim 1/3$ for a black hole (Figure~\ref{fig:lxlo_corr}), but a low albedo can account for 100\% of optical emission as originating from the irradiated disc. The optical emission from the disc in a high inclination ($i \gtrsim 70^\circ$) system could appear fainter ($<1/6$ contribution) because only its edge would be visible (as a black body). In such a case, the jet would need to contribute significantly to the optical emission. Next, we fit the spectral energy distribution for X9 to further constrain the relative contributions of the jet and disc.

\subsubsection{Broadband spectra}
\label{sec:mods}

Following \citet{2007ApJ...670..600G}, we start by fitting two different physically motivated {\sc Xspec} models to the broadband spectrum: a broken power-law and a black body (to represent the jet and disc, respectively), and a double black body (to represent the outer disc and the inner accretion flow). In addition, we also fit an irradiated disc model. The model parameters are listed in Table~\ref{tab:models}. 

For the {\it broken power-law + black body} model, we fix the spectral index of the optically-thin part of the jet to $\alpha = -0.8$ \citep{2007ApJ...670..600G}. We find the projected size of the disc is only $R_{\rm d, model} = 3 \times 10^9$\,cm (compared to the expected $R_{\rm d, NS} = 24 \times 10^9$\,cm for a neutron star and a disc filling up 90\% of its Roche lobe). While the emission in the optical band is mostly dominated by the jet, the projected size of the black body component requires an edge-on system.

In the {\it double black body} model, the size of the large, cool component ($R_2 \approx 2.5 \times 10^{10}$\,cm) identified with the outer disc indicates either a face-on disc around a neutron star, or a black hole seen at intermediate inclination angles. The small, hot component could be the viscously-heated inner disc, an ADAF, the base of the jet, or the disk/stream impact point. Both of these simple fits show that both a neutron star and black hole are compatible with the observed broadband spectrum. 

\begin{table}
\centering
\begin{minipage}{\columnwidth}
	\centering
	\caption{Parameters for the best fits to the broadband spectrum, using three different models. $R_{1,2}$ and $T_{1,2}$ are the projected sizes of the black body components and their respective temperatures. For {\tt diskir}, $T_1$ is the temperature at the inner truncation radius $R_1$. In all cases, $R_2$ is the radius of the outer edge of the disc. $\alpha$ is the spectral index of the optically thick jet emission. For each fit, we show the $\chi^2$ statistic of the fit and the degrees of freedom (ndf) for the optical range only. We compare the size of the disc from these models, $R_{\rm d, model} = R_2 / (\cos i)^{1/2}$ (assuming $i = 60^{\circ}$) with the theoretical disc size for a neutron star and a black hole (for a disc that extends out to 90\% of the primary's Roche lobe; \citealp{2012A&A...537A.104V}).}
	\begin{tabular}{llll}
	\hline
	Parameter & {\tt pow}\,$+$\,{\tt bbody} & $2\times$\,{\tt bbody} & {\tt diskir} \\
	\hline
	\hline
	$T_1$ ($10^4$ K) & -- & 2.6 & 17.4 \\
	$R_{1}$ ($10^8$ cm) & -- & $34\pm14$ & $1.0\pm0.02$ \\
	$T_2$ ($10^4$ K) & 2.6 & 0.6 & -- \\
	$R_{2}$ ($10^8$ cm) & $33\pm5$ & $250\pm46$ & $770\pm190$ \\
	$\alpha$ & $0.02\pm0.01$ & -- & -- \\
	\hline
	$\rm \chi^2/ndf$ & $1576/1305$ & $1570/1306$ & $1573/1304$ \\
	$R_{\rm d, model}/R_{\rm d, NS}$ & $0.30\pm0.04$ & $2.3\pm0.4$& $4.9\pm1.2$\\
	$R_{\rm d, model}/R_{\rm d, BH}$ & $0.16\pm0.04$ & $1.2\pm0.4$& $2.6\pm0.6$\\
	\hline
	\hline
	\end{tabular}
	\label{tab:models}
\end{minipage}
\end{table}

Assuming the optical emission in X9 originates from X-ray reprocessing alone, we can fit an irradiated disc model ({\tt diskir}; \citealp{2008MNRAS.388..753G}) to the 3,000--8,500\,{\AA} spectrum reported in this work, assuming a 1--10\,keV power-law ($\Gamma = 1.1$) X-ray spectrum with a luminosity $L_{\rm 1-10\,keV} = 10^{34}$\,erg\,s$^{-1}$ representing the unobscured corona. We do not include the broadband optical and FUV measurements reported by \citet{2008ApJ...683.1006K} because they have not been taken simultaneously with our data, and variability (especially in the FUV band) might compromise the fit.

We fix the radius of the Compton-illuminated disc $r_{\rm irr} = 1.2$ (relative to the inner disc radius), the electron temperature $kT_{\rm e} = 100$\,keV, and the fraction of luminosity in the Compton tail that is thermalised in the inner disc $f_{\rm in} = 0.1$. These values have been chosen as typical for X-ray binaries in the hard state \citep{1997MNRAS.292L..21P}. In addition, we fix the temperature of the inner disc $T_{\rm in} = 0.015$\,keV, chosen so that the soft X-ray emission from the inner disc does not become visible in the X-rays.

We find the irradiated disc model to be a good representation of the observed broadband spectrum. Repeating the fit with different values of $kT_{\rm e}$ and $T_{\rm in}$ does not change these results significantly. We find that for an inclination angle $i = 60^{\circ}$, the data are best described by a disc extending out to $R_{\rm d} = (8\pm2) \times 10^{10}$\,cm, or factors of $R_{\rm d,model}/R_{\rm d, NS} = 4.9\pm 1.2$ and $R_{\rm d,model}/R_{\rm d, BH} = 2.3 \pm 0.5$ times larger than expected for a neutron star or black hole respectively. An explanation for this discrepancy is the inclination angle. For a face-on system, the parameters of best fit would correspond to disc radii $R_{\rm d,model}/R_{\rm d, NS} = 2.2\pm 0.5$ and $R_{\rm d,model}/R_{\rm d, BH} = 1.2 \pm 0.3$. In this case, a neutron star is disfavoured, but a black hole is compatible, as long as $i \lesssim 30^{\circ}$. 

In summary, we find that the optical emission can originate from the following sources: a jet and a disc ($i \approx 90^\circ$, regardless of the nature of the primary), a disc and an inner accretion flow ($i \lesssim 30^\circ$ for a neutron star, or $i \approx 60^\circ$ for a black hole), or X-ray reprocessing by the disc (only for a $i \lesssim 30^\circ$ black hole). 

\subsection{C\,{\sc iv} emission}
\label{sec:civ}

We have found the optical spectrum of X9 to be featureless. In the FUV, however, \citet{2008ApJ...683.1006K} have unambiguously detected C\,{\sc iv} emission. Because of its double-peaked profile, they attributed it to a rotating medium, possibly the accretion disc or disc wind. Absorption towards the system (intrinsic or extrinsic), however, could result in a similar shape of the profile. For example, \citet{2002AJ....124.3348H} have attributed the double-peaked profile of the C\,{\sc iv} doublet in the carbon-rich ultra-compact X-ray binary 4U 1626--67 to absorption near the line centres, superposed on the emission line. The lack of absorption lines in the FUV spectrum of X9, in contrast to their presence in 4U 1626--67 \citep{2002AJ....124.3348H}, shows that the double-peaked nature of the lines is more readily explained by a rotating medium in X9.

\begin{figure}
	\includegraphics[width=\columnwidth]{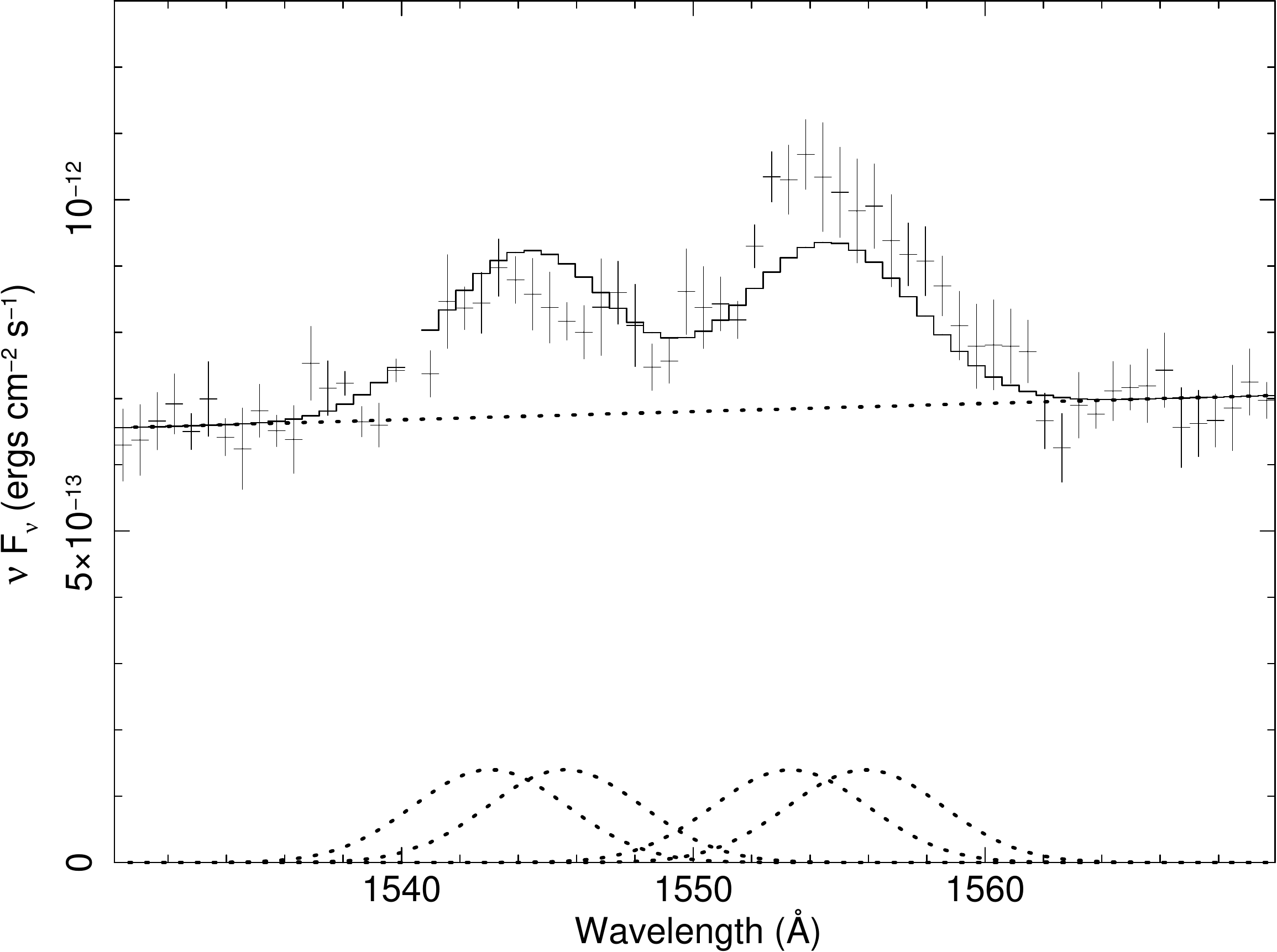}
	\caption{The C\,{\sc iv} $\lambda\lambda 1549,1551$ doublet in X9, fit with two sets of two symmetrical Gaussian line profiles added to the local continuum. Each model component is displayed with a dotted line, and the total model is shown in the full line.}
	\label{fig:civ_fit}
\end{figure}

The FWHM of the C\,{\sc iv} doublet and the orbital period can be used to estimate the mass of the primary object by using the FWHM -- $K_2$ (where $K_2$ is the projected velocity amplitude of the donor; \citealp{2015ApJ...808...80C}) and the DP/FWHM--q (where DP is the double-peak separation; \citealp{2016ApJ...822...99C}) relations. Following the same procedure as \citet{2015ApJ...808...80C} and \citet{2016ApJ...822...99C}, we fit the C\,{\sc iv} doublet with two symmetrical Gaussians (one for each line in the doublet) to measure its FWHM, and with $2 \times 2$ symmetrical Gaussians to measure the double-peak separation. We find FWHM = $3600 \pm 140$\,km\,s$^{-1}$, and DP = $2070\pm70$\,km\,s$^{-1}$ (Figure~\ref{fig:civ_fit}). We note the reason behind the poor fit ($\chi_\nu = 1.50$) in Figure~\ref{fig:civ_fit} is forcing the two peaks to have the same normalization, to follow the same procedure as \citet{2016ApJ...822...99C}. Leaving the relative intensities of the blue and red peaks free to vary leads to a statistically better fit ($\chi_\nu = 1.13$), but to a similar value DP = $2190 \pm 60$\,km\,s$^{-1}$. Changing the relative intensities within the doublet to the theoretical ratio $I_{\lambda 1548} / I_{\lambda 1550}$ = 2:1 \citep{1982MNRAS.201P..39F} does not change the FWHM and DP significantly. With a total exposure time for the FUV spectrum of 82,200\,s \citep{2008ApJ...683.1006K}, i.e. almost 50 binary orbits, we assume the measured C\,{\sc iv} profile is representative of the average emission in the system. We do not correct the FWHM for the instrumental resolution because of its small value ($\Delta\lambda = 1.2$\,{\AA}). 

Compared to the H\,$\alpha$ line in quiescent X-ray binaries, the FWHM of the C\,{\sc iv} doublet in X9 is one of the highest observed \citep{2015ApJ...808...80C}. The FWHM of the C\,{\sc iv} doublet lies in the tentatively defined forbidden region for cataclysmic variables ($\rm FWHM_{H\,\alpha} \gtrsim 2600$\,km\,s$^{-1}$; \citealp{2015ApJ...808...80C}), further strengthening the evidence against such an object. Because the C\,{\sc iv} doublet has higher energy than the H\,$\alpha$ line, the former might originate from closer in to the primary object in an X-ray binary, where the disc is hotter. In contrast to this idea, the FUV and optical spectra (albeit non-simultaneous) of A0620--00 in quiescence show that the emission features of C\,{\sc iv} and H\,$\alpha$ have similar FWHM ($\approx 2000$\,km\,s$^{-1}$; \citealp{1994MNRAS.266..137M, 2011ApJ...743...26F}). Even in outburst, the optical spectrum of the black hole X-ray binary MAXI J1659--152 includes many emission lines between 3,000--10,000\,{\AA} with similar FWHM $= 1500-2000$\,km\,s$^{-1}$ \citep{2012ApJ...746L..23K}, indicating that at a given luminosity, emission lines likely originate from similar characteristic radii, regardless of their species. It is unknown, however, if the relationships of \citet{2015ApJ...808...80C} and \citet{2016ApJ...822...99C} hold for the extreme mass ratios in ultra-compact X-ray binaries because the sample of \citet{2015ApJ...808...80C} includes only H-rich X-ray binaries with orbital periods longer than $P_{\rm orb} = 0.17$\,days.

The accreting neutron star 4U\,1626--67 is a good candidate for testing if the Casares relations hold for the C\,{\sc iv} line in ultra-compacts. It has strong C\,{\sc iv} emission \citep{2002AJ....124.3348H}, a short orbital period (41\,min; \citealp{1981ApJ...244.1001M}) and a low-mass donor ($M_2 \approx 0.02\,M_\odot$), seen at an inclination angle $i \lesssim 33^{\circ}$; \citealp{1990A&A...234..195V, 1998ApJ...492..342C}). In this source, the C\,{\sc iv} FWHM $\approx 1500$\,km\,s$^{-1}$. Using $K_2 = (0.233 \pm 0.013) \, \times$\,FWHM \citep{2015ApJ...808...80C}, we predict $K_2 \approx 350$\,km\,s$^{-1}$. Using the primary mass function, $f(M_1) = P_{\rm orb} K_2^3 / 2 \pi G = M_1 \sin^3 i / (1+q)^2$, we find a primary mass $M_1 \gtrsim 0.8\,M_\odot$, which is compatible with the canonical mass for a neutron star ($M_1 = 1.4\,M_\odot$). While this value supports the same FWHM--$K_2$ relationship for the C\,{\sc iv} line in ultra-compacts as the H\,$\alpha$ line in typical X-ray binaries, we cannot be confident about this interpretation because of the poorly known inclination angle for 4U\,1626--67, and the lack of other test sources.

\begin{figure}
	\includegraphics[width=\columnwidth]{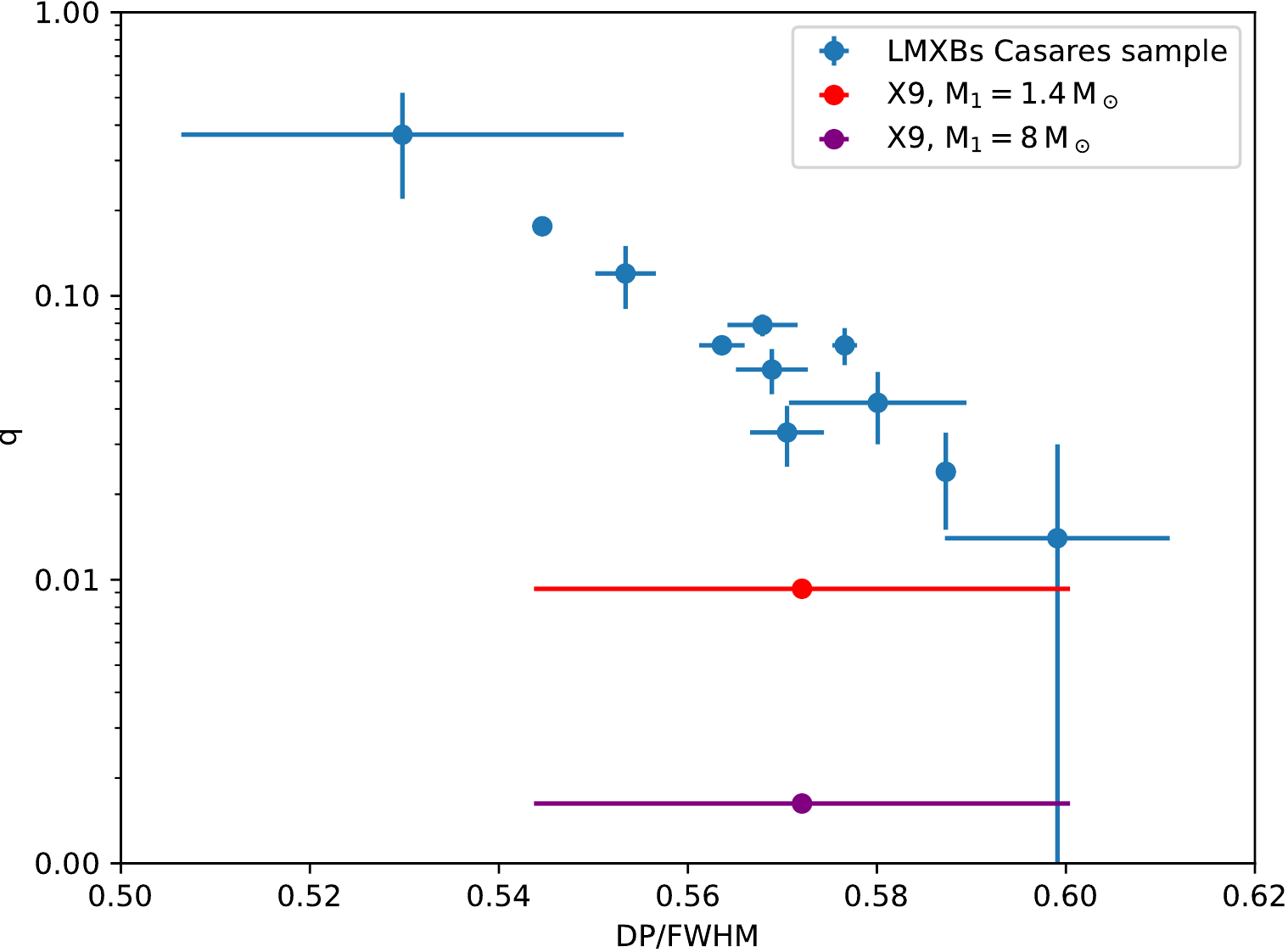}
	\caption{Mass ratio (q) versus DP/FWHM of the C\,{\sc iv} doublet of X9, compared with the H\,$\alpha$ line for the low-mass X-ray binary (LMXB) sample of \citet{2016ApJ...822...99C}. The X9 data are consistent with the system hosting either a neutron star or a black hole, but the signal to noise ratio of the C\,{\sc iv} doublet in X9 is too low to distinguish between the two scenarios.}
	\label{fig:q_vs_dpfwhm}
\end{figure}

Applied to X9, we predict $K_2 = 850 \pm 60$\,km\,s$^{-1}$ using the FWHM--$K_2$ relation. Similarly, we can use the empirically-found ratio between the outer disc velocity ($K_{\rm d}$) and the orbital velocity of the donor star ($K_2$) to reach a similar conclusion. This approach has previously been used to constrain the nature of the X-ray transient Swift\,J1357.2--0933 as a black hole \citep{2013Sci...339.1048C}. This ratio has been found to lie in the range $K_{\rm d}/K_2 = 1.1 - 1.2$ \citep{1994ApJ...436..848O}. Estimating $K_2$ based on DP is also more suitable than the FWHM-based method in the case of edge-on systems, where all of the line emission comes from the outer disc. In X9, DP = $2070\pm70$\,km\,s$^{-1}$ implies $K_2 = K_{\rm d} / 1.15 = ({\rm DP}/2) / 1.15 = 900 \pm 30$\,km\,s$^{-1}$. Both methods (FWHM--$K_2$ and DP--$K_2$) give similar $K_2$ values. The mass function implies a minimum primary mass $M_1 \geq 1\,M_\odot$ (for an edge-on system, $i = 90^\circ$, and a negligible donor mass), compatible with both a neutron star and a black hole. Alternatively, comparing the mass ratio and DP/FWHM in X9 with those of other X-ray binaries, we find that a neutron star is a slightly better fit than a black hole (Figure~\ref{fig:q_vs_dpfwhm}). The low DP/FWHM values in X9 could be explained by double peaks that originate further out in the Roche lobe than assumed for other X-ray binaries; in quiescent ultra-compact X-ray binaries, the outer disc is likely to truncate at a larger radius relative to the Roche lobe \citep{2012A&A...537A.104V}.

The C\,{\sc iv} emission could alternatively originate from the bright spot at the outer edge of the disc. The line luminosity emitted by the C\,{\sc iv} doublet is $L_{{\rm C}\textsc{iv}} \approx 6\times10^{30}$\,erg\,s$^{-1}$. In addition, the predicted luminosity of the stream/disc impact point was $L_{\rm impact} < 3 \times 10^{32}$\,erg\,s$^{-1}$ (Section~\ref{sec:source_emission}), a factor of 50 larger than the line luminosity, so it is also possible for the C\,{\sc iv} emission to originate from there as well.

For an inclination angle $i \gtrsim 30^\circ$ (Section~\ref{sec:orb_mod}), we restrict the primary mass to the range $M_1 \lesssim 10\,M_\odot$. The threshold inclination angle (for $M_1 = 3\,M_\odot$, between a neutron star and a black hole) is $i \approx 50^\circ$. We highlight that these estimates are based on the assumption that there is no C\,{\sc iv} absorption, the double-peaked C\,{\sc iv} emission is due to motion of the accretion disc, and that it originates from the same characteristic radius as the H\,$\alpha$ line in typical X-ray binaries. The mass and inclination angle estimates presented above could therefore be subject to large systematic errors, and should therefore only be used as rough guides.

In summary, the C\,{\sc iv} emission suggests $i \gtrsim 50^\circ$ for a neutron star, or $i \lesssim 50^\circ$ for a black hole. Using these results to constrain the model predictions from Section~\ref{sec:mods}, we find the optical emission in a neutron star to most likely come both from the jet and disc. For a black hole, the emission would come from X-ray reprocessing in the disc.


\subsection{X-ray spectrum}
\label{sec:x-spec}

One of the peculiarities of this source is that its X-ray spectrum is considerably harder than the spectra of other X-ray binaries with either a black hole or neutron star primary, at similar X-ray luminosities. This may, in fact, be a natural consequence of the abundances of the accreted material, and if this is the case, there are important implications for the likelihood that the accretor is a black hole, and for understanding emission mechanisms from accreting black holes at low accretion rates.

In the context of ADAF models \citep{1995ApJ...452..710N}, there are three major processes that can contribute to the X-ray emission: synchrotron emission, inverse compton emission, and bremsstrahlung emission. The seed photons for the Compton scattering may be the synchrotron photons, the bremsstrahlung photons, or, in the case of neutron star accretors, the photons produced in the boundary layer where the excess energy carried by the flow is dissipated near the surface of the neutron star. Ordinarily, the bremsstrahlung emission dominates only in the very outer parts of the emission region. In an ultra-compact binary, however, the bremsstrahlung emission should become considerably more important, since bremsstrahlung emissivity scales as $\langle Z^2 \rangle$, where $Z$ is the charge per nucleus. The bremsstrahlung emission in a system with a C/O white dwarf mass donor should be $\sim50$ times more important than in a system with a H donor. Systems with He white dwarf donor stars should have intermediate properties. 

A full treatment of this effect is well beyond the scope of this paper, but we can examine Figures 6 \& 7 of \citet{1995ApJ...452..710N} to obtain some intuition about how important the effect is likely to be. In their Figure 6(b), it is clear that, for neutron stars, the Compton emission dominates heavily over bremsstrahlung across the range of radii where nearly all the emission is produced. Therefore, if we assume that the reason for the hard X-ray spectrum is the relative importance of bremsstrahlung emission, a neutron star accretor is disfavoured. Indeed, neutron star ultra-compact systems have X-ray spectra as soft as those seen in other neutron star systems (e.g. for the persistent C/O-rich 2S\,0918--549 and 4U\,0614+091, $\Gamma \approx 2.1$; \citealp{2005A&A...441..675I}, and $\Gamma \approx 2.2$; \citealp{2010ApJ...710..117M}, respectively). There is no other obvious mechanism, either, for getting an especially hard X-ray spectrum from an accretion disk corona (ADC) source. For black hole accretors, on the other hand, Figure 6(a) of \citet{1995ApJ...452..710N} shows that, at the critical accretion rate, the bremsstrahlung emission is a factor of about 100 below the synchrotron level at the inner edge of the accretion disc, and falls off much more slowly with distance from the black hole. Furthermore, from their Figure 7, we see that, as the accretion rate drops, the crossover radius between bremsstrahlung and synchrotron / synchrotron self-Compton emission moves inwards. 

We propose that the harder X-ray spectrum of X9 relative to other low-accretion-rate systems is due to its C/O-dominated chemical composition, which boosts the bremsstrahlung component of the X-ray emission in the case of a black hole accretor. If so, high X-ray spectral hardness will be a key selection criterion to discovering more such objects in large surveys. A more detailed modelling of how the bremsstrahlung / synchrotron ratio changes with metallicity and nature of the primary is a promising avenue to uncover if X9 is a neutron star or black hole.

\subsection{Origin of orbital modulations}
\label{sec:orb_mod}

In the X-ray timing analysis of X9, \citet{2017MNRAS.467.2199B} suggested an orbital period of 28.2 minutes. This orbital modulation is characterised by long ($\sim30$\% of the orbit) and shallow drops in the X-ray light curve ($\approx 15$\% of flux), while the 27.2 min FUV signal is shallower ($\approx 5$\% of flux). 

In X-rays, orbital modulations can arise from the obscuration of the X-ray source, the reflection of X-rays by the companion star, or from changes in the mass accretion rate. Obscuration can be provided by cold clouds in the disc, a flared or warped disc, the stream/disc impact point (X-ray dips), or by the donor star (total or partial eclipses). Since the donor is too small to be obscuring the X-ray source (as already noted in \citealp{2017MNRAS.467.2199B}), the alternative is that X9 is similar to an X-ray dipper, where clumps in the disc are blocking a fraction of the X-rays. Indeed, periodic X-ray dips reflective of the binary period have been observed in a number of high-inclination X-ray binaries such as XB\,1916--053 \citep{1982ApJ...253L..61W}. Such dips are produced by a bulge in the disc from the collision of the accretion stream with the disc, and can appear at the outer disc edge \citep{1982ApJ...257..318W}, at the circularization radius \citep{1987A&A...178..137F}, or between the circularization radius and outer disc \citep{1996ApJ...470.1024A}. Regardless of where these impact bulges are located in the disc, they are stationary in the binary frame. Such bulges can stretch over a large fraction of the disc, potentially explaining the large fraction of the orbital period X9 spends in partial obscuration. An inner warped disk, which is likely in a dynamically-formed system where the spin of the primary is misaligned with respect to the orbital plane, could potentially supply the required obscuration even in more face-on configurations.

The modulation could also potentially arise from X-rays reflected off the donor. If the primary is a neutron star ($M_1 = 1.4\,M_\odot$), then $a = 0.34\,R_\odot$. For a radius $R_2 = 0.04\,R_\odot$, the donor occupies $f = 1/4 \, (R_2/a)^2 \approx 0.3$\% of the sky as seen from the primary, meaning that up to 0.3\% of the X-ray flux is reflected by the donor (less in the case of a heavier primary). The source of the orbital modulation (in X-rays) is therefore unlikely to be associated with reflection from the donor. The last possibility is that the orbital modulations are caused by an oscillating mass accretion rate. \citet{2009MNRAS.399.1633Z} found this to be unlikely in low-mass X-ray binaries because the viscous time-scale of the disc is much longer than the orbital period, dampening any variations in the mass accretion at the edge of the disc. We therefore conclude the orbital modulation in the X-ray band is most likely caused by obscuring material in the disc. Next, we attempt to constrain the inclination angle of the system.

For a donor of mass $M_2 = 0.013\,M_{\odot}$, the lack of X-ray eclipses in X9 restricts the inclination angle to $i \lesssim 85^\circ$ (assuming a point source for the X-ray emission, and a disc much thinner than the size of the donor; \citealp{1987A&A...178..137F}). Alternatively, the disc could fully shadow the donor star (a viable option for a disc opening angle $\theta \lesssim 5^\circ$; Section~\ref{sec:sorb}), or the signal to noise ratio of the phased X-ray orbital modulation might be too low to show eclipses, in which case edge-on systems ($i = 90^\circ$) are also allowed. Our interpretation of the orbital modulation as obscuration by features in the disc suggests $i \gtrsim 60^\circ$ \citep{1987A&A...178..137F}. Comparing the orbital modulation profile of X9 with other sources whose inclination is known is not straightforward. For many systems, the inclination angle is only an upper or lower limit based on the presence or lack of dips or eclipses in their light curves. There are a handful of examples of sources showing orbital modulations in X-rays, for which better estimates for their inclination angles exist. For Sco\,X--1, which has an orbital inclination of $i \approx 40^\circ - 50^\circ$ \citep{2002ApJ...568..273S}, there is evidence for a 1\% orbital modulation in the X-ray band \citep{2003PASP..115..739V}, although its physical origin is unclear. 4U\,1820--30 has a 685\,s orbital modulation with a 1.5\% amplitude \citep{1987ApJ...312L..17S}. The inclination of this system is $i \approx 20^\circ - 50^\circ$ \citep{1997ApJ...482L..69A, 2004ApJ...602L.105B}.

Based on the presence of orbital modulations in the X-ray light curve of X9, its inclination is likely to be in the range $i \gtrsim 60^\circ$, but could be as low as $i \approx 30^\circ$. These values are compatible with the acceptable spectral fits of \citet{2017MNRAS.467.2199B} in the 1--79\,keV band ($i < 68^\circ$) and in the 0.4--79\,keV band ($i < 66^\circ$). \citet{2017MNRAS.467.2199B} warn, however, that their estimates of the inclination angle could be inaccurate, since they rely on spectral models which do not take into account the metallicity of the reflecting disc.

\subsection{Origin of superorbital period}
\label{sec:sorb}

We have recovered the seven-day superorbital period of \citet{2017MNRAS.467.2199B} in the $V$ and $I$ bands, confirming the suggestion of a periodic X-ray modulation \citep{2017MNRAS.467.2199B}. The three significant periods in X9 (28.2 min, 27.2 min and seven days) are not consistent with being related by a beat relation. Specifically, we have $\Delta f \approx 1.95$ cycles per day (between the two short periods), but the long period of seven-days corresponds to 0.14 cycles per day.

Superorbital modulations can occur due to a change in the mass accretion rate, obscuration of the source, or a change in the inclination. These can be induced by precession (nodal -- of the inclination of disc plane, or apsidal -- of an eccentric disc in a fixed plane), which can have a variety of physical origins.

The semi-amplitude of the superorbital modulations is a factor of 1.9 in the X-rays, whereas in the optical it is only 5\%. For reprocessed X-rays, the optical emission should vary as $L_{\rm opt} \propto L_{\rm X}^{\beta}$, where $\beta = 1/2$ \citep{1994A&A...290..133V}. For X9, one would then naively expect a semi-amplitude of a factor of $\approx 1.4$ in the optical band. Similarly, if the optical emission is dominated by the viscously heated disc, $\beta = 0.17-0.5$, or if the jet is the the source of emission, $\beta = 0.7-1.4$, in all cases predicting much larger (by at least a factor of two) superorbital modulations than what is observed. The two ways for a steadily accreting source to have such a large disparity between the modulations of the two bands is if a non-variable optical source dominates the optical emission in X9, or if these modulations are not driven by changes in mass accretion. The former scenario can occur by emission from the donor, or if an additional stellar companion, or another unassociated cluster star dominate the optical luminosity. If that were the case, however, the star would need to be at least as bright as the accretion flow. The spectral energy distribution would then be closer to a white dwarf atmosphere (with absorption lines), and the superorbital modulations would likely be different between the $V$ and $I$ bands (set by the different temperatures of the star and accretion flow). We do not observe any of these three features, so it is unlikely that a stellar object (donor or otherwise) drowns out the superorbital variability from the accretion flow. 

Geometric effects such as occultation or changes of projected surface area are more likely to be causing the observed modulations. In the case of viewing angle effects and assuming $F_{\rm X} \propto (\cos i)^{1/2}$ (if X-rays are emitted from close to the disc surface), a face-on disc ($i = 0$) at maximum flux would require an almost edge-on disc ($i \approx 85^\circ$) at minimum flux. It is therefore difficult to invoke simple oscillations in the projected disc area as an explanation. We therefore find periodic obscuration to be the most likely interpretation for the superorbital period. In that case, the intrinsic bolometric luminosity of X9 needs to be at least as large as the maximum observed bolometric luminosity ($L_{\rm bol} \gtrsim 5 \times 10^{34}$\,erg\,s$^{-1}$). 

If mass transfer is driven solely by gravitational wave radiation, the maximum luminosity allowed by mass transfer to a neutron star or black hole is $L_{\rm bol, NS} \lesssim 2\times 10^{35}$\,erg\,s$^{-1}$ or $L_{\rm bol, BH} \lesssim 2 \times 10^{34}$\,erg\,s$^{-1}$, respectively (Section~\ref{sec:mdot}). The former value is compatible with the estimated bolometric luminosity, but the latter value is a factor of $\sim$2.5 smaller than required for a black hole. However, the uncertainties associated with calculating the mass accretion rate may be as large as an order of magnitude, so we cannot rule out a black hole.

One explanation for the high measured radio to X-ray luminosity ratio, relative to neutron stars \citep{2015MNRAS.453.3918M, 2017MNRAS.467.2199B}, could be that X9 is an ADC source, which would reduce the observed X-ray luminosity due to obscuration of the inner disc by a bulge in the disc. This shows that if X9 is an ADC source, its intrinsic X-ray ($L_{\rm 1-10\,keV} \sim 10^{34}$\,erg\,s$^{-1}$) and radio luminosities would be in line with other neutron stars. Otherwise, a black hole would be the most radio-faint at that X-ray luminosity.

The large difference between the observed superorbital amplitudes in the X-ray and optical bands is either caused by a more variable obscuration fraction in the X-rays, or by a larger absorption cross-section at short wavelengths. Depending on the geometry of the system, different obscuration factors for the X-ray and optical sources can be achieved. If $i \approx 90^\circ$, the obscuration of the optical source (be it the jet or edge of the disc, but not the surface of the disc) varies little as the disc precesses, but how much of the X-ray emission is absorbed depends on the geometry of the outer disc. If $i < 80^\circ$, and the optical emission is dominated by emission at the disc surface, only the inner regions of the disc may periodically occult the central X-ray source (otherwise the superorbital modulations would be prominent in the optical band too).

Next, we investigate whether the outer disc can reach the required height to obscure a significant part of the X-ray corona. In the simple case of a thin disc, the height of the disc at the outer edge is given by $H \approx c_{\rm s} P_{\rm orb} / (2 \pi)$ \citep{2002apa..book.....F}, where $c_{\rm s}$ is the local sound speed, $c_{\rm s} = \sqrt{k_{\rm B} T / (\mu m_{\rm H})}$ ($k_{\rm B}$ is Boltzmann's constant, $T$ is the temperature at the outer disc, $\mu$ is the mean atomic weight of the accreted gas, and $m_{\rm H}$ is the atomic weight for atomic H). Taking $T = 22,000$\,K (the estimated upper limit for the donor; Section~\ref{sec:donor}) and $\mu = 7$ (for a singly-ionized C/O mixture), we find $H \approx 10^8$\,cm, or a disc opening angle $\theta < 1^\circ$. Such a small opening angle cannot produce the observed superorbital modulations. However, energy deposited by the accretion stream on the outer edge of the disc can thicken it \citep{1982ApJ...257..318W}. In typical low-mass X-ray binaries, the opening (semi-)angle of the disc is $\theta \approx 12^\circ$ \citep{1996A&A...314..484D}, but metal-rich discs can be a factor of $\approx$2 thinner (keeping all other disc parameters the same; \citealp{2006AstL...32..257D}). Coupled with a weak dependence of the aspect ratio on the radius ($H/R \propto R^{1/8}$; \citealp{2002apa..book.....F}), the outer disc in X9 is likely to be on the order of $\theta \approx 5^\circ$. The surface layer of the accretion disc is likely to be optically thin, so we take the above value as an upper limit to the height of the optically-thick region. The disc obscures the X-ray corona and inner disc for inclination angles $i \gtrsim 85^\circ$.

In view of the broadband modelling (Section~\ref{sec:mods}) and C\,{\sc iv} emission (Section~\ref{sec:civ}) results, the X-ray corona can either be obscured by the inner or outer discs in the case of a neutron star ($i > 60^\circ$, disc and jet emission), but only by the inner disc in the case of a black hole ($i < 30^\circ$, disc emission only).

Next, we investigate the potential mechanisms behind disc precession, which are detailed in Appendix~\ref{sec:mechs}. We find that the seven-day superorbital period in the X9 system could either be driven by tidal effects, or the Kozai mechanism (in a hierarchical triple system) in the case of a neutron star. In the case of a black hole, we have not found a mechanism that could give rise to the observed superorbital period. We cannot readily find a physical explanation to the short FUV period (27.2 min, in conflict with the 28.2 min X-ray period), which, at just over 2$\sigma$ significance, could be spurious. The broadband models and the C\,{\sc iv} emission cannot be used to distinguish between a neutron star and a black hole. Given the lack of an explanation for the observed superorbital period in the case of a black hole, but the presence of a hard X-ray spectrum ($\Gamma \approx 1.1$), which is more likely to be produced in a black hole system, the nature of the primary remains unsolved.

\subsection{Formation and evolution}
\label{sec:simul}

\begin{figure}
	\includegraphics[width=\columnwidth]{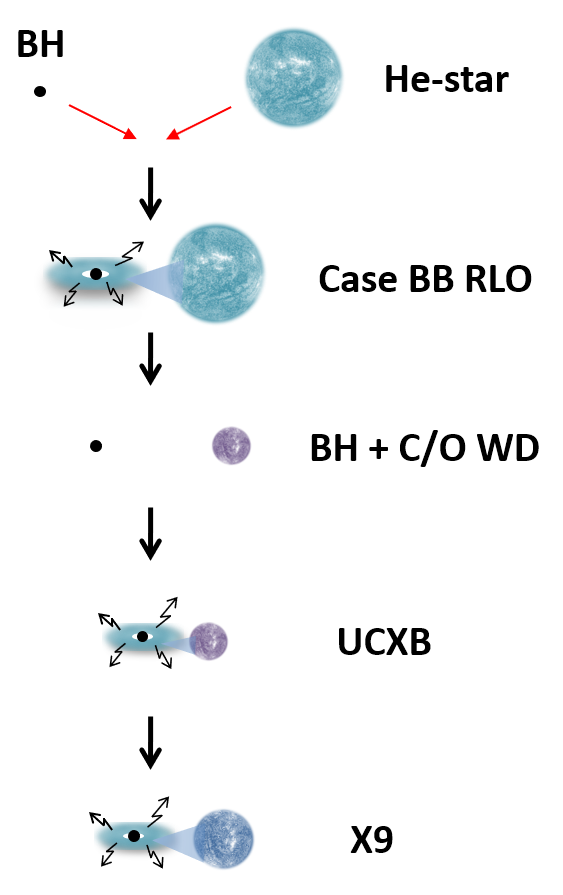}
	\caption{Illustration showing the evolutionary sequence of a $2.0\,M_\odot$ He-star in orbit with a $5.0\,M_\odot$ black hole (BH). The binary loses angular momentum via gravitational wave radiation, leading to case BB mass transfer (after core He is exhausted, before carbon ignition), and eventually to a black hole and C/O white dwarf (WD) binary. The next stage of this system is an ultra-compact X-ray binary (UCXB) resembling X9.}
	\label{fig:cartoon}
\end{figure}

Given that there are already many binary evolution calculations in the literature for neutron star ultra-compact X-ray binaries, but essentially none for ultra-compacts with black hole primaries, we decided to explore at least one simple channel that might produce such a system (illustrated in Figure~\ref{fig:cartoon}). 

We evolved a $2.0\,M_\odot$ He-star and a $5\,M_\odot$ black hole with an initial orbital period of 0.1673 days (4.015 hr). These initial masses were chosen to reflect the likely properties of the components of X9: a C/O white dwarf donor (progenitor of a He-star of mass $M < 2.1\,M_\odot$; \citealp{2002MNRAS.331.1027D}), and a light black hole; see Section~\ref{sec:civ}). Such a He-star may originate from a $5-6\,M_\odot$ star formed in the globular cluster 10--12 Gyr ago, which subsequently lost its outer layers of H in a stellar collision or mass-transfer episode. For this simulation, we used the {\tt BEC} binary evolution code, originally developed by \citet{1997PhDT........55B} on the basis of a single-star code \citep{1998A&A...329..551L} which solves the hydrodynamic structure and evolution equations \citep{1990sse..book.....K} using Lagrangian methods. More details about an updated version of this code applied to X-ray binaries can be found in \citet{2012MNRAS.425.1601T}, \citet{2014A&A...571A..45I} and \citet{2015MNRAS.451.2123T}. 

\begin{figure}
	\begin{subfigure}[b]{\columnwidth}
	\includegraphics[width=\textwidth]{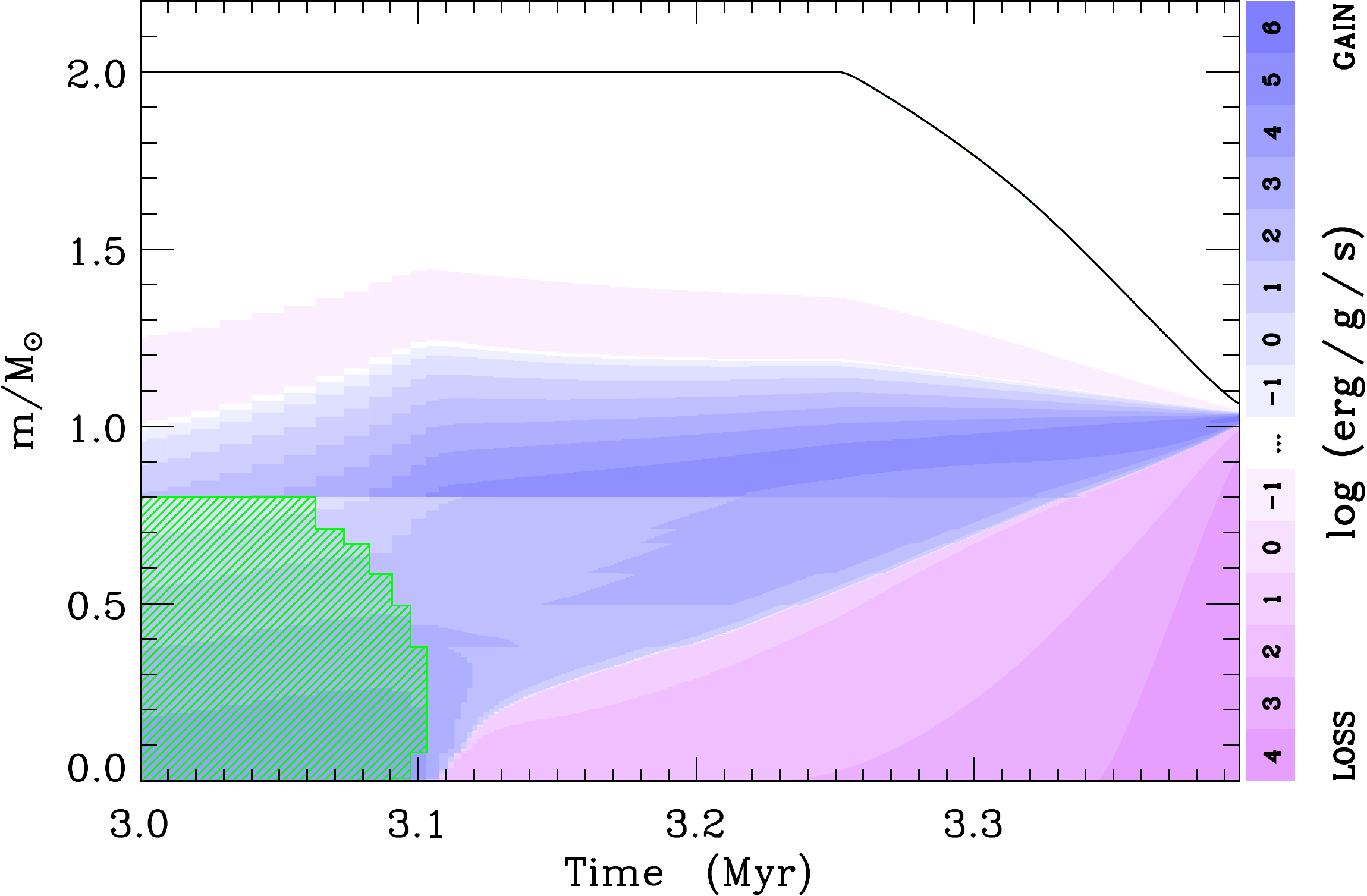}
	\caption{Kippenhahn diagram}
	\label{fig:kipp}
	\end{subfigure}
	\hfill
	\begin{subfigure}[b]{\columnwidth}
	\includegraphics[trim={0 0cm 0cm 1.0cm},clip,width=\textwidth]{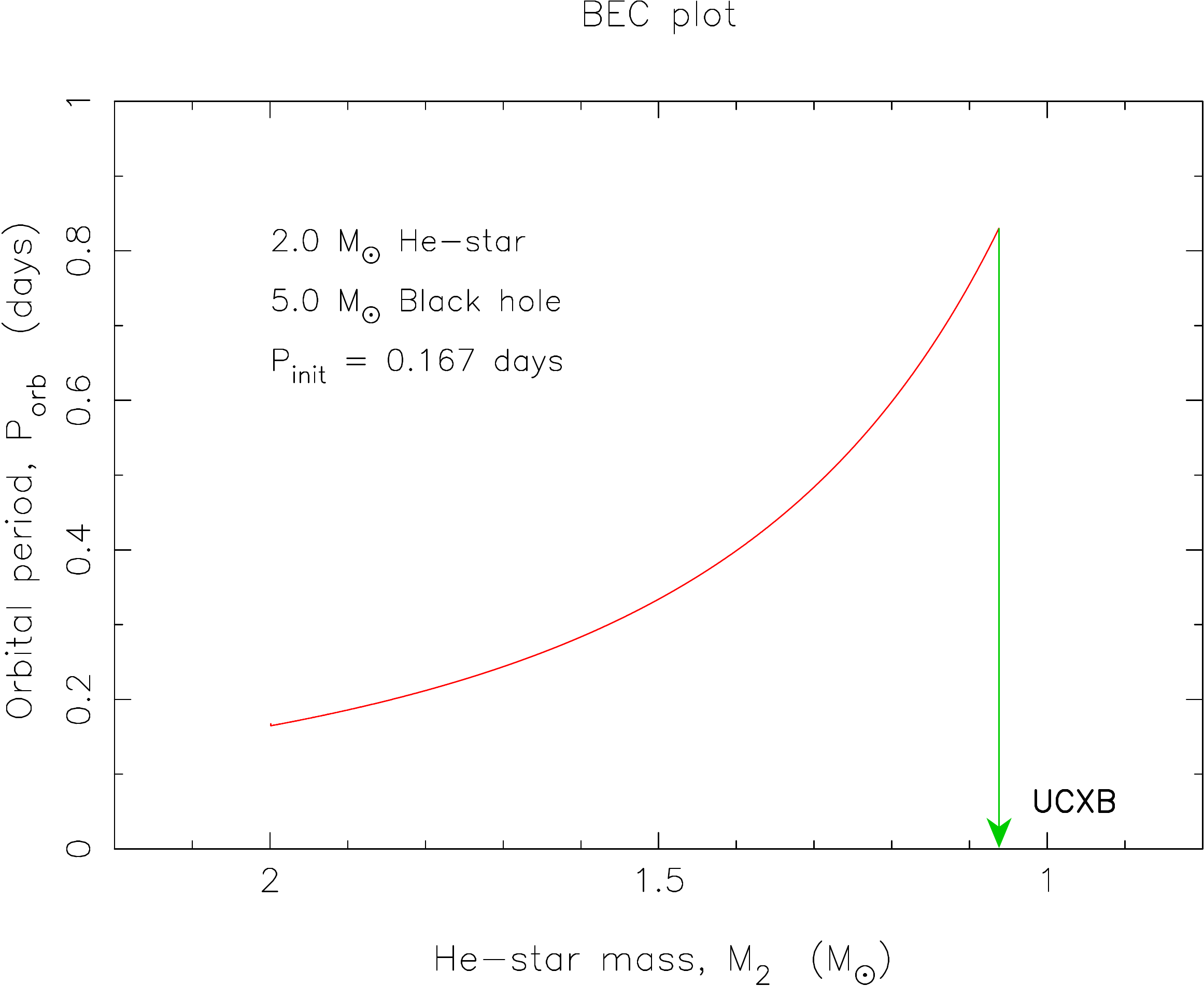}
	\caption{Evolution of the orbital period}
	\label{fig:p_evo}
	\end{subfigure}
	\hfill
	\begin{subfigure}[b]{\columnwidth}
	\includegraphics[width=\textwidth]{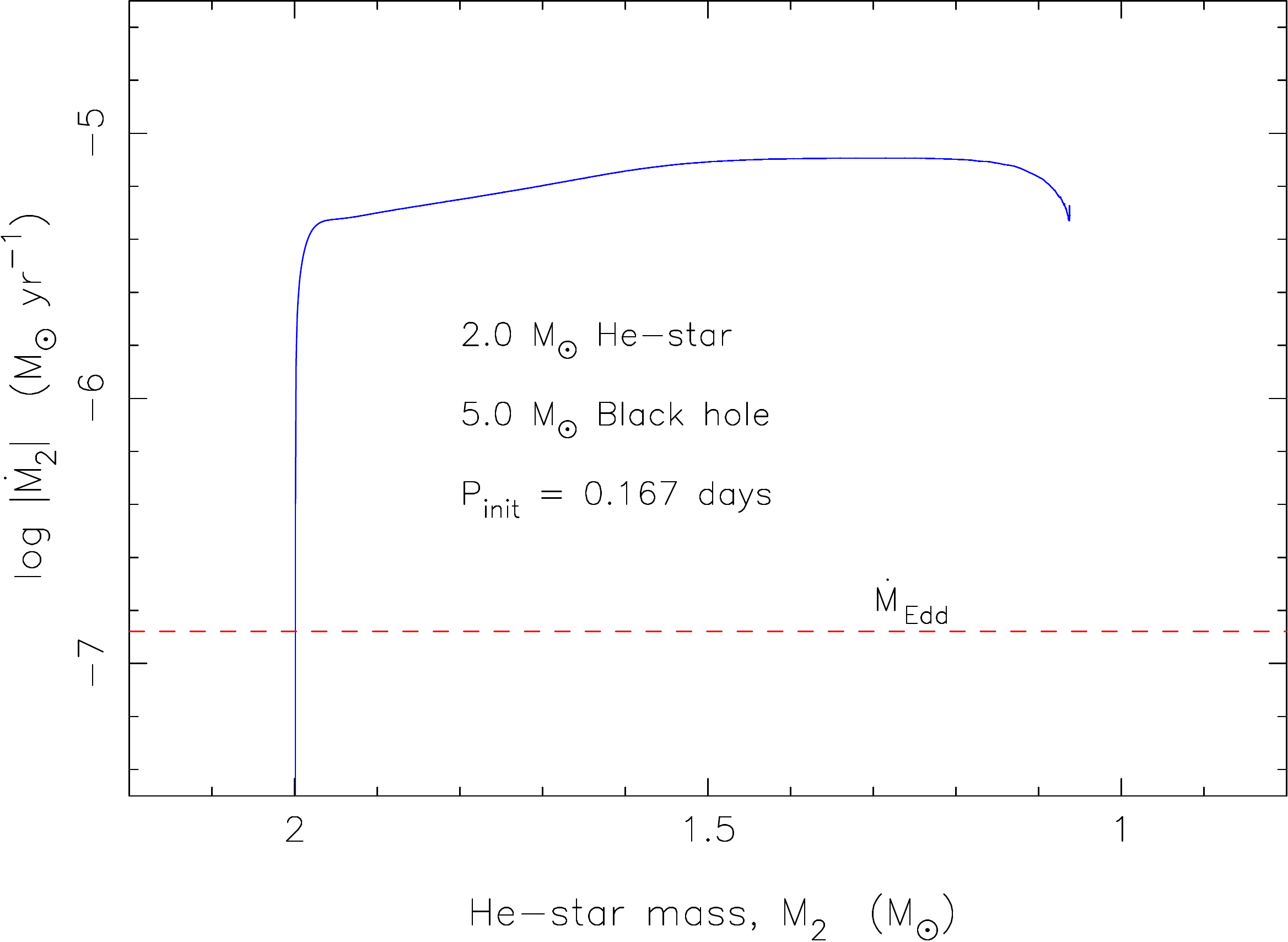}
	\caption{Evolution of the mass-transfer rate}
	\label{fig:mdot_evo}
	\end{subfigure}
\caption{The evolution of a $2.0\,M_\odot$  He-star + $5\,M_\odot$ black hole binary, starting at an initial orbital period $P_{\rm orb} = 0.167$\,d. (\subref{fig:kipp}) Kippenhahn diagram (mass coordinate against time) of the donor. At the beginning of the simulation, He burns in a convective core (green hatched area) surrounded by a radiative shell. Once central He is exhausted, the star continues to burn He in a shell around the degenerate C/O core (pink area), driving the star to expand, leading to RLO (starting at 3.25\,Myr). (\subref{fig:p_evo}) Orbital expansion as a result of mass transfer from the donor star (evolution from left to right). (\subref{fig:mdot_evo}) Mass-loss rate from the donor star ($\dot{M}_2 \approx 10^{-6} - 10^{-5}\,M_\odot$\,yr$^{-1}$) exceeding the Eddington accretion rate of the black hole ($\dot{M}_2 \approx 10^{-7}\,M_\odot$\,yr$^{-1}$, red dashed line), leading to isotropic re-emission of matter from the system.} 
\label{fig:evo}
\end{figure}

We assume the binary to be losing orbital angular momentum via gravitational radiation only (no significant magnetic braking is expected to occur because the He-star does not have an outer convective zone; \citealp{2002ApJ...565.1107P}). The binary shrinks its orbit due to gravitational wave radiation until the He-star fills its Roche lobe and initiates mass transfer. For the first 3.1 Myr of the simulation, before Roche-lobe overflow (RLO) starts, the donor is burning He in a convective core. After 3.1 Myr, the C/O core (which is not massive enough to fuse heavier elements) becomes degenerate, and  He-burning continues in the radiative shell (Figure~\ref{fig:kipp}). Due to the subsequent radial expansion of the star, mass transfer is initiated after 3.25~Myr (Case BB RLO; \citealp{2015MNRAS.451.2123T}), at a rate of $\dot{M}_2 \approx 10^{-6}-10^{-5}\,M_\odot$yr$^{-1}$ (Figure~\ref{fig:mdot_evo}). In this super-Eddington regime, excess material is blown away from the vicinity of the black hole (isotropic re-emission). This rapid mass loss leads to an almost naked C/O core (with a tiny $0.04\,M_\odot$ surface layer of He) which will detach after $1.5\times 10^5\;{\rm yr}$ and produce a C/O white dwarf. The outcome is thus a $1.062\,M_\odot$ C/O white dwarf orbiting a $5.016\,M_\odot$ black hole with $P_{\rm orb}=0.830$ days (19.9 hr). As a result of gravitational wave radiation, this system will become an ultra-compact X-ray binary (i.e. a progenitor to X9) after 9.81~Gyr (less than the age of the cluster, $\approx 11$ Gyr; \citealp{2003A&A...408..529G}), thereby quickly losing the leftover white dwarf surface layer of He. Adjusting the initial orbital period can lead to an ultra-compact X-ray binary after 1--13 Gyr. Hence, we find that an initial binary made up of a black hole and a He-star can lead to the formation of an ultra-compact X-ray binary with a C/O white dwarf donor within the lifetime, or in the recent history of 47~Tuc. This formation path need not necessarily have been followed by X9 (if it is a black hole), and we only suggest it as a possible scenario. It is also possible that any initial binary would exchange energy with surrounding stars/binaries via dynamical interactions in the cluster \citep{2010ApJ...717..948I}, especially in the early stages of its evolution.

Previously, similar simulations of He-star and neutron star binaries leading to the formation of ultra-compact systems have been performed \citep[e.g.][]{2013A&A...552A..69V, 2013ApJ...768..184H}. In certain cases, these calculations show rapid contraction of the orbit while the He-star is transferring mass and burning He, only switching to orbital expansion at later stages. This is in contrast to our simulation, where the binary undergoes orbital expansion at all stages of the mass transfer (Figure~\ref{fig:p_evo}). This difference is the result of applying different component masses and initial orbital periods. In particular, \citet{2013A&A...552A..69V} followed low mass ($0.35-1.0$ $M_\odot$) He-stars transferring mass to a 1.4 $M_\odot$ neutron star and assuming initial orbital periods between 17--48~min. In our simulation, the component masses are 2.0 $M_\odot$ and 5.0 $M_\odot$, and the initial orbital period is 96~min. Despite the larger masses, the long orbital period strongly decreases the efficiency of loss of orbital angular momentum from gravitational wave radiation. Also, a more massive  He-star evolves much faster, and the initial mass ratio in our case is $q = 0.4$, whereas in the work of \citet{2013A&A...552A..69V} the applied mass ratio was often closer to unity. Lastly, our He-star initiates RLO while it is still undergoing He shell burning (hence, a (sub)giant donor star), and thus the evolutionary time-scale during mass transfer is relatively small. All of these effects will cause our system to widen once RLO starts.

We can also consider the plausibility of the black hole scenario in light of the already existing evidence for at least one ultra-compact X-ray binary with a black hole primary observed as a ULX in an extragalactic globular cluster. Although at least some ULXs ($L_{\rm X} > 10^{39}$\,erg\,$^{-1}$) host neutron stars \citep{2014Natur.514..202B, 2017Sci...355..817I, 2017MNRAS.466L..48I}, the majority of globular cluster ULXs may still host black hole primaries. This is because ULXs powered by accretion on to neutron stars are thought to require high surface magnetic fields ($B > 10^{11}$\,G; \citealp{2016MNRAS.458L..10K, 2015MNRAS.448L..40E, 2016MNRAS.457.1101T}). In globular clusters, however, pulsars typically have much weaker magnetic fields ($B \approx 10^9$\,G), with only four out of $>$140 known globular cluster pulsars reaching $B \approx 10^{11}-10^{12}$\,G \citep{2011ApJ...742...51B}.

The best candidate for a black hole with a white dwarf donor in a ULX is XMMU\,J122939.7+075333 \citep{2007Natur.445..183M}, associated with the globular cluster RZ\,2109 in the Virgo galaxy NGC\,4472 \citep{2001AJ....121..210R}. It has an X-ray luminosity of about $5\times10^{39}$ erg\,s$^{-1}$ and shows clear evidence for strong forbidden oxygen emission lines, with stringent upper limits for H or He emission lines \citep{2007ApJ...669L..69Z, 2008ApJ...683L.139Z, 2014ApJ...785..147S}. The lack of any other globular cluster in NGC\,4472 with similar emission lines \citep{2012ApJ...752...90P} strongly argues against anisotropic emission being responsible for the brightness of the source, since the emission from the forbidden lines should be isotropic. Another system, CXOU\,J1229410+075744, in the same galaxy, shows variability of a factor of four with a peak luminosity of about $2\times10^{39}$ erg\,s$^{-1}$, but is persistent over a time-scale of more than a decade \citep{2011MNRAS.410.1655M}. This persistently bright emission could be explained through either a long period binary with a highly evolved donor \citep[e.g.,][]{2006MNRAS.368L..25T}, or an ultra-compact binary. Another candidate is CXOKMZ\,J033831.7--353058 in NGC\,1399 in the Fornax cluster \citep{2010ApJ...721..323S}.

It is not straightforward to determine the number of globular clusters observed by Chandra in which one could have established the presence of a source which varies by significantly more than the Eddington luminosity of a neutron star. However, simply by adding the numbers of clusters detected by {\it HST} in the Virgo and Fornax cluster surveys, we are likely to incur uncertainties no larger than the uncertainties that come from basing our population statistics on the mere presence of XMMU\,J122939.7+075333. In the Virgo cluster, there are about 13,000 globular clusters in {\it HST} data \citep{2009ApJS..180...54J}, and in the Fornax cluster survey, there are about 9,000 \citep{2015ApJS..221...13J}. To be clear, there are other galaxies which have been well observed by Chandra for which there are catalogues of globular clusters from either {\it HST} or ground-based data; and there are, as well, wider-field ground-based data for globular clusters in Virgo. On the other hand, many of the smaller Virgo Cluster galaxies have X-ray data only from the AMUSE project \citep{2008ApJ...680..154G}. It is reasonable to estimate that about 20,000 clusters have been well-enough observed by Chandra to allow for the identification of strong ultra-compact black hole candidates, and that $\sim$1 of them has been discovered. Thus, we can estimate that at least $10^{-5}$ of clusters contain bright ultra-compact black hole X-ray binaries, and use this number to determine whether it is plausible to find a much fainter one in the Milky Way's globular cluster system.
 
Next, we look at the evolutionary time-scales of ultra-compact black hole X-ray binaries \citep{2012A&A...537A.104V}. If we take XMMU\,J122939.7+075333 to have an orbital period of about 5 minutes, as would be expected given its X-ray luminosity, its lifetime should be just under $10^5$ years. For a source with an orbital period of half an hour, the expected characteristic age is just under $10^8$ years. There should, thus, be $\approx1000$ black hole X-ray binaries at an orbital period of half an hour for every one at an orbital period of 5 minutes. In turn, assuming the $10^{-5}$ fraction from above, and the number of globular clusters in the Milky Way ($> 100$; \citealp{1996AJ....112.1487H}) we expect $\sim$1 Milky Way cluster to have a black hole X-ray binary at an orbital period of less than half an hour. It is thus reasonable that we have seen such an object. Furthermore, assuming that these systems formed mostly relatively early in the lifetimes of the clusters ($\sim 10^{10}$ years), we may expect $\sim100$ times as many ultra-compact black hole X-ray binaries at any orbital period (i.e. of order one per cluster, but with a higher concentration expected in the clusters with higher stellar interaction rates). Similarly, \citet{2017ApJ...843L..30I} estimate globular clusters should contain about one faint black hole\,--\,white dwarf binary per $10^5\,M_\odot$. These additional systems should be at longer periods, be considerably fainter, and exhibit transient outbursts. In quiescence, some of them could be nearly indistinguishable from cataclysmic variables, except in observations using radio emission or optical/UV spectroscopy, and they may easily be among the sources classified as quiescent dwarf novae in existing Chandra work. However, the majority of these old, longer-period ($\sim$2\,h) systems might be too faint to be detectable.

\section{Conclusions}

We acquired a broadband optical (3,000--10,000\,{\AA}) spectrum of X9, finding a blue continuum with no identifiable emission (or absorption) lines. In particular, we place a $3\sigma$ upper limit on the H\,$\alpha$ line in emission of $\rm -EW < 14$\,{\AA}. This limit is below that of H-rich X-ray binaries observed at similar X-ray luminosities ($L_{\rm 2-10\,keV} = 10^{33}-10^{34}$\,erg\,s$^{-1}$), which typically have $\rm -EW_{H\,\alpha} \sim 50$\,{\AA}. This result strengthens the evidence for a C/O white dwarf donor as identified by \citet{2017MNRAS.467.2199B}.

Performing a timing analysis on archival photometric $V$ and $I$-band data, we find the most significant period is $P \approx 7$ days, consistent with the superorbital period suggested by \citet{2017MNRAS.467.2199B} based on X-ray data. In these bands there is no evidence for any signal at the orbital period. Timing analysis of archival FUV data returns one marginally significant period (at less than $3\sigma$ significance): 27.2 min. This value is close to, but not equal to the 28.2 min orbital period identified by \citet{2017MNRAS.467.2199B} (who estimated a significance of $> 5\sigma$ for this period). 

The superorbital modulations are weak in the optical bands (5\% semi-amplitude), and strong in the X-ray band (factor of 3 semi-amplitude), suggesting the occultation of an extended X-ray corona is probably driving this variability. If this interpretation is correct, the intrinsic luminosity of X9 could be stable at a level of $L_{\rm 2-10\,keV} = 10^{34}-10^{35}$\,erg\,s$^{-1}$. In this case, its radio luminosity makes a neutron star a more likely interpretation. Otherwise, it would be the radio-faintest known black hole at this X-ray luminosity. The C\,{\sc iv} doublet and the broadband spectrum indicate that if X9 is a neutron star, it is observed at a high inclination ($i > 60^\circ$) and the optical emission is dominated by the jet and disc. For a black hole, the emission would be dominated by reprocessing of X-rays by the disc, and is seen at a low inclination ($i < 30^\circ$). While the seven-day superorbital modulation can be explained by the precession of the disc due to tidal torques exerted by the companion star to a neutron star, we cannot find a way to produce the same modulation for a black hole. The estimate for the inclination angle based on the presence of orbital modulations in the X-ray band ($i \gtrsim 60^\circ$, favouring a neutron star) is slightly at odds with the estimation based on the X-ray spectrum ($i \lesssim 45^\circ$ to $i \lesssim 68^\circ$ depending on the model, favouring a black hole). The hardness of the X-ray spectrum is easier to explain in the context of a black hole. Presently, not all the observables of X9 can easily be explained by either a neutron star or black hole. 

While we know of neutron star ultra-compact systems and theoretical pathways to produce such objects, the latter pathways have not been proposed for black hole ultra-compacts. We therefore performed binary evolution calculations of a 2.0 $M_\odot$ He-star and a 5.0 $M_\odot$ black hole binary, to show that if X9 does host a black hole, such a system could indeed be produced in a globular cluster. We note that while this work was under review, \citet{2017ApJ...851L...4C} found that based on hydrodynamic simulations, C/O white dwarf donors in globular clusters can form X-ray binaries with stable mass transfer only if their primaries are black holes.

The easiest way to confirm the nature of this system would be if it went into a bright outburst; pulsations or thermonuclear bursts would then verify a neutron star nature. Otherwise, phase-resolved FUV spectroscopy, through the C\,{\sc iv} or Mg\,{\sc ii} lines, might be used to solve the system parameters if they arise from the heated face of the donor. Unfortunately, the Bowen fluorescence lines (driven by He\,{\sc ii} emission) typically used for this purpose will not be produced in a C/O donor. Another way to confirm the parameters of X9 might be the detection of its gravitational waves. The strain amplitude for a binary with a circular orbit is given by \citep{2004MNRAS.349..181N}:
\begin{equation}
h = 5 \times 10^{-22} \left( \frac{\mathcal{M}}{M_\odot} \right)^{5/3} \left( \frac{P_{\rm orb}}{\rm 1 hr} \right)^{-2/3} \left( \frac{d}{\rm 1 kpc} \right)^{-1} ,
\end{equation}
where $\mathcal{M} = (M_1 M_2)^{3/5} / (M_1 + M_2)^{1/5}$ is the chirp mass of the system. For $M_1 = 8\,M_\odot$, $M_2 = 0.016\,M_\odot$, $h = 1\times10^{-23}$. At the gravitational wave frequency $\nu_{\rm GW} = 1.2 \times 10^{-3}$\,Hz (twice the orbital frequency), X9 will not be detectable to the gravitational wave detector \textit{LISA} even if it hosts a black hole, as it falls below the Galactic foreground noise from double white dwarf \citep{2004MNRAS.349..181N}.

\section*{Acknowledgements}
We thank the anonymous referee for the helpful comments. VT acknowledges a CSIRS scholarship from Curtin University. JCAM-J is the recipient of an Australian Research Council Future Fellowship (FT140101082). JS and AB acknowledge support from the Packard Foundation and NSF grant AST--1308124. COH and GRS are supported by NSERC Discovery Grants, and COH by a Discovery Accelerator Supplement. RMP acknowledges support from Curtin University through the Peter Curran Memorial Fellowship. FB received funding from the European Union's Horizon 2020 research and innovation programme under the Marie Sklodowska-Curie grant agreement no. 664931. We thank M. D{\'{\i}}az Trigo for discussions on the ADC scenario, S. Lockwood for help with the {\it STIS} calibration and R. Urquhart for help with {\sc Xspec}. Based on observations made with the NASA/ESA {\it Hubble Space Telescope}, obtained at the Space Telescope Science Institute, which is operated by the Association of Universities for Research in Astronomy, Inc., under NASA contract NAS 5--26555. Support for Program number HST--GO--14203.002 was provided by NASA through a grant from the Space Telescope Science Institute. The International Centre for Radio Astronomy Research is a joint venture between Curtin University and the University of Western Australia, funded by the state government of Western Australia and the joint venture partners. This research has made use of NASA's Astrophysics Data System. STSDAS and PyRAF are products of the Space Telescope Science Institute.



\bibliographystyle{mnras}
\bibliography{x9-hst}

\begin{thebibliography}{}
\makeatletter
\relax
\def\mn@urlcharsother{\let\do\@makeother \do\$\do\&\do\#\do\^\do\_\do\%\do\~}
\def\mn@doi{\begingroup\mn@urlcharsother \@ifnextchar [ {\mn@doi@}
  {\mn@doi@[]}}
\def\mn@doi@[#1]#2{\def\@tempa{#1}\ifx\@tempa\@empty \href
  {http://dx.doi.org/#2} {doi:#2}\else \href {http://dx.doi.org/#2} {#1}\fi
  \endgroup}
\def\mn@eprint#1#2{\mn@eprint@#1:#2::\@nil}
\def\mn@eprint@arXiv#1{\href {http://arxiv.org/abs/#1} {{\tt arXiv:#1}}}
\def\mn@eprint@dblp#1{\href {http://dblp.uni-trier.de/rec/bibtex/#1.xml}
  {dblp:#1}}
\def\mn@eprint@#1:#2:#3:#4\@nil{\def\@tempa {#1}\def\@tempb {#2}\def\@tempc
  {#3}\ifx \@tempc \@empty \let \@tempc \@tempb \let \@tempb \@tempa \fi \ifx
  \@tempb \@empty \def\@tempb {arXiv}\fi \@ifundefined
  {mn@eprint@\@tempb}{\@tempb:\@tempc}{\expandafter \expandafter \csname
  mn@eprint@\@tempb\endcsname \expandafter{\@tempc}}}

\bibitem[\protect\citeauthoryear{{Albrow}, {Gilliland}, {Brown}, {Edmonds},
  {Guhathakurta}  \& {Sarajedini}}{{Albrow} et~al.}{2001}]{2001ApJ...559.1060A}
{Albrow} M.~D.,  {Gilliland} R.~L.,  {Brown} T.~M.,  {Edmonds} P.~D.,
  {Guhathakurta} P.,   {Sarajedini} A.,  2001, \mn@doi [\apj] {10.1086/322353},
  \href {http://adsabs.harvard.edu/abs/2001ApJ...559.1060A} {559, 1060}

\bibitem[\protect\citeauthoryear{{Allard}, {Homeier}  \& {Freytag}}{{Allard}
  et~al.}{2012}]{2012RSPTA.370.2765A}
{Allard} F.,  {Homeier} D.,   {Freytag} B.,  2012, \mn@doi [Philosophical
  Transactions of the Royal Society of London Series A]
  {10.1098/rsta.2011.0269}, \href
  {http://adsabs.harvard.edu/abs/2012RSPTA.370.2765A} {370, 2765}

\bibitem[\protect\citeauthoryear{{Anderson}, {Margon}, {Deutsch}, {Downes}  \&
  {Allen}}{{Anderson} et~al.}{1997}]{1997ApJ...482L..69A}
{Anderson} S.~F.,  {Margon} B.,  {Deutsch} E.~W.,  {Downes} R.~A.,   {Allen}
  R.~G.,  1997, \mn@doi [\apjl] {10.1086/310672}, \href
  {http://adsabs.harvard.edu/abs/1997ApJ...482L..69A} {482, L69}

\bibitem[\protect\citeauthoryear{{Armitage} \& {Livio}}{{Armitage} \&
  {Livio}}{1996}]{1996ApJ...470.1024A}
{Armitage} P.~J.,  {Livio} M.,  1996, \mn@doi [\apj] {10.1086/177928}, \href
  {http://adsabs.harvard.edu/abs/1996ApJ...470.1024A} {470, 1024}

\bibitem[\protect\citeauthoryear{{Arnaud}}{{Arnaud}}{1996}]{1996ASPC..101...17A}
{Arnaud} K.~A.,  1996, in {Jacoby} G.~H.,  {Barnes} J.,  eds,  Astronomical
  Society of the Pacific Conference Series Vol. 101, Astronomical Data Analysis
  Software and Systems V. p.~17

\bibitem[\protect\citeauthoryear{{Bachetti} et~al.,}{{Bachetti}
  et~al.}{2014}]{2014Natur.514..202B}
{Bachetti} M.,  et~al., 2014, \mn@doi [\nat] {10.1038/nature13791}, \href
  {http://adsabs.harvard.edu/abs/2014Natur.514..202B} {514, 202}

\bibitem[\protect\citeauthoryear{{Bahramian} et~al.,}{{Bahramian}
  et~al.}{2014}]{2014ApJ...780..127B}
{Bahramian} A.,  et~al., 2014, \mn@doi [\apj] {10.1088/0004-637X/780/2/127},
  \href {http://adsabs.harvard.edu/abs/2014ApJ...780..127B} {780, 127}

\bibitem[\protect\citeauthoryear{{Bahramian} et~al.,}{{Bahramian}
  et~al.}{2017}]{2017MNRAS.467.2199B}
{Bahramian} A.,  et~al., 2017, \mn@doi [\mnras] {10.1093/mnras/stx166}, \href
  {http://adsabs.harvard.edu/abs/2017MNRAS.467.2199B} {467, 2199}

\bibitem[\protect\citeauthoryear{{Ballantyne} \& {Strohmayer}}{{Ballantyne} \&
  {Strohmayer}}{2004}]{2004ApJ...602L.105B}
{Ballantyne} D.~R.,  {Strohmayer} T.~E.,  2004, \mn@doi [\apjl]
  {10.1086/382703}, \href {http://adsabs.harvard.edu/abs/2004ApJ...602L.105B}
  {602, L105}

\bibitem[\protect\citeauthoryear{{Bardeen} \& {Petterson}}{{Bardeen} \&
  {Petterson}}{1975}]{1975ApJ...195L..65B}
{Bardeen} J.~M.,  {Petterson} J.~A.,  1975, \mn@doi [\apjl] {10.1086/181711},
  \href {http://adsabs.harvard.edu/abs/1975ApJ...195L..65B} {195, L65}

\bibitem[\protect\citeauthoryear{{Bernardini}, {Russell}, {Kolojonen},
  {Stella}, {Hynes}  \& {Corbel}}{{Bernardini}
  et~al.}{2016}]{2016ApJ...826..149B}
{Bernardini} F.,  {Russell} D.~M.,  {Kolojonen} K.~I.~I.,  {Stella} L.,
  {Hynes} R.~I.,   {Corbel} S.,  2016, \mn@doi [\apj]
  {10.3847/0004-637X/826/2/149}, \href
  {http://adsabs.harvard.edu/abs/2016ApJ...826..149B} {826, 149}

\bibitem[\protect\citeauthoryear{{Bildsten}}{{Bildsten}}{2002}]{2002ApJ...577L..27B}
{Bildsten} L.,  2002, \mn@doi [\apjl] {10.1086/344085}, \href
  {http://adsabs.harvard.edu/abs/2002ApJ...577L..27B} {577, L27}

\bibitem[\protect\citeauthoryear{{Bogdanov}, {Heinke}, {{\"O}zel}  \&
  {G{\"u}ver}}{{Bogdanov} et~al.}{2016}]{2016ApJ...831..184B}
{Bogdanov} S.,  {Heinke} C.~O.,  {{\"O}zel} F.,   {G{\"u}ver} T.,  2016,
  \mn@doi [\apj] {10.3847/0004-637X/831/2/184}, \href
  {http://adsabs.harvard.edu/abs/2016ApJ...831..184B} {831, 184}

\bibitem[\protect\citeauthoryear{{Boyles}, {Lorimer}, {Turk}, {Mnatsakanov},
  {Lynch}, {Ransom}, {Freire}  \& {Belczynski}}{{Boyles}
  et~al.}{2011}]{2011ApJ...742...51B}
{Boyles} J.,  {Lorimer} D.~R.,  {Turk} P.~J.,  {Mnatsakanov} R.,  {Lynch}
  R.~S.,  {Ransom} S.~M.,  {Freire} P.~C.,   {Belczynski} K.,  2011, \mn@doi
  [\apj] {10.1088/0004-637X/742/1/51}, \href
  {http://adsabs.harvard.edu/abs/2011ApJ...742...51B} {742, 51}

\bibitem[\protect\citeauthoryear{{Brassington} et~al.,}{{Brassington}
  et~al.}{2010}]{2010ApJ...725.1805B}
{Brassington} N.~J.,  et~al., 2010, \mn@doi [\apj]
  {10.1088/0004-637X/725/2/1805}, \href
  {http://adsabs.harvard.edu/abs/2010ApJ...725.1805B} {725, 1805}

\bibitem[\protect\citeauthoryear{{Braun}}{{Braun}}{1997}]{1997PhDT........55B}
{Braun} H.,  1997, PhD thesis, , Ludwig-Maximilians-Univ.~M{\"u}nchen, (1997)

\bibitem[\protect\citeauthoryear{{Casares}}{{Casares}}{2015}]{2015ApJ...808...80C}
{Casares} J.,  2015, \mn@doi [\apj] {10.1088/0004-637X/808/1/80}, \href
  {http://adsabs.harvard.edu/abs/2015ApJ...808...80C} {808, 80}

\bibitem[\protect\citeauthoryear{{Casares}}{{Casares}}{2016}]{2016ApJ...822...99C}
{Casares} J.,  2016, \mn@doi [\apj] {10.3847/0004-637X/822/2/99}, \href
  {http://adsabs.harvard.edu/abs/2016ApJ...822...99C} {822, 99}

\bibitem[\protect\citeauthoryear{{Chakrabarty}}{{Chakrabarty}}{1998}]{1998ApJ...492..342C}
{Chakrabarty} D.,  1998, \mn@doi [\apj] {10.1086/305035}, \href
  {http://adsabs.harvard.edu/abs/1998ApJ...492..342C} {492, 342}

\bibitem[\protect\citeauthoryear{{Chomiuk}, {Strader}, {Maccarone},
  {Miller-Jones}, {Heinke}, {Noyola}, {Seth}  \& {Ransom}}{{Chomiuk}
  et~al.}{2013}]{2013ApJ...777...69C}
{Chomiuk} L.,  {Strader} J.,  {Maccarone} T.~J.,  {Miller-Jones} J.~C.~A.,
  {Heinke} C.,  {Noyola} E.,  {Seth} A.~C.,   {Ransom} S.,  2013, \mn@doi
  [\apj] {10.1088/0004-637X/777/1/69}, \href
  {http://adsabs.harvard.edu/abs/2013ApJ...777...69C} {777, 69}

\bibitem[\protect\citeauthoryear{{Chou}, {Grindlay}  \& {Bloser}}{{Chou}
  et~al.}{2001}]{2001ApJ...549.1135C}
{Chou} Y.,  {Grindlay} J.~E.,   {Bloser} P.~F.,  2001, \mn@doi [\apj]
  {10.1086/319443}, \href {http://adsabs.harvard.edu/abs/2001ApJ...549.1135C}
  {549, 1135}

\bibitem[\protect\citeauthoryear{Church, Strader, Davies  \& Bobrick}{Church
  et~al.}{2017}]{2017ApJ...851L...4C}
Church R.~P.,  Strader J.,  Davies M.~B.,   Bobrick A.,  2017, The
  Astrophysical Journal Letters, 851, L4

\bibitem[\protect\citeauthoryear{{Clark}, {Markert}  \& {Li}}{{Clark}
  et~al.}{1975}]{1975ApJ...199L..93C}
{Clark} G.~W.,  {Markert} T.~H.,   {Li} F.~K.,  1975, \mn@doi [\apjl]
  {10.1086/181856}, \href {http://adsabs.harvard.edu/abs/1975ApJ...199L..93C}
  {199, L93}

\bibitem[\protect\citeauthoryear{{Corral-Santana}, {Casares},
  {Mu{\~n}oz-Darias}, {Rodr{\'{\i}}guez-Gil}, {Shahbaz}, {Torres}, {Zurita}  \&
  {Tyndall}}{{Corral-Santana} et~al.}{2013}]{2013Sci...339.1048C}
{Corral-Santana} J.~M.,  {Casares} J.,  {Mu{\~n}oz-Darias} T.,
  {Rodr{\'{\i}}guez-Gil} P.,  {Shahbaz} T.,  {Torres} M.~A.~P.,  {Zurita} C.,
  {Tyndall} A.~A.,  2013, \mn@doi [Science] {10.1126/science.1228222}, \href
  {http://adsabs.harvard.edu/abs/2013Sci...339.1048C} {339, 1048}

\bibitem[\protect\citeauthoryear{{Deloye} \& {Bildsten}}{{Deloye} \&
  {Bildsten}}{2003}]{2003ApJ...598.1217D}
{Deloye} C.~J.,  {Bildsten} L.,  2003, \mn@doi [\apj] {10.1086/379063}, \href
  {http://adsabs.harvard.edu/abs/2003ApJ...598.1217D} {598, 1217}

\bibitem[\protect\citeauthoryear{{Deutsch}, {Anderson}, {Margon}  \&
  {Downes}}{{Deutsch} et~al.}{1996}]{1996ApJ...472L..97D}
{Deutsch} E.~W.,  {Anderson} S.~F.,  {Margon} B.,   {Downes} R.~A.,  1996,
  \mn@doi [\apjl] {10.1086/310373}, \href
  {http://adsabs.harvard.edu/abs/1996ApJ...472L..97D} {472, L97}

\bibitem[\protect\citeauthoryear{{Deutsch}, {Margon}  \& {Anderson}}{{Deutsch}
  et~al.}{2000}]{2000ApJ...530L..21D}
{Deutsch} E.~W.,  {Margon} B.,   {Anderson} S.~F.,  2000, \mn@doi [\apjl]
  {10.1086/312486}, \href {http://adsabs.harvard.edu/abs/2000ApJ...530L..21D}
  {530, L21}

\bibitem[\protect\citeauthoryear{{Dewi}, {Pols}, {Savonije}  \& {van den
  Heuvel}}{{Dewi} et~al.}{2002}]{2002MNRAS.331.1027D}
{Dewi} J.~D.~M.,  {Pols} O.~R.,  {Savonije} G.~J.,   {van den Heuvel} E.~P.~J.,
   2002, \mn@doi [\mnras] {10.1046/j.1365-8711.2002.05257.x}, \href
  {http://adsabs.harvard.edu/abs/2002MNRAS.331.1027D} {331, 1027}

\bibitem[\protect\citeauthoryear{{Dufour}, {Fontaine}, {Liebert}, {Schmidt}  \&
  {Behara}}{{Dufour} et~al.}{2008}]{2008ApJ...683..978D}
{Dufour} P.,  {Fontaine} G.,  {Liebert} J.,  {Schmidt} G.~D.,   {Behara} N.,
  2008, \mn@doi [\apj] {10.1086/589855}, \href
  {http://adsabs.harvard.edu/abs/2008ApJ...683..978D} {683, 978}

\bibitem[\protect\citeauthoryear{{Dunkel}, {Chluba}  \& {Sunyaev}}{{Dunkel}
  et~al.}{2006}]{2006AstL...32..257D}
{Dunkel} J.,  {Chluba} J.,   {Sunyaev} R.~A.,  2006, \mn@doi [Astronomy
  Letters] {10.1134/S1063773706040062}, \href
  {http://adsabs.harvard.edu/abs/2006AstL...32..257D} {32, 257}

\bibitem[\protect\citeauthoryear{{Eastman}, {Siverd}  \& {Gaudi}}{{Eastman}
  et~al.}{2010}]{2010PASP..122..935E}
{Eastman} J.,  {Siverd} R.,   {Gaudi} B.~S.,  2010, \mn@doi [\pasp]
  {10.1086/655938}, \href {http://adsabs.harvard.edu/abs/2010PASP..122..935E}
  {122, 935}

\bibitem[\protect\citeauthoryear{{Edmonds}, {Gilliland}, {Heinke}  \&
  {Grindlay}}{{Edmonds} et~al.}{2003a}]{2003ApJ...596.1177E}
{Edmonds} P.~D.,  {Gilliland} R.~L.,  {Heinke} C.~O.,   {Grindlay} J.~E.,
  2003a, \mn@doi [\apj] {10.1086/378193}, \href
  {http://adsabs.harvard.edu/abs/2003ApJ...596.1177E} {596, 1177}

\bibitem[\protect\citeauthoryear{{Edmonds}, {Gilliland}, {Heinke}  \&
  {Grindlay}}{{Edmonds} et~al.}{2003b}]{2003ApJ...596.1197E}
{Edmonds} P.~D.,  {Gilliland} R.~L.,  {Heinke} C.~O.,   {Grindlay} J.~E.,
  2003b, \mn@doi [\apj] {10.1086/378194}, \href
  {http://adsabs.harvard.edu/abs/2003ApJ...596.1197E} {596, 1197}

\bibitem[\protect\citeauthoryear{{Eggleton}}{{Eggleton}}{1983}]{1983ApJ...268..368E}
{Eggleton} P.~P.,  1983, \mn@doi [\apj] {10.1086/160960}, \href
  {http://adsabs.harvard.edu/abs/1983ApJ...268..368E} {268, 368}

\bibitem[\protect\citeauthoryear{{Ek{\c s}i}, {Anda{\c c}}, {{\c
  C}{\i}k{\i}nto{\u g}lu}, {Gen{\c c}ali}, {G{\"u}ng{\"o}r}  \&
  {{\"O}ztekin}}{{Ek{\c s}i} et~al.}{2015}]{2015MNRAS.448L..40E}
{Ek{\c s}i} K.~Y.,  {Anda{\c c}} {\.I}.~C.,  {{\c C}{\i}k{\i}nto{\u g}lu} S.,
  {Gen{\c c}ali} A.~A.,  {G{\"u}ng{\"o}r} C.,   {{\"O}ztekin} F.,  2015,
  \mn@doi [\mnras] {10.1093/mnrasl/slu199}, \href
  {http://adsabs.harvard.edu/abs/2015MNRAS.448L..40E} {448, L40}

\bibitem[\protect\citeauthoryear{{Fender}, {Russell}, {Knigge}, {Soria},
  {Hynes}  \& {Goad}}{{Fender} et~al.}{2009}]{2009MNRAS.393.1608F}
{Fender} R.~P.,  {Russell} D.~M.,  {Knigge} C.,  {Soria} R.,  {Hynes} R.~I.,
  {Goad} M.,  2009, \mn@doi [\mnras] {10.1111/j.1365-2966.2008.14299.x}, \href
  {http://adsabs.harvard.edu/abs/2009MNRAS.393.1608F} {393, 1608}

\bibitem[\protect\citeauthoryear{{Flower}, {Penn}  \& {Seaton}}{{Flower}
  et~al.}{1982}]{1982MNRAS.201P..39F}
{Flower} D.~R.,  {Penn} C.~J.,   {Seaton} M.~J.,  1982, \mn@doi [\mnras]
  {10.1093/mnras/201.1.39P}, \href
  {http://adsabs.harvard.edu/abs/1982MNRAS.201P..39F} {201, 39P}

\bibitem[\protect\citeauthoryear{{Ford}, {Kozinsky}  \& {Rasio}}{{Ford}
  et~al.}{2000}]{2000ApJ...535..385F}
{Ford} E.~B.,  {Kozinsky} B.,   {Rasio} F.~A.,  2000, \mn@doi [\apj]
  {10.1086/308815}, \href {http://adsabs.harvard.edu/abs/2000ApJ...535..385F}
  {535, 385}

\bibitem[\protect\citeauthoryear{{Fragile}, {Mathews}  \& {Wilson}}{{Fragile}
  et~al.}{2001}]{2001ApJ...553..955F}
{Fragile} P.~C.,  {Mathews} G.~J.,   {Wilson} J.~R.,  2001, \mn@doi [\apj]
  {10.1086/320990}, \href {http://adsabs.harvard.edu/abs/2001ApJ...553..955F}
  {553, 955}

\bibitem[\protect\citeauthoryear{{Fragile}, {Blaes}, {Anninos}  \&
  {Salmonson}}{{Fragile} et~al.}{2007}]{2007ApJ...668..417F}
{Fragile} P.~C.,  {Blaes} O.~M.,  {Anninos} P.,   {Salmonson} J.~D.,  2007,
  \mn@doi [\apj] {10.1086/521092}, \href
  {http://adsabs.harvard.edu/abs/2007ApJ...668..417F} {668, 417}

\bibitem[\protect\citeauthoryear{{Frank}, {King}  \& {Lasota}}{{Frank}
  et~al.}{1987}]{1987A&A...178..137F}
{Frank} J.,  {King} A.~R.,   {Lasota} J.-P.,  1987, \aap, \href
  {http://adsabs.harvard.edu/abs/1987A%26A...178..137F} {178, 137}

\bibitem[\protect\citeauthoryear{{Frank}, {King}  \& {Raine}}{{Frank}
  et~al.}{2002}]{2002apa..book.....F}
{Frank} J.,  {King} A.,   {Raine} D.~J.,  2002, {Accretion Power in
  Astrophysics: Third Edition}

\bibitem[\protect\citeauthoryear{{Froning} et~al.,}{{Froning}
  et~al.}{2011}]{2011ApJ...743...26F}
{Froning} C.~S.,  et~al., 2011, \mn@doi [\apj] {10.1088/0004-637X/743/1/26},
  \href {http://adsabs.harvard.edu/abs/2011ApJ...743...26F} {743, 26}

\bibitem[\protect\citeauthoryear{{Gallo}, {Migliari}, {Markoff}, {Tomsick},
  {Bailyn}, {Berta}, {Fender}  \& {Miller-Jones}}{{Gallo}
  et~al.}{2007}]{2007ApJ...670..600G}
{Gallo} E.,  {Migliari} S.,  {Markoff} S.,  {Tomsick} J.~A.,  {Bailyn} C.~D.,
  {Berta} S.,  {Fender} R.,   {Miller-Jones} J.~C.~A.,  2007, \mn@doi [\apj]
  {10.1086/521524}, \href {http://adsabs.harvard.edu/abs/2007ApJ...670..600G}
  {670, 600}

\bibitem[\protect\citeauthoryear{{Gallo}, {Treu}, {Jacob}, {Woo}, {Marshall}
  \& {Antonucci}}{{Gallo} et~al.}{2008}]{2008ApJ...680..154G}
{Gallo} E.,  {Treu} T.,  {Jacob} J.,  {Woo} J.-H.,  {Marshall} P.~J.,
  {Antonucci} R.,  2008, \mn@doi [\apj] {10.1086/588012}, \href
  {http://adsabs.harvard.edu/abs/2008ApJ...680..154G} {680, 154}

\bibitem[\protect\citeauthoryear{{Gierli{\'n}ski}, {Done}  \&
  {Page}}{{Gierli{\'n}ski} et~al.}{2008}]{2008MNRAS.388..753G}
{Gierli{\'n}ski} M.,  {Done} C.,   {Page} K.,  2008, \mn@doi [\mnras]
  {10.1111/j.1365-2966.2008.13431.x}, \href
  {http://adsabs.harvard.edu/abs/2008MNRAS.388..753G} {388, 753}

\bibitem[\protect\citeauthoryear{{Giersz} \& {Heggie}}{{Giersz} \&
  {Heggie}}{2011}]{2011MNRAS.410.2698G}
{Giersz} M.,  {Heggie} D.~C.,  2011, \mn@doi [\mnras]
  {10.1111/j.1365-2966.2010.17648.x}, \href
  {http://adsabs.harvard.edu/abs/2011MNRAS.410.2698G} {410, 2698}

\bibitem[\protect\citeauthoryear{{Gilliland} et~al.,}{{Gilliland}
  et~al.}{2000}]{2000ApJ...545L..47G}
{Gilliland} R.~L.,  et~al., 2000, \mn@doi [\apjl] {10.1086/317334}, \href
  {http://adsabs.harvard.edu/abs/2000ApJ...545L..47G} {545, L47}

\bibitem[\protect\citeauthoryear{{Gratton}, {Bragaglia}, {Carretta},
  {Clementini}, {Desidera}, {Grundahl}  \& {Lucatello}}{{Gratton}
  et~al.}{2003}]{2003A&A...408..529G}
{Gratton} R.~G.,  {Bragaglia} A.,  {Carretta} E.,  {Clementini} G.,  {Desidera}
  S.,  {Grundahl} F.,   {Lucatello} S.,  2003, \mn@doi [\aap]
  {10.1051/0004-6361:20031003}, \href
  {http://adsabs.harvard.edu/abs/2003A%26A...408..529G} {408, 529}

\bibitem[\protect\citeauthoryear{{Grindlay}, {Bailyn}, {Cohn}, {Lugger},
  {Thorstensen}  \& {Wegner}}{{Grindlay} et~al.}{1988}]{1988ApJ...334L..25G}
{Grindlay} J.~E.,  {Bailyn} C.~D.,  {Cohn} H.,  {Lugger} P.~M.,  {Thorstensen}
  J.~R.,   {Wegner} G.,  1988, \mn@doi [\apjl] {10.1086/185305}, \href
  {http://adsabs.harvard.edu/abs/1988ApJ...334L..25G} {334, L25}

\bibitem[\protect\citeauthoryear{{Grindlay}, {Heinke}, {Edmonds}  \&
  {Murray}}{{Grindlay} et~al.}{2001}]{2001Sci...292.2290G}
{Grindlay} J.~E.,  {Heinke} C.,  {Edmonds} P.~D.,   {Murray} S.~S.,  2001,
  \mn@doi [Science] {10.1126/science.1061135}, \href
  {http://adsabs.harvard.edu/abs/2001Sci...292.2290G} {292, 2290}

\bibitem[\protect\citeauthoryear{{Harris}}{{Harris}}{1996}]{1996AJ....112.1487H}
{Harris} W.~E.,  1996, \mn@doi [\aj] {10.1086/118116}, \href
  {http://adsabs.harvard.edu/abs/1996AJ....112.1487H} {112, 1487}

\bibitem[\protect\citeauthoryear{{Heinke}, {Ivanova}, {Engel}, {Pavlovskii},
  {Sivakoff}, {Cartwright}  \& {Gladstone}}{{Heinke}
  et~al.}{2013}]{2013ApJ...768..184H}
{Heinke} C.~O.,  {Ivanova} N.,  {Engel} M.~C.,  {Pavlovskii} K.,  {Sivakoff}
  G.~R.,  {Cartwright} T.~F.,   {Gladstone} J.~C.,  2013, \mn@doi [\apj]
  {10.1088/0004-637X/768/2/184}, \href
  {http://adsabs.harvard.edu/abs/2013ApJ...768..184H} {768, 184}

\bibitem[\protect\citeauthoryear{{Hills}}{{Hills}}{1984}]{1984AJ.....89.1811H}
{Hills} J.~G.,  1984, \mn@doi [\aj] {10.1086/113691}, \href
  {http://adsabs.harvard.edu/abs/1984AJ.....89.1811H} {89, 1811}

\bibitem[\protect\citeauthoryear{{Hirose} \& {Osaki}}{{Hirose} \&
  {Osaki}}{1990}]{1990PASJ...42..135H}
{Hirose} M.,  {Osaki} Y.,  1990, \pasj, \href
  {http://adsabs.harvard.edu/abs/1990PASJ...42..135H} {42, 135}

\bibitem[\protect\citeauthoryear{{Homer}, {Anderson}, {Wachter}  \&
  {Margon}}{{Homer} et~al.}{2002}]{2002AJ....124.3348H}
{Homer} L.,  {Anderson} S.~F.,  {Wachter} S.,   {Margon} B.,  2002, \mn@doi
  [\aj] {10.1086/344600}, \href
  {http://adsabs.harvard.edu/abs/2002AJ....124.3348H} {124, 3348}

\bibitem[\protect\citeauthoryear{{Irwin}, {Brink}, {Bregman}  \&
  {Roberts}}{{Irwin} et~al.}{2010}]{2010ApJ...712L...1I}
{Irwin} J.~A.,  {Brink} T.~G.,  {Bregman} J.~N.,   {Roberts} T.~P.,  2010,
  \mn@doi [\apjl] {10.1088/2041-8205/712/1/L1}, \href
  {http://adsabs.harvard.edu/abs/2010ApJ...712L...1I} {712, L1}

\bibitem[\protect\citeauthoryear{{Israel} et~al.,}{{Israel}
  et~al.}{2017a}]{2017Sci...355..817I}
{Israel} G.~L.,  et~al., 2017a, \mn@doi [Science] {10.1126/science.aai8635},
  \href {http://adsabs.harvard.edu/abs/2017Sci...355..817I} {355, 817}

\bibitem[\protect\citeauthoryear{{Israel} et~al.,}{{Israel}
  et~al.}{2017b}]{2017MNRAS.466L..48I}
{Israel} G.~L.,  et~al., 2017b, \mn@doi [\mnras] {10.1093/mnrasl/slw218}, \href
  {http://adsabs.harvard.edu/abs/2017MNRAS.466L..48I} {466, L48}

\bibitem[\protect\citeauthoryear{{Istrate}, {Tauris}  \& {Langer}}{{Istrate}
  et~al.}{2014}]{2014A&A...571A..45I}
{Istrate} A.~G.,  {Tauris} T.~M.,   {Langer} N.,  2014, \mn@doi [\aap]
  {10.1051/0004-6361/201424680}, \href
  {http://adsabs.harvard.edu/abs/2014A%26A...571A..45I} {571, A45}

\bibitem[\protect\citeauthoryear{{Ivanova}}{{Ivanova}}{2008}]{2008msah.conf..101I}
{Ivanova} N.,  2008, in {Hubrig} S.,  {Petr-Gotzens} M.,   {Tokovinin} A.,
  eds, Multiple Stars Across the H-R Diagram. p.~101,
  \mn@doi{10.1007/978-3-540-74745-1_14}

\bibitem[\protect\citeauthoryear{{Ivanova}, {Rasio}, {Lombardi}, {Dooley}  \&
  {Proulx}}{{Ivanova} et~al.}{2005}]{2005ApJ...621L.109I}
{Ivanova} N.,  {Rasio} F.~A.,  {Lombardi} Jr. J.~C.,  {Dooley} K.~L.,
  {Proulx} Z.~F.,  2005, \mn@doi [\apjl] {10.1086/429220}, \href
  {http://adsabs.harvard.edu/abs/2005ApJ...621L.109I} {621, L109}

\bibitem[\protect\citeauthoryear{{Ivanova}, {Heinke}, {Rasio}, {Belczynski}  \&
  {Fregeau}}{{Ivanova} et~al.}{2008}]{2008MNRAS.386..553I}
{Ivanova} N.,  {Heinke} C.~O.,  {Rasio} F.~A.,  {Belczynski} K.,   {Fregeau}
  J.~M.,  2008, \mn@doi [\mnras] {10.1111/j.1365-2966.2008.13064.x}, \href
  {http://adsabs.harvard.edu/abs/2008MNRAS.386..553I} {386, 553}

\bibitem[\protect\citeauthoryear{{Ivanova}, {Chaichenets}, {Fregeau}, {Heinke},
  {Lombardi}  \& {Woods}}{{Ivanova} et~al.}{2010}]{2010ApJ...717..948I}
{Ivanova} N.,  {Chaichenets} S.,  {Fregeau} J.,  {Heinke} C.~O.,  {Lombardi}
  Jr. J.~C.,   {Woods} T.~E.,  2010, \mn@doi [\apj]
  {10.1088/0004-637X/717/2/948}, \href
  {http://adsabs.harvard.edu/abs/2010ApJ...717..948I} {717, 948}

\bibitem[\protect\citeauthoryear{{Ivanova}, {da Rocha}, {Van}  \&
  {Nandez}}{{Ivanova} et~al.}{2017}]{2017ApJ...843L..30I}
{Ivanova} N.,  {da Rocha} C.~A.,  {Van} K.~X.,   {Nandez} J.~L.~A.,  2017,
  \mn@doi [\apjl] {10.3847/2041-8213/aa7b76}, \href
  {http://adsabs.harvard.edu/abs/2017ApJ...843L..30I} {843, L30}

\bibitem[\protect\citeauthoryear{{Jord{\'a}n} et~al.,}{{Jord{\'a}n}
  et~al.}{2009}]{2009ApJS..180...54J}
{Jord{\'a}n} A.,  et~al., 2009, \mn@doi [\apjs] {10.1088/0067-0049/180/1/54},
  \href {http://adsabs.harvard.edu/abs/2009ApJS..180...54J} {180, 54}

\bibitem[\protect\citeauthoryear{{Jord{\'a}n}, {Peng}, {Blakeslee},
  {C{\^o}t{\'e}}, {Eyheramendy}  \& {Ferrarese}}{{Jord{\'a}n}
  et~al.}{2015}]{2015ApJS..221...13J}
{Jord{\'a}n} A.,  {Peng} E.~W.,  {Blakeslee} J.~P.,  {C{\^o}t{\'e}} P.,
  {Eyheramendy} S.,   {Ferrarese} L.,  2015, \mn@doi [\apjs]
  {10.1088/0067-0049/221/1/13}, \href
  {http://adsabs.harvard.edu/abs/2015ApJS..221...13J} {221, 13}

\bibitem[\protect\citeauthoryear{{Kalemci}, {Din{\c c}er}, {Tomsick}, {Buxton},
  {Bailyn}  \& {Chun}}{{Kalemci} et~al.}{2013}]{2013ApJ...779...95K}
{Kalemci} E.,  {Din{\c c}er} T.,  {Tomsick} J.~A.,  {Buxton} M.~M.,  {Bailyn}
  C.~D.,   {Chun} Y.~Y.,  2013, \mn@doi [\apj] {10.1088/0004-637X/779/2/95},
  \href {http://adsabs.harvard.edu/abs/2013ApJ...779...95K} {779, 95}

\bibitem[\protect\citeauthoryear{{Kaur} et~al.,}{{Kaur}
  et~al.}{2012}]{2012ApJ...746L..23K}
{Kaur} R.,  et~al., 2012, \mn@doi [\apjl] {10.1088/2041-8205/746/2/L23}, \href
  {http://adsabs.harvard.edu/abs/2012ApJ...746L..23K} {746, L23}

\bibitem[\protect\citeauthoryear{{Khargharia}, {Froning}  \&
  {Robinson}}{{Khargharia} et~al.}{2010}]{2010ApJ...716.1105K}
{Khargharia} J.,  {Froning} C.~S.,   {Robinson} E.~L.,  2010, \mn@doi [\apj]
  {10.1088/0004-637X/716/2/1105}, \href
  {http://adsabs.harvard.edu/abs/2010ApJ...716.1105K} {716, 1105}

\bibitem[\protect\citeauthoryear{{Kimmig}, {Seth}, {Ivans}, {Strader},
  {Caldwell}, {Anderton}  \& {Gregersen}}{{Kimmig}
  et~al.}{2015}]{2015AJ....149...53K}
{Kimmig} B.,  {Seth} A.,  {Ivans} I.~I.,  {Strader} J.,  {Caldwell} N.,
  {Anderton} T.,   {Gregersen} D.,  2015, \mn@doi [\aj]
  {10.1088/0004-6256/149/2/53}, \href
  {http://adsabs.harvard.edu/abs/2015AJ....149...53K} {149, 53}

\bibitem[\protect\citeauthoryear{{King} \& {Lasota}}{{King} \&
  {Lasota}}{2016}]{2016MNRAS.458L..10K}
{King} A.,  {Lasota} J.-P.,  2016, \mn@doi [\mnras] {10.1093/mnrasl/slw011},
  \href {http://adsabs.harvard.edu/abs/2016MNRAS.458L..10K} {458, L10}

\bibitem[\protect\citeauthoryear{{Kippenhahn} \& {Weigert}}{{Kippenhahn} \&
  {Weigert}}{1990}]{1990sse..book.....K}
{Kippenhahn} R.,  {Weigert} A.,  1990, {Stellar Structure and Evolution}

\bibitem[\protect\citeauthoryear{{Knevitt}, {Wynn}, {Vaughan}  \&
  {Watson}}{{Knevitt} et~al.}{2014}]{2014MNRAS.437.3087K}
{Knevitt} G.,  {Wynn} G.~A.,  {Vaughan} S.,   {Watson} M.~G.,  2014, \mn@doi
  [\mnras] {10.1093/mnras/stt2008}, \href
  {http://adsabs.harvard.edu/abs/2014MNRAS.437.3087K} {437, 3087}

\bibitem[\protect\citeauthoryear{{Knigge}, {Zurek}, {Shara}  \&
  {Long}}{{Knigge} et~al.}{2002}]{2002ApJ...579..752K}
{Knigge} C.,  {Zurek} D.~R.,  {Shara} M.~M.,   {Long} K.~S.,  2002, \mn@doi
  [\apj] {10.1086/342835}, \href
  {http://adsabs.harvard.edu/abs/2002ApJ...579..752K} {579, 752}

\bibitem[\protect\citeauthoryear{{Knigge}, {Zurek}, {Shara}, {Long}  \&
  {Gilliland}}{{Knigge} et~al.}{2003}]{2003ApJ...599.1320K}
{Knigge} C.,  {Zurek} D.~R.,  {Shara} M.~M.,  {Long} K.~S.,   {Gilliland}
  R.~L.,  2003, \mn@doi [\apj] {10.1086/379609}, \href
  {http://adsabs.harvard.edu/abs/2003ApJ...599.1320K} {599, 1320}

\bibitem[\protect\citeauthoryear{{Knigge}, {Dieball}, {Ma{\'{\i}}z
  Apell{\'a}niz}, {Long}, {Zurek}  \& {Shara}}{{Knigge}
  et~al.}{2008}]{2008ApJ...683.1006K}
{Knigge} C.,  {Dieball} A.,  {Ma{\'{\i}}z Apell{\'a}niz} J.,  {Long} K.~S.,
  {Zurek} D.~R.,   {Shara} M.~M.,  2008, \mn@doi [\apj] {10.1086/589987}, \href
  {http://adsabs.harvard.edu/abs/2008ApJ...683.1006K} {683, 1006}

\bibitem[\protect\citeauthoryear{{Kotze} \& {Charles}}{{Kotze} \&
  {Charles}}{2012}]{2012MNRAS.420.1575K}
{Kotze} M.~M.,  {Charles} P.~A.,  2012, \mn@doi [\mnras]
  {10.1111/j.1365-2966.2011.20146.x}, \href
  {http://adsabs.harvard.edu/abs/2012MNRAS.420.1575K} {420, 1575}

\bibitem[\protect\citeauthoryear{{Kozai}}{{Kozai}}{1962}]{1962AJ.....67..591K}
{Kozai} Y.,  1962, \mn@doi [\aj] {10.1086/108790}, \href
  {http://adsabs.harvard.edu/abs/1962AJ.....67..591K} {67, 591}

\bibitem[\protect\citeauthoryear{{Krauss}, {Schulz}, {Chakrabarty}, {Juett}  \&
  {Cottam}}{{Krauss} et~al.}{2007}]{2007ApJ...660..605K}
{Krauss} M.~I.,  {Schulz} N.~S.,  {Chakrabarty} D.,  {Juett} A.~M.,   {Cottam}
  J.,  2007, \mn@doi [\apj] {10.1086/513592}, \href
  {http://adsabs.harvard.edu/abs/2007ApJ...660..605K} {660, 605}

\bibitem[\protect\citeauthoryear{{Kulkarni}, {Hut}  \& {McMillan}}{{Kulkarni}
  et~al.}{1993}]{1993Natur.364..421K}
{Kulkarni} S.~R.,  {Hut} P.,   {McMillan} S.,  1993, \mn@doi [\nat]
  {10.1038/364421a0}, \href {http://adsabs.harvard.edu/abs/1993Natur.364..421K}
  {364, 421}

\bibitem[\protect\citeauthoryear{{Landau} \& {Lifshitz}}{{Landau} \&
  {Lifshitz}}{1975}]{1975ctf..book.....L}
{Landau} L.~D.,  {Lifshitz} E.~M.,  1975, {The classical theory of fields}

\bibitem[\protect\citeauthoryear{{Langer}}{{Langer}}{1998}]{1998A&A...329..551L}
{Langer} N.,  1998, \aap, \href
  {http://adsabs.harvard.edu/abs/1998A%26A...329..551L} {329, 551}

\bibitem[\protect\citeauthoryear{{Larwood}}{{Larwood}}{1998}]{1998MNRAS.299L..32L}
{Larwood} J.,  1998, \mn@doi [\mnras] {10.1046/j.1365-8711.1998.01978.x}, \href
  {http://adsabs.harvard.edu/abs/1998MNRAS.299L..32L} {299, L32}

\bibitem[\protect\citeauthoryear{{Lense} \& {Thirring}}{{Lense} \&
  {Thirring}}{1918}]{LT18}
{Lense} J.,  {Thirring} H.,  1918, Phys. Z., 156, 19

\bibitem[\protect\citeauthoryear{{Maccarone}}{{Maccarone}}{2003}]{2003A&A...409..697M}
{Maccarone} T.~J.,  2003, \mn@doi [\aap] {10.1051/0004-6361:20031146}, \href
  {http://adsabs.harvard.edu/abs/2003A%26A...409..697M} {409, 697}

\bibitem[\protect\citeauthoryear{{Maccarone} \& {Knigge}}{{Maccarone} \&
  {Knigge}}{2007}]{2007A&G....48e..12M}
{Maccarone} T.,  {Knigge} C.,  2007, \mn@doi [Astronomy and Geophysics]
  {10.1111/j.1468-4004.2007.48512.x}, \href
  {http://adsabs.harvard.edu/abs/2007A%26G....48e..12M} {48, 5.12}

\bibitem[\protect\citeauthoryear{{Maccarone}, {Kundu}, {Zepf}  \&
  {Rhode}}{{Maccarone} et~al.}{2007}]{2007Natur.445..183M}
{Maccarone} T.~J.,  {Kundu} A.,  {Zepf} S.~E.,   {Rhode} K.~L.,  2007, \mn@doi
  [\nat] {10.1038/nature05434}, \href
  {http://adsabs.harvard.edu/abs/2007Natur.445..183M} {445, 183}

\bibitem[\protect\citeauthoryear{{Maccarone}, {Kundu}, {Zepf}  \&
  {Rhode}}{{Maccarone} et~al.}{2010}]{2010MNRAS.409L..84M}
{Maccarone} T.~J.,  {Kundu} A.,  {Zepf} S.~E.,   {Rhode} K.~L.,  2010, \mn@doi
  [\mnras] {10.1111/j.1745-3933.2010.00953.x}, \href
  {http://adsabs.harvard.edu/abs/2010MNRAS.409L..84M} {409, L84}

\bibitem[\protect\citeauthoryear{{Maccarone}, {Kundu}, {Zepf}  \&
  {Rhode}}{{Maccarone} et~al.}{2011}]{2011MNRAS.410.1655M}
{Maccarone} T.~J.,  {Kundu} A.,  {Zepf} S.~E.,   {Rhode} K.~L.,  2011, \mn@doi
  [\mnras] {10.1111/j.1365-2966.2010.17547.x}, \href
  {http://adsabs.harvard.edu/abs/2011MNRAS.410.1655M} {410, 1655}

\bibitem[\protect\citeauthoryear{{Malumuth} et~al.,}{{Malumuth}
  et~al.}{2003}]{2003PASP..115..218M}
{Malumuth} E.~M.,  et~al., 2003, \mn@doi [\pasp] {10.1086/345913}, \href
  {http://adsabs.harvard.edu/abs/2003PASP..115..218M} {115, 218}

\bibitem[\protect\citeauthoryear{{Marsh}, {Robinson}  \& {Wood}}{{Marsh}
  et~al.}{1994}]{1994MNRAS.266..137M}
{Marsh} T.~R.,  {Robinson} E.~L.,   {Wood} J.~H.,  1994, \mn@doi [\mnras]
  {10.1093/mnras/266.1.137}, \href
  {http://adsabs.harvard.edu/abs/1994MNRAS.266..137M} {266, 137}

\bibitem[\protect\citeauthoryear{{Mazeh} \& {Shaham}}{{Mazeh} \&
  {Shaham}}{1979}]{1979A&A....77..145M}
{Mazeh} T.,  {Shaham} J.,  1979, \aap, \href
  {http://adsabs.harvard.edu/abs/1979A%26A....77..145M} {77, 145}

\bibitem[\protect\citeauthoryear{{Menou} \& {McClintock}}{{Menou} \&
  {McClintock}}{2001}]{2001ApJ...557..304M}
{Menou} K.,  {McClintock} J.~E.,  2001, \mn@doi [\apj] {10.1086/321665}, \href
  {http://adsabs.harvard.edu/abs/2001ApJ...557..304M} {557, 304}

\bibitem[\protect\citeauthoryear{{Middleditch}, {Mason}, {Nelson}  \&
  {White}}{{Middleditch} et~al.}{1981}]{1981ApJ...244.1001M}
{Middleditch} J.,  {Mason} K.~O.,  {Nelson} J.~E.,   {White} N.~E.,  1981,
  \mn@doi [\apj] {10.1086/158772}, \href
  {http://adsabs.harvard.edu/abs/1981ApJ...244.1001M} {244, 1001}

\bibitem[\protect\citeauthoryear{{Migliari} et~al.,}{{Migliari}
  et~al.}{2010}]{2010ApJ...710..117M}
{Migliari} S.,  et~al., 2010, \mn@doi [\apj] {10.1088/0004-637X/710/1/117},
  \href {http://adsabs.harvard.edu/abs/2010ApJ...710..117M} {710, 117}

\bibitem[\protect\citeauthoryear{{Miller-Jones} et~al.,}{{Miller-Jones}
  et~al.}{2015}]{2015MNRAS.453.3918M}
{Miller-Jones} J.~C.~A.,  et~al., 2015, \mn@doi [\mnras]
  {10.1093/mnras/stv1869}, \href
  {http://adsabs.harvard.edu/abs/2015MNRAS.453.3918M} {453, 3918}

\bibitem[\protect\citeauthoryear{{Morscher}, {Pattabiraman}, {Rodriguez},
  {Rasio}  \& {Umbreit}}{{Morscher} et~al.}{2015}]{2015ApJ...800....9M}
{Morscher} M.,  {Pattabiraman} B.,  {Rodriguez} C.,  {Rasio} F.~A.,   {Umbreit}
  S.,  2015, \mn@doi [\apj] {10.1088/0004-637X/800/1/9}, \href
  {http://adsabs.harvard.edu/abs/2015ApJ...800....9M} {800, 9}

\bibitem[\protect\citeauthoryear{{Narayan} \& {McClintock}}{{Narayan} \&
  {McClintock}}{2008}]{2008NewAR..51..733N}
{Narayan} R.,  {McClintock} J.~E.,  2008, \mn@doi [\nar]
  {10.1016/j.newar.2008.03.002}, \href
  {http://adsabs.harvard.edu/abs/2008NewAR..51..733N} {51, 733}

\bibitem[\protect\citeauthoryear{{Narayan} \& {Yi}}{{Narayan} \&
  {Yi}}{1994}]{1994ApJ...428L..13N}
{Narayan} R.,  {Yi} I.,  1994, \mn@doi [\apjl] {10.1086/187381}, \href
  {http://adsabs.harvard.edu/abs/1994ApJ...428L..13N} {428, L13}

\bibitem[\protect\citeauthoryear{{Narayan} \& {Yi}}{{Narayan} \&
  {Yi}}{1995}]{1995ApJ...452..710N}
{Narayan} R.,  {Yi} I.,  1995, \mn@doi [\apj] {10.1086/176343}, \href
  {http://adsabs.harvard.edu/abs/1995ApJ...452..710N} {452, 710}

\bibitem[\protect\citeauthoryear{{Nelemans} \& {Jonker}}{{Nelemans} \&
  {Jonker}}{2010}]{2010NewAR..54...87N}
{Nelemans} G.,  {Jonker} P.~G.,  2010, \mn@doi [\nar]
  {10.1016/j.newar.2010.09.021}, \href
  {http://adsabs.harvard.edu/abs/2010NewAR..54...87N} {54, 87}

\bibitem[\protect\citeauthoryear{{Nelemans}, {Jonker}, {Marsh}  \& {van der
  Klis}}{{Nelemans} et~al.}{2004a}]{2004MNRAS.348L...7N}
{Nelemans} G.,  {Jonker} P.~G.,  {Marsh} T.~R.,   {van der Klis} M.,  2004a,
  \mn@doi [\mnras] {10.1111/j.1365-2966.2004.07486.x}, \href
  {http://adsabs.harvard.edu/abs/2004MNRAS.348L...7N} {348, L7}

\bibitem[\protect\citeauthoryear{{Nelemans}, {Yungelson}  \& {Portegies
  Zwart}}{{Nelemans} et~al.}{2004b}]{2004MNRAS.349..181N}
{Nelemans} G.,  {Yungelson} L.~R.,   {Portegies Zwart} S.~F.,  2004b, \mn@doi
  [\mnras] {10.1111/j.1365-2966.2004.07479.x}, \href
  {http://adsabs.harvard.edu/abs/2004MNRAS.349..181N} {349, 181}

\bibitem[\protect\citeauthoryear{{Nelemans}, {Jonker}  \& {Steeghs}}{{Nelemans}
  et~al.}{2006}]{2006MNRAS.370..255N}
{Nelemans} G.,  {Jonker} P.~G.,   {Steeghs} D.,  2006, \mn@doi [\mnras]
  {10.1111/j.1365-2966.2006.10496.x}, \href
  {http://adsabs.harvard.edu/abs/2006MNRAS.370..255N} {370, 255}

\bibitem[\protect\citeauthoryear{{Nelson} \& {Papaloizou}}{{Nelson} \&
  {Papaloizou}}{2000}]{2000MNRAS.315..570N}
{Nelson} R.~P.,  {Papaloizou} J.~C.~B.,  2000, \mn@doi [\mnras]
  {10.1046/j.1365-8711.2000.03478.x}, \href
  {http://adsabs.harvard.edu/abs/2000MNRAS.315..570N} {315, 570}

\bibitem[\protect\citeauthoryear{{Nelson} \& {Rappaport}}{{Nelson} \&
  {Rappaport}}{2003}]{2003ApJ...598..431N}
{Nelson} L.~A.,  {Rappaport} S.,  2003, \mn@doi [\apj] {10.1086/378798}, \href
  {http://adsabs.harvard.edu/abs/2003ApJ...598..431N} {598, 431}

\bibitem[\protect\citeauthoryear{{Nelson}, {Rappaport}  \& {Joss}}{{Nelson}
  et~al.}{1986}]{1986ApJ...304..231N}
{Nelson} L.~A.,  {Rappaport} S.~A.,   {Joss} P.~C.,  1986, \mn@doi [\apj]
  {10.1086/164156}, \href {http://adsabs.harvard.edu/abs/1986ApJ...304..231N}
  {304, 231}

\bibitem[\protect\citeauthoryear{{Ogilvie} \& {Dubus}}{{Ogilvie} \&
  {Dubus}}{2001}]{2001MNRAS.320..485O}
{Ogilvie} G.~I.,  {Dubus} G.,  2001, \mn@doi [\mnras]
  {10.1046/j.1365-8711.2001.04011.x}, \href
  {http://adsabs.harvard.edu/abs/2001MNRAS.320..485O} {320, 485}

\bibitem[\protect\citeauthoryear{{Olech}, {Rutkowski}  \&
  {Schwarzenberg-Czerny}}{{Olech} et~al.}{2009}]{2009MNRAS.399..465O}
{Olech} A.,  {Rutkowski} A.,   {Schwarzenberg-Czerny} A.,  2009, \mn@doi
  [\mnras] {10.1111/j.1365-2966.2009.15298.x}, \href
  {http://adsabs.harvard.edu/abs/2009MNRAS.399..465O} {399, 465}

\bibitem[\protect\citeauthoryear{{Orosz}, {Bailyn}, {Remillard}, {McClintock}
  \& {Foltz}}{{Orosz} et~al.}{1994}]{1994ApJ...436..848O}
{Orosz} J.~A.,  {Bailyn} C.~D.,  {Remillard} R.~A.,  {McClintock} J.~E.,
  {Foltz} C.~B.,  1994, \mn@doi [\apj] {10.1086/174962}, \href
  {http://adsabs.harvard.edu/abs/1994ApJ...436..848O} {436, 848}

\bibitem[\protect\citeauthoryear{{Osaki}}{{Osaki}}{1985}]{1985A&A...144..369O}
{Osaki} Y.,  1985, \aap, \href
  {http://adsabs.harvard.edu/abs/1985A%26A...144..369O} {144, 369}

\bibitem[\protect\citeauthoryear{{{\"O}zel}, {Psaltis}, {Narayan}  \&
  {McClintock}}{{{\"O}zel} et~al.}{2010}]{2010ApJ...725.1918O}
{{\"O}zel} F.,  {Psaltis} D.,  {Narayan} R.,   {McClintock} J.~E.,  2010,
  \mn@doi [\apj] {10.1088/0004-637X/725/2/1918}, \href
  {http://adsabs.harvard.edu/abs/2010ApJ...725.1918O} {725, 1918}

\bibitem[\protect\citeauthoryear{{Paczynski}}{{Paczynski}}{1977}]{1977ApJ...216..822P}
{Paczynski} B.,  1977, \mn@doi [\apj] {10.1086/155526}, \href
  {http://adsabs.harvard.edu/abs/1977ApJ...216..822P} {216, 822}

\bibitem[\protect\citeauthoryear{{Papaloizou} \& {Pringle}}{{Papaloizou} \&
  {Pringle}}{1977}]{1977MNRAS.181..441P}
{Papaloizou} J.,  {Pringle} J.~E.,  1977, \mn@doi [\mnras]
  {10.1093/mnras/181.3.441}, \href
  {http://adsabs.harvard.edu/abs/1977MNRAS.181..441P} {181, 441}

\bibitem[\protect\citeauthoryear{{Paresce}, {de Marchi}  \&
  {Ferraro}}{{Paresce} et~al.}{1992}]{1992Natur.360...46P}
{Paresce} F.,  {de Marchi} G.,   {Ferraro} F.~R.,  1992, \mn@doi [\nat]
  {10.1038/360046a0}, \href {http://adsabs.harvard.edu/abs/1992Natur.360...46P}
  {360, 46}

\bibitem[\protect\citeauthoryear{{Patterson} et~al.,}{{Patterson}
  et~al.}{2005}]{2005PASP..117.1204P}
{Patterson} J.,  et~al., 2005, \mn@doi [\pasp] {10.1086/447771}, \href
  {http://adsabs.harvard.edu/abs/2005PASP..117.1204P} {117, 1204}

\bibitem[\protect\citeauthoryear{{Peacock}, {Zepf}  \& {Maccarone}}{{Peacock}
  et~al.}{2012}]{2012ApJ...752...90P}
{Peacock} M.~B.,  {Zepf} S.~E.,   {Maccarone} T.~J.,  2012, \mn@doi [\apj]
  {10.1088/0004-637X/752/2/90}, \href
  {http://adsabs.harvard.edu/abs/2012ApJ...752...90P} {752, 90}

\bibitem[\protect\citeauthoryear{{Pearson}}{{Pearson}}{2003}]{2003MNRAS.346L..21P}
{Pearson} K.~J.,  2003, \mn@doi [\mnras] {10.1046/j.1365-2966.2003.07290.x},
  \href {http://adsabs.harvard.edu/abs/2003MNRAS.346L..21P} {346, L21}

\bibitem[\protect\citeauthoryear{{Peters}}{{Peters}}{1964}]{1964PhRv..136.1224P}
{Peters} P.~C.,  1964, \mn@doi [Physical Review] {10.1103/PhysRev.136.B1224},
  \href {http://adsabs.harvard.edu/abs/1964PhRv..136.1224P} {136, 1224}

\bibitem[\protect\citeauthoryear{{Peuten}, {Zocchi}, {Gieles}, {Gualandris}  \&
  {H{\'e}nault-Brunet}}{{Peuten} et~al.}{2016}]{2016MNRAS.462.2333P}
{Peuten} M.,  {Zocchi} A.,  {Gieles} M.,  {Gualandris} A.,
  {H{\'e}nault-Brunet} V.,  2016, \mn@doi [\mnras] {10.1093/mnras/stw1726},
  \href {http://adsabs.harvard.edu/abs/2016MNRAS.462.2333P} {462, 2333}

\bibitem[\protect\citeauthoryear{{Podsiadlowski}, {Rappaport}  \&
  {Pfahl}}{{Podsiadlowski} et~al.}{2002}]{2002ApJ...565.1107P}
{Podsiadlowski} P.,  {Rappaport} S.,   {Pfahl} E.~D.,  2002, \mn@doi [\apj]
  {10.1086/324686}, \href {http://adsabs.harvard.edu/abs/2002ApJ...565.1107P}
  {565, 1107}

\bibitem[\protect\citeauthoryear{{Poutanen}, {Krolik}  \& {Ryde}}{{Poutanen}
  et~al.}{1997}]{1997MNRAS.292L..21P}
{Poutanen} J.,  {Krolik} J.~H.,   {Ryde} F.,  1997, \mn@doi [\mnras]
  {10.1093/mnras/292.1.L21}, \href
  {http://adsabs.harvard.edu/abs/1997MNRAS.292L..21P} {292, L21}

\bibitem[\protect\citeauthoryear{{Priedhorsky} \& {Verbunt}}{{Priedhorsky} \&
  {Verbunt}}{1988}]{1988ApJ...333..895P}
{Priedhorsky} W.~C.,  {Verbunt} F.,  1988, \mn@doi [\apj] {10.1086/166798},
  \href {http://adsabs.harvard.edu/abs/1988ApJ...333..895P} {333, 895}

\bibitem[\protect\citeauthoryear{{Quataert} \& {Narayan}}{{Quataert} \&
  {Narayan}}{1999}]{1999ApJ...520..298Q}
{Quataert} E.,  {Narayan} R.,  1999, \mn@doi [\apj] {10.1086/307439}, \href
  {http://adsabs.harvard.edu/abs/1999ApJ...520..298Q} {520, 298}

\bibitem[\protect\citeauthoryear{{Retter}, {Chou}, {Bedding}  \&
  {Naylor}}{{Retter} et~al.}{2002}]{2002MNRAS.330L..37R}
{Retter} A.,  {Chou} Y.,  {Bedding} T.~R.,   {Naylor} T.,  2002, \mn@doi
  [\mnras] {10.1046/j.1365-8711.2002.05280.x}, \href
  {http://adsabs.harvard.edu/abs/2002MNRAS.330L..37R} {330, L37}

\bibitem[\protect\citeauthoryear{{Rhode} \& {Zepf}}{{Rhode} \&
  {Zepf}}{2001}]{2001AJ....121..210R}
{Rhode} K.~L.,  {Zepf} S.~E.,  2001, \mn@doi [\aj] {10.1086/318039}, \href
  {http://adsabs.harvard.edu/abs/2001AJ....121..210R} {121, 210}

\bibitem[\protect\citeauthoryear{{Russell}, {Fender}, {Hynes}, {Brocksopp},
  {Homan}, {Jonker}  \& {Buxton}}{{Russell} et~al.}{2006}]{2006MNRAS.371.1334R}
{Russell} D.~M.,  {Fender} R.~P.,  {Hynes} R.~I.,  {Brocksopp} C.,  {Homan} J.,
   {Jonker} P.~G.,   {Buxton} M.~M.,  2006, \mn@doi [\mnras]
  {10.1111/j.1365-2966.2006.10756.x}, \href
  {http://adsabs.harvard.edu/abs/2006MNRAS.371.1334R} {371, 1334}

\bibitem[\protect\citeauthoryear{{Russell} et~al.,}{{Russell}
  et~al.}{2013}]{2013MNRAS.429..815R}
{Russell} D.~M.,  et~al., 2013, \mn@doi [\mnras] {10.1093/mnras/sts377}, \href
  {http://adsabs.harvard.edu/abs/2013MNRAS.429..815R} {429, 815}

\bibitem[\protect\citeauthoryear{{Salaris}, {Held}, {Ortolani}, {Gullieuszik}
  \& {Momany}}{{Salaris} et~al.}{2007}]{2007A&A...476..243S}
{Salaris} M.,  {Held} E.~V.,  {Ortolani} S.,  {Gullieuszik} M.,   {Momany} Y.,
  2007, \mn@doi [\aap] {10.1051/0004-6361:20078445}, \href
  {http://adsabs.harvard.edu/abs/2007A%26A...476..243S} {476, 243}

\bibitem[\protect\citeauthoryear{{Sengar}, {Tauris}, {Langer}  \&
  {Istrate}}{{Sengar} et~al.}{2017}]{2017MNRAS.470L...6S}
{Sengar} R.,  {Tauris} T.~M.,  {Langer} N.,   {Istrate} A.~G.,  2017, \mn@doi
  [\mnras] {10.1093/mnrasl/slx064}, \href
  {http://adsabs.harvard.edu/abs/2017MNRAS.470L...6S} {470, L6}

\bibitem[\protect\citeauthoryear{{Shahbaz}, {Smale}, {Naylor}, {Charles}, {van
  Paradijs}, {Hassall}  \& {Callanan}}{{Shahbaz}
  et~al.}{1996}]{1996MNRAS.282.1437S}
{Shahbaz} T.,  {Smale} A.~P.,  {Naylor} T.,  {Charles} P.~A.,  {van Paradijs}
  J.,  {Hassall} B.~J.~M.,   {Callanan} P.,  1996, \mn@doi [\mnras]
  {10.1093/mnras/282.4.1437}, \href
  {http://adsabs.harvard.edu/abs/1996MNRAS.282.1437S} {282, 1437}

\bibitem[\protect\citeauthoryear{{Shahbaz}, {Watson}  \& {Dhillon}}{{Shahbaz}
  et~al.}{2014}]{2014MNRAS.440..504S}
{Shahbaz} T.,  {Watson} C.~A.,   {Dhillon} V.~S.,  2014, \mn@doi [\mnras]
  {10.1093/mnras/stu267}, \href
  {http://adsabs.harvard.edu/abs/2014MNRAS.440..504S} {440, 504}

\bibitem[\protect\citeauthoryear{{Shih}, {Kundu}, {Maccarone}, {Zepf}  \&
  {Joseph}}{{Shih} et~al.}{2010}]{2010ApJ...721..323S}
{Shih} I.~C.,  {Kundu} A.,  {Maccarone} T.~J.,  {Zepf} S.~E.,   {Joseph} T.~D.,
   2010, \mn@doi [\apj] {10.1088/0004-637X/721/1/323}, \href
  {http://adsabs.harvard.edu/abs/2010ApJ...721..323S} {721, 323}

\bibitem[\protect\citeauthoryear{{Sigurdsson} \& {Hernquist}}{{Sigurdsson} \&
  {Hernquist}}{1993}]{1993Natur.364..423S}
{Sigurdsson} S.,  {Hernquist} L.,  1993, \mn@doi [\nat] {10.1038/364423a0},
  \href {http://adsabs.harvard.edu/abs/1993Natur.364..423S} {364, 423}

\bibitem[\protect\citeauthoryear{{Sippel} \& {Hurley}}{{Sippel} \&
  {Hurley}}{2013}]{2013MNRAS.430L..30S}
{Sippel} A.~C.,  {Hurley} J.~R.,  2013, \mn@doi [\mnras]
  {10.1093/mnrasl/sls044}, \href
  {http://adsabs.harvard.edu/abs/2013MNRAS.430L..30S} {430, L30}

\bibitem[\protect\citeauthoryear{{Steeghs} \& {Casares}}{{Steeghs} \&
  {Casares}}{2002}]{2002ApJ...568..273S}
{Steeghs} D.,  {Casares} J.,  2002, \mn@doi [\apj] {10.1086/339224}, \href
  {http://adsabs.harvard.edu/abs/2002ApJ...568..273S} {568, 273}

\bibitem[\protect\citeauthoryear{{Steele}, {Zepf}, {Maccarone}, {Kundu},
  {Rhode}  \& {Salzer}}{{Steele} et~al.}{2014}]{2014ApJ...785..147S}
{Steele} M.~M.,  {Zepf} S.~E.,  {Maccarone} T.~J.,  {Kundu} A.,  {Rhode} K.~L.,
    {Salzer} J.~J.,  2014, \mn@doi [\apj] {10.1088/0004-637X/785/2/147}, \href
  {http://adsabs.harvard.edu/abs/2014ApJ...785..147S} {785, 147}

\bibitem[\protect\citeauthoryear{{Stella}, {Priedhorsky}  \& {White}}{{Stella}
  et~al.}{1987}]{1987ApJ...312L..17S}
{Stella} L.,  {Priedhorsky} W.,   {White} N.~E.,  1987, \mn@doi [\apjl]
  {10.1086/184811}, \href {http://adsabs.harvard.edu/abs/1987ApJ...312L..17S}
  {312, L17}

\bibitem[\protect\citeauthoryear{{Strader}, {Chomiuk}, {Maccarone},
  {Miller-Jones}  \& {Seth}}{{Strader} et~al.}{2012}]{2012Natur.490...71S}
{Strader} J.,  {Chomiuk} L.,  {Maccarone} T.~J.,  {Miller-Jones} J.~C.~A.,
  {Seth} A.~C.,  2012, \mn@doi [\nat] {10.1038/nature11490}, \href
  {http://adsabs.harvard.edu/abs/2012Natur.490...71S} {490, 71}

\bibitem[\protect\citeauthoryear{{Tauris}, {Langer}  \& {Kramer}}{{Tauris}
  et~al.}{2012}]{2012MNRAS.425.1601T}
{Tauris} T.~M.,  {Langer} N.,   {Kramer} M.,  2012, \mn@doi [\mnras]
  {10.1111/j.1365-2966.2012.21446.x}, \href
  {http://adsabs.harvard.edu/abs/2012MNRAS.425.1601T} {425, 1601}

\bibitem[\protect\citeauthoryear{{Tauris}, {Langer}  \&
  {Podsiadlowski}}{{Tauris} et~al.}{2015}]{2015MNRAS.451.2123T}
{Tauris} T.~M.,  {Langer} N.,   {Podsiadlowski} P.,  2015, \mn@doi [\mnras]
  {10.1093/mnras/stv990}, \href
  {http://adsabs.harvard.edu/abs/2015MNRAS.451.2123T} {451, 2123}

\bibitem[\protect\citeauthoryear{{Tetarenko}, {Sivakoff}, {Heinke}  \&
  {Gladstone}}{{Tetarenko} et~al.}{2016}]{2016ApJS..222...15T}
{Tetarenko} B.~E.,  {Sivakoff} G.~R.,  {Heinke} C.~O.,   {Gladstone} J.~C.,
  2016, \mn@doi [\apjs] {10.3847/0067-0049/222/2/15}, \href
  {http://adsabs.harvard.edu/abs/2016ApJS..222...15T} {222, 15}

\bibitem[\protect\citeauthoryear{{Tody}}{{Tody}}{1993}]{1993ASPC...52..173T}
{Tody} D.,  1993, in {Hanisch} R.~J.,  {Brissenden} R.~J.~V.,   {Barnes} J.,
  eds,  Astronomical Society of the Pacific Conference Series Vol. 52,
  Astronomical Data Analysis Software and Systems II. p.~173

\bibitem[\protect\citeauthoryear{{Truss} \& {Done}}{{Truss} \&
  {Done}}{2006}]{2006MNRAS.368L..25T}
{Truss} M.,  {Done} C.,  2006, \mn@doi [\mnras]
  {10.1111/j.1745-3933.2006.00149.x}, \href
  {http://adsabs.harvard.edu/abs/2006MNRAS.368L..25T} {368, L25}

\bibitem[\protect\citeauthoryear{{Tsygankov}, {Mushtukov}, {Suleimanov}  \&
  {Poutanen}}{{Tsygankov} et~al.}{2016}]{2016MNRAS.457.1101T}
{Tsygankov} S.~S.,  {Mushtukov} A.~A.,  {Suleimanov} V.~F.,   {Poutanen} J.,
  2016, \mn@doi [\mnras] {10.1093/mnras/stw046}, \href
  {http://ukads.nottingham.ac.uk/abs/2016MNRAS.457.1101T} {457, 1101}

\bibitem[\protect\citeauthoryear{{Vanderlinde}, {Levine}  \&
  {Rappaport}}{{Vanderlinde} et~al.}{2003}]{2003PASP..115..739V}
{Vanderlinde} K.~W.,  {Levine} A.~M.,   {Rappaport} S.~A.,  2003, \mn@doi
  [\pasp] {10.1086/375388}, \href
  {http://adsabs.harvard.edu/abs/2003PASP..115..739V} {115, 739}

\bibitem[\protect\citeauthoryear{{Vanderplas}, {Connolly}, {Ivezi{\'c}}  \&
  {Gray}}{{Vanderplas} et~al.}{2012}]{astroML}
{Vanderplas} J.,  {Connolly} A.,  {Ivezi{\'c}} {\v Z}.,   {Gray} A.,  2012, in
  Conference on Intelligent Data Understanding (CIDU). pp 47 --54,
  \mn@doi{10.1109/CIDU.2012.6382200}

\bibitem[\protect\citeauthoryear{{Verbunt}}{{Verbunt}}{1987}]{1987ApJ...312L..23V}
{Verbunt} F.,  1987, \mn@doi [\apjl] {10.1086/184812}, \href
  {http://adsabs.harvard.edu/abs/1987ApJ...312L..23V} {312, L23}

\bibitem[\protect\citeauthoryear{{Verbunt}}{{Verbunt}}{1993}]{1993ARA&A..31...93V}
{Verbunt} F.,  1993, \mn@doi [\araa] {10.1146/annurev.aa.31.090193.000521},
  \href {http://adsabs.harvard.edu/abs/1993ARA%26A..31...93V} {31, 93}

\bibitem[\protect\citeauthoryear{{Verbunt} \& {Rappaport}}{{Verbunt} \&
  {Rappaport}}{1988}]{1988ApJ...332..193V}
{Verbunt} F.,  {Rappaport} S.,  1988, \mn@doi [\apj] {10.1086/166645}, \href
  {http://adsabs.harvard.edu/abs/1988ApJ...332..193V} {332, 193}

\bibitem[\protect\citeauthoryear{{Verbunt}, {Wijers}  \& {Burm}}{{Verbunt}
  et~al.}{1990}]{1990A&A...234..195V}
{Verbunt} F.,  {Wijers} R.~A.~M.~J.,   {Burm} H.~M.~G.,  1990, \aap, \href
  {http://adsabs.harvard.edu/abs/1990A%26A...234..195V} {234, 195}

\bibitem[\protect\citeauthoryear{{Verbunt}, {Bunk}, {Hasinger}  \&
  {Johnston}}{{Verbunt} et~al.}{1995}]{1995A&A...300..732V}
{Verbunt} F.,  {Bunk} W.,  {Hasinger} G.,   {Johnston} H.~M.,  1995, \aap,
  \href {http://adsabs.harvard.edu/abs/1995A%26A...300..732V} {300, 732}

\bibitem[\protect\citeauthoryear{{Vrtilek}, {Raymond}, {Garcia}, {Verbunt},
  {Hasinger}  \& {Kurster}}{{Vrtilek} et~al.}{1990}]{1990A&A...235..162V}
{Vrtilek} S.~D.,  {Raymond} J.~C.,  {Garcia} M.~R.,  {Verbunt} F.,  {Hasinger}
  G.,   {Kurster} M.,  1990, \aap, \href
  {http://adsabs.harvard.edu/abs/1990A%26A...235..162V} {235, 162}

\bibitem[\protect\citeauthoryear{{Wang} \& {Chakrabarty}}{{Wang} \&
  {Chakrabarty}}{2010}]{2010ApJ...712..653W}
{Wang} Z.,  {Chakrabarty} D.,  2010, \mn@doi [\apj]
  {10.1088/0004-637X/712/1/653}, \href
  {http://adsabs.harvard.edu/abs/2010ApJ...712..653W} {712, 653}

\bibitem[\protect\citeauthoryear{{Werner}, {Nagel}, {Rauch}, {Hammer}  \&
  {Dreizler}}{{Werner} et~al.}{2006}]{2006A&A...450..725W}
{Werner} K.,  {Nagel} T.,  {Rauch} T.,  {Hammer} N.~J.,   {Dreizler} S.,  2006,
  \mn@doi [\aap] {10.1051/0004-6361:20053768}, \href
  {http://adsabs.harvard.edu/abs/2006A%26A...450..725W} {450, 725}

\bibitem[\protect\citeauthoryear{{White} \& {Holt}}{{White} \&
  {Holt}}{1982}]{1982ApJ...257..318W}
{White} N.~E.,  {Holt} S.~S.,  1982, \mn@doi [\apj] {10.1086/159991}, \href
  {http://adsabs.harvard.edu/abs/1982ApJ...257..318W} {257, 318}

\bibitem[\protect\citeauthoryear{{White} \& {Swank}}{{White} \&
  {Swank}}{1982}]{1982ApJ...253L..61W}
{White} N.~E.,  {Swank} J.~H.,  1982, \mn@doi [\apjl] {10.1086/183737}, \href
  {http://adsabs.harvard.edu/abs/1982ApJ...253L..61W} {253, L61}

\bibitem[\protect\citeauthoryear{{Whitehurst}}{{Whitehurst}}{1988}]{1988MNRAS.232...35W}
{Whitehurst} R.,  1988, \mn@doi [\mnras] {10.1093/mnras/232.1.35}, \href
  {http://adsabs.harvard.edu/abs/1988MNRAS.232...35W} {232, 35}

\bibitem[\protect\citeauthoryear{{Whitehurst} \& {King}}{{Whitehurst} \&
  {King}}{1991}]{1991MNRAS.249...25W}
{Whitehurst} R.,  {King} A.,  1991, \mn@doi [\mnras] {10.1093/mnras/249.1.25},
  \href {http://adsabs.harvard.edu/abs/1991MNRAS.249...25W} {249, 25}

\bibitem[\protect\citeauthoryear{{Wijers} \& {Pringle}}{{Wijers} \&
  {Pringle}}{1999}]{1999MNRAS.308..207W}
{Wijers} R.~A.~M.~J.,  {Pringle} J.~E.,  1999, \mn@doi [\mnras]
  {10.1046/j.1365-8711.1999.02720.x}, \href
  {http://adsabs.harvard.edu/abs/1999MNRAS.308..207W} {308, 207}

\bibitem[\protect\citeauthoryear{{Wood}, {Thomas}  \& {Simpson}}{{Wood}
  et~al.}{2009}]{2009MNRAS.398.2110W}
{Wood} M.~A.,  {Thomas} D.~M.,   {Simpson} J.~C.,  2009, \mn@doi [\mnras]
  {10.1111/j.1365-2966.2009.15252.x}, \href
  {http://adsabs.harvard.edu/abs/2009MNRAS.398.2110W} {398, 2110}

\bibitem[\protect\citeauthoryear{{Woodgate} et~al.,}{{Woodgate}
  et~al.}{1998}]{1998PASP..110.1183W}
{Woodgate} B.~E.,  et~al., 1998, \mn@doi [\pasp] {10.1086/316243}, \href
  {http://adsabs.harvard.edu/abs/1998PASP..110.1183W} {110, 1183}

\bibitem[\protect\citeauthoryear{{Zdziarski}, {Wen}  \&
  {Gierli{\'n}ski}}{{Zdziarski} et~al.}{2007}]{2007MNRAS.377.1006Z}
{Zdziarski} A.~A.,  {Wen} L.,   {Gierli{\'n}ski} M.,  2007, \mn@doi [\mnras]
  {10.1111/j.1365-2966.2007.11686.x}, \href
  {http://adsabs.harvard.edu/abs/2007MNRAS.377.1006Z} {377, 1006}

\bibitem[\protect\citeauthoryear{{Zdziarski}, {Kawabata}  \&
  {Mineshige}}{{Zdziarski} et~al.}{2009}]{2009MNRAS.399.1633Z}
{Zdziarski} A.~A.,  {Kawabata} R.,   {Mineshige} S.,  2009, \mn@doi [\mnras]
  {10.1111/j.1365-2966.2009.15386.x}, \href
  {http://adsabs.harvard.edu/abs/2009MNRAS.399.1633Z} {399, 1633}

\bibitem[\protect\citeauthoryear{{Zechmeister} \& {K{\"u}rster}}{{Zechmeister}
  \& {K{\"u}rster}}{2009}]{2009A&A...496..577Z}
{Zechmeister} M.,  {K{\"u}rster} M.,  2009, \mn@doi [\aap]
  {10.1051/0004-6361:200811296}, \href
  {http://adsabs.harvard.edu/abs/2009A%26A...496..577Z} {496, 577}

\bibitem[\protect\citeauthoryear{{Zepf}, {Maccarone}, {Bergond}, {Kundu},
  {Rhode}  \& {Salzer}}{{Zepf} et~al.}{2007}]{2007ApJ...669L..69Z}
{Zepf} S.~E.,  {Maccarone} T.~J.,  {Bergond} G.,  {Kundu} A.,  {Rhode} K.~L.,
  {Salzer} J.~J.,  2007, \mn@doi [\apjl] {10.1086/523797}, \href
  {http://adsabs.harvard.edu/abs/2007ApJ...669L..69Z} {669, L69}

\bibitem[\protect\citeauthoryear{{Zepf} et~al.,}{{Zepf}
  et~al.}{2008}]{2008ApJ...683L.139Z}
{Zepf} S.~E.,  et~al., 2008, \mn@doi [\apjl] {10.1086/591937}, \href
  {http://adsabs.harvard.edu/abs/2008ApJ...683L.139Z} {683, L139}

\bibitem[\protect\citeauthoryear{{de Jong}, {van Paradijs}  \&
  {Augusteijn}}{{de Jong} et~al.}{1996}]{1996A&A...314..484D}
{de Jong} J.~A.,  {van Paradijs} J.,   {Augusteijn} T.,  1996, \aap, \href
  {http://adsabs.harvard.edu/abs/1996A%26A...314..484D} {314, 484}

\bibitem[\protect\citeauthoryear{{in't Zand}, {Cumming}, {van der Sluys},
  {Verbunt}  \& {Pols}}{{in't Zand} et~al.}{2005}]{2005A&A...441..675I}
{in't Zand} J.~J.~M.,  {Cumming} A.,  {van der Sluys} M.~V.,  {Verbunt} F.,
  {Pols} O.~R.,  2005, \mn@doi [\aap] {10.1051/0004-6361:20053002}, \href
  {http://adsabs.harvard.edu/abs/2005A%26A...441..675I} {441, 675}

\bibitem[\protect\citeauthoryear{{in't Zand}, {Jonker}  \& {Markwardt}}{{in't
  Zand} et~al.}{2007}]{2007A&A...465..953I}
{in't Zand} J.~J.~M.,  {Jonker} P.~G.,   {Markwardt} C.~B.,  2007, \mn@doi
  [\aap] {10.1051/0004-6361:20066678}, \href
  {http://adsabs.harvard.edu/abs/2007A%26A...465..953I} {465, 953}

\bibitem[\protect\citeauthoryear{{van Haaften}, {Nelemans}, {Voss}, {Wood}  \&
  {Kuijpers}}{{van Haaften} et~al.}{2012}]{2012A&A...537A.104V}
{van Haaften} L.~M.,  {Nelemans} G.,  {Voss} R.,  {Wood} M.~A.,   {Kuijpers}
  J.,  2012, \mn@doi [\aap] {10.1051/0004-6361/201117880}, \href
  {http://adsabs.harvard.edu/abs/2012A%26A...537A.104V} {537, A104}

\bibitem[\protect\citeauthoryear{{van Haaften}, {Nelemans}, {Voss}, {Toonen},
  {Portegies Zwart}, {Yungelson}  \& {van der Sluys}}{{van Haaften}
  et~al.}{2013}]{2013A&A...552A..69V}
{van Haaften} L.~M.,  {Nelemans} G.,  {Voss} R.,  {Toonen} S.,  {Portegies
  Zwart} S.~F.,  {Yungelson} L.~R.,   {van der Sluys} M.~V.,  2013, \mn@doi
  [\aap] {10.1051/0004-6361/201220552}, \href
  {http://adsabs.harvard.edu/abs/2013A%26A...552A..69V} {552, A69}

\bibitem[\protect\citeauthoryear{{van Paradijs} \& {McClintock}}{{van Paradijs}
  \& {McClintock}}{1994}]{1994A&A...290..133V}
{van Paradijs} J.,  {McClintock} J.~E.,  1994, \aap, \href
  {http://adsabs.harvard.edu/abs/1994A%26A...290..133V} {290, 133}

\makeatother
\end{thebibliography}


\appendix
\section{Potential mechanisms behind the superorbital modulation}
\label{sec:mechs}

\subsection{Superhumps}

In cataclysmic variables, the accretion disc is often tidally unstable, becoming eccentric and precessing. The periodically variable tidal dissipation in the outer parts of the disc gives rise to light curve modulations known as superhumps, which have time-scales slightly longer (by a few percent) than the orbital period \citep{1988MNRAS.232...35W, 1990PASJ...42..135H}.
Similar to cataclysmic variable systems, the 28.2 min period could be related to the superhump mechanism (as already noted by \citealp{2017MNRAS.467.2199B}), and the 27.2 min period could be the orbital period. Superhumps have also been proposed for a number of X-ray binaries like 4U 1820--30 \citep{2010ApJ...712..653W} and XB\,1916--053 \citep{2002MNRAS.330L..37R}, both ultra-compacts. Calculating the period excess as is typically done for superhumps, $\epsilon = (P_{\rm sh} - P_{\rm o}) / P_{\rm o}$, we obtain $\epsilon \approx 0.04$. Taking the calibration of \citet{2005PASP..117.1204P} for superhumps in cataclysmic variables, $\epsilon = 0.18q + 0.29q^2$, we estimate $q = 0.17$ for X9 (if the same relationship holds for all mass ratios in X-ray binaries), which is much larger than the expected $q < 0.01$. If the donor mass is $M_2 \approx 0.02\,M_\odot$, $q = 0.17$ gives $M_1 \approx 0.1\,M_\odot$, much lower than expected for a neutron star. Instead, the orbital period could be longer than the superhump period (negative superhumps). In this case, observations \citep{2002MNRAS.330L..37R, 2009MNRAS.399..465O} and simulations \citep{2009MNRAS.398.2110W} show that for very small $q$ values ($q \lesssim 0.1$), the negative and positive superhumps should have similar values. We therefore rule out the 27.2 min signal as a classical superhump.

Superhumps typically seen in cataclysmic variables are due to the 3:1 resonance (for each orbit of the secondary, the material in the disc orbits three times), and are encountered in systems with mass ratio $q \lesssim 0.25$. For more extreme mass ratios ($q \lesssim 0.025$), the 2:1 resonance can be reached, because the disc can be larger (as a fraction of the orbital separation; \citealp{1991MNRAS.249...25W}).

In quiescent systems with mass ratios $q \lesssim 0.25$, accretion discs are thought to extend up to the 3:1 resonance radius \citep{1990PASJ...42..135H}. During outbursts, however, the outer disc can expand to reach the tidal truncation radius \citep{1977ApJ...216..822P, 1977MNRAS.181..441P, 1988MNRAS.232...35W}. In addition, the discs of systems with extreme mass ratios ($q \lesssim 0.01$) in quiescence are likely to occupy a large fraction of the Roche lobe (extending beyond the 2:1 resonance radius), while still being constrained by it \citep{1988ApJ...333..895P, 2012A&A...537A.104V}.

In X9, for a donor mass $M_2 = 0.010 - 0.016\,M_\odot$, the value $P_{\rm long}/P_{\rm orb} = 357$ requires $M_1 = 1.8 - 2.8\,M_\odot$ if the long period is caused by precession of the 2:1 resonance radius \citep{2003MNRAS.346L..21P}. The beat (long) period between the superhump and orbital periods is not usually observed in cataclysmic variables. The large amplitude of the long-period modulation in X-rays is therefore surprising, but could arise in some circumstances. For example, an elliptical disc with a given aspect ratio $H/R$ will reach maximum thickness at its apoapsis, and minimum thickness at periapsis \citep{1982ApJ...257..318W, 1985A&A...144..369O}. If the X-ray emitting corona is of similar size to the maximum disc thickness, the fraction occulted by the edge of the disc will vary sinusoidally on the precession period.

\subsection{Other tidal effects}

To check if the superorbital modulation could be caused by nodal precession from tidal forces, we use the relationship of \citet[Equation~5]{1998MNRAS.299L..32L}, between the ratio of orbital to precession periods, and system parameters:
\begin{equation}
\frac{P_{\rm orb}}{P_{\rm long}} = \frac{3}{7} \beta^{3/2} \frac{q \: r_{\rm L}^{3/2} \cos \delta}{(1+q)^{1/2}},
\end{equation}
where $P_{\rm long}$ is the precession period, $\beta$ is the disc size as a fraction of the primary's Roche lobe, $\delta$ is the tilt between the planes of the disc and the orbit. The minimum precession period occurs when the inner and outer disc are aligned ($\delta = 0$), for the maximum mass ratio ($q = 0.016/1.4$). We find that for $P = 28.2$\,min and $\beta = 0.9$ \citep{2012A&A...537A.104V}, the minimum nodal precession period is $P_{\rm long} \gtrsim 7.8$\,d, only slightly larger than the measured 7\,d period. This suggests that tidal precession could be the source of the superorbital modulation if the primary is a neutron star (similar to the 2:1 superhump interpretation above). The requirement of a small misalignment angle, and the large amplitude of this modulation, suggest a nearly edge-on view of the system, as suggested also by the C\,{\sc iv} doublet in the case of a neutron star.

\citet{1999MNRAS.308..207W} provide descriptions for forced precession (by tidal forces from the donor star) and for radiatively-driven precession. Their conclusion is that the superorbital periods observed in X-ray binaries can be explained by radiatively driven precession (with a factor of 2 scatter between the predicted and observed periods, owing to unknowns in system parameters). In the case of X9, however, the low mass ratio ($q < 0.02$) and low X-ray luminosity (relative to persistent and soft states) suggest tidal disc precession is more likely \citep{2001MNRAS.320..485O, 2012MNRAS.420.1575K}. As described by \citet{2001MNRAS.320..485O}, radiation-driven warping can only be important in soft X-ray transients during outbursts. Using the formulations of \citet{1999MNRAS.308..207W}, the period associated with forced tidal precession respectively is given by:
\begin{equation}
\left( \frac{P_{\rm long}}{1\rm~d} \right) = 7.64\times 10^{-6} \pi \left(\frac{P_{\rm orb}}{1 {\rm\,min}}\right)^2 \left(\frac{1+q}{q}\right) M_{1.4}^{1/2} R_{11}^{-3/2},
\end{equation}
where $M_{1.4}$ is the primary mass in units of $1.4\,M_\odot$, and $R_{11}$ is the disc radius in units of $10^{11}$\,cm. We find $P_{\rm long} \approx 190$ days for a black hole and $P_{\rm tidal} = 40$ days for a neutron star.

Summarizing, we find the tidal precession model of \citet{1998MNRAS.299L..32L} may be able to explain the superorbital period in X9 for an edge-on neutron star.

\subsection{Lense-Thirring precession}

In the vicinity of neutron stars and black holes, accretion discs experience general relativistic frame-dragging effects. In particular, the Lense--Thirring effect \citep{LT18} is the nodal precession of an accretion disc due to the spin of the accreting object. When the orbital plane of an X-ray binary is tilted with respect to the spin axis of an accreting compact object (as would be expected in a dynamically-formed system), the accretion disc can warp, so that the outer disc lies in the orbital plane, but the inner disc is aligned with the spin axis of the compact object \citep{1975ApJ...195L..65B}. When this occurs, the Lense-Thirring precession of the inner disc may be observed through modulations in the X-ray light curve of the binary.

For a disc that precesses like a solid body, the Lense--Thirring precession is given by \citet[their Equation 43]{2007ApJ...668..417F}. For constant surface density, and $R_{\rm out} \gg R_{\rm in}$:
\begin{equation}
\label{eq:LT}
P_{\rm long} = \frac{\pi G M_1}{5 c^3} \frac{r_{\rm out}^{\frac{5}{2}} r_{\rm in}^{\frac{1}{2}}}{a},
\end{equation}
where in this case $a$ is the dimensionless spin of the black hole, $r_{\rm out}$ and $r_{\rm in}$ are the outer and inner disc radii in units of gravitational radii. Taking a slow-spinning $a=0.1$, $8\,M_\odot$ black hole, and $r_{\rm in} = 500\,r_{\rm g}$, Equation~{\ref{eq:LT} requires $r_{\rm out} \approx 1600$\,$r_{\rm g}$ for the precession period to match the observed $P_{\rm long}=$ 7 days. The transition between the the inner, precessing disc, and outer disc (the Bardeen--Petterson radius) could occur at a radius of up to a few hundred gravitational radii \citep{1975ApJ...195L..65B, 2000MNRAS.315..570N, 2001ApJ...553..955F}. We find, however, that the precessing disc in our test case needs to extend beyond $10^3$\,$r_{\rm g}$, where the misalignment between the disc plane and the orbital plane is likely to be very small (a few degrees), regardless of the spin axis of the black hole. Other warping mechanisms or a neutron star are equally unlikely to provide a larger tilt; low-mass X-ray binaries are only prone to radiation warping during outbursts \citep{2001MNRAS.320..485O}. For the X-ray superorbital period to have the large amplitude that is observed, the disc would need to be seen nearly edge-on. As discussed in Section~\ref{sec:civ}, however, the double-peak separation of the C\,{\sc iv} doublet is incompatible with a black hole system seen edge-on.

\subsection{Hierarchical triple}

A hierarchical triple system can drive the eccentricity and inclination of the inner binary to oscillate via the Kozai mechanism \citep{1962AJ.....67..591K}. This mechanism is thought to be the main channel for creating black hole -- white dwarf binaries in globular clusters \citep{2010ApJ...717..948I} and can explain the superorbital modulations of some ultra-compact X-ray binaries \citep{2007MNRAS.377.1006Z, 2010MNRAS.409L..84M}. The Kozai mechanism could explain the difference between the short optical and X-ray periods, and the superorbital modulation. 

In the core of 47~Tuc ($\sigma = 11$\,km\,s$^{-1}$), double neutron star binaries with semi-major axes smaller than $\approx 20$ A.U. ($P_{\rm orb} < 50$ years; angular separation $<0.01$\arcsec) are stable against dynamical disruption by neighbouring cluster stars \citep{1984AJ.....89.1811H}, but are susceptible to eccentricities from fly-by encounters. Low mass main-sequence stars are less likely as a third body because they require shorter periods, making the triple less stable. The maximum mass for a main-sequence star that can fit in within the broadband spectrum of X9 is $M_3 \approx 0.5\,M_\odot$, but the third companion could instead be a heavy white dwarf or a neutron star. This scenario is not unlikely; \citet{2011MNRAS.410.2698G} find that half of the mass in the core of 47~Tuc is likely to be made up of white dwarfs. According to the globular cluster simulations of \citet[see their Table 1]{2008msah.conf..101I}, the fractions of hierarchical triples with main sequence stars and white dwarfs as outer companions are 79\% and 20\% respectively. Theoretically, $\approx$5\% of all binaries in the core of a globular cluster as dense as 47 Tuc are expected to form stable hierarchical triples in one Gyr \citep{2008msah.conf..101I}.

For XB\,1916--053, such a system was invoked to explain a 1\% difference between the optical and X-ray periods (50\,min) seen in conjunction with a 199\,day superorbital period\footnote{It was later shown that negative superhumps could explain the optical--X-ray period excess in XB\,1916--053, without the need to invoke a third body \citep{2002MNRAS.330L..37R}.}\citep{1988ApJ...334L..25G, 2001ApJ...549.1135C}. In that case, the beat between the optical and X-ray periods is the period of the outer companion. In the case of X9, the outer companion would have a period of $P_3 \approx 13$ hours (the beat between the optical 27.2 min and X-ray 28.2 min). For $M_2 \ll M_1$, and a nearly circular inner orbit, the Kozai period is approximately $P_{\rm long} \approx P_3^2 (M_1 + M_3) M_3^{-1} P_1^{-1}$ \citep{1979A&A....77..145M, 2000ApJ...535..385F}. For $P_1 = 28.2$ min and $P_3 = 13$ hours, we obtain a lower limit for the Kozai period of $P_{\rm long} \gtrsim 15$ days (for $M_3 \gg M_1$), in contrast to the found period of $P_{\rm long} \approx 7$ days. This argues against the Kozai mechanism as the source of period excess between the optical and X-ray periods, and of the superorbital period. Disregarding the short FUV period (which has low significance), the Kozai mechanism can still explain the superorbital period as a variation of the inclination angle of the inner binary (and thus of the disc). Again, this scenario requires an edge-on view of the system, making a neutron star more likely.

\bsp	
\label{lastpage}
\end{document}